\documentclass[12pt]{article}
\usepackage{amsfonts}
\usepackage{amssymb}
\usepackage{graphicx}
\usepackage{amsmath}
\usepackage{makeidx}
\usepackage{indentfirst}
\usepackage[T1]{fontenc}
\usepackage[utf8]{inputenc}

\newcounter{resultnum}[section]
\newcounter{conclusionnum}[section]
\newcounter{conditionnum}[section]
\newcounter{conjecturenum}[section]
\newcounter{examplenum}[section]
\newcounter{exercisenum}[section]
\newcounter{lemmanum}[section]
\newcounter{notationnum}[section]
\newcounter{theoremnum}[section]
\newcounter{definitionnum}[section]
\newcounter{corollarynum}[section]
\newcounter{remarknum}[section]
\newcounter{propositionnum}[section]
\newcounter{acknowledgementnum}[section]
\newcounter{algorithmnum}[section]
\newcounter{axiomnum}[section]
\newcounter{casenum}[section]
\newcounter{claimnum}[section]
\newcounter{summarynum}[section]
\newcounter{problemnum}[section]
\begin{document}

\title{Black Holes with MDRs and Bekenstein-Hawking and Perelman Entropies
for Finsler-Lagrange-Hamilton Spaces\\
\vspace{.1 in}}
\date{February 19, 2019}
\author{ \vspace{.1 in}{\ \textbf{Lauren\c{t}iu Bubuianu}}\thanks{%
email: laurentiu.bubuianu@tvr.ro } \\
%EndAName
{\small \textit{TVR Ia\c{s}i, \ 33 Lasc\v{a}r Catargi street; and University
Apollonia, 2 Muzicii street, 700107 Ia\c{s}i, Romania }}\\
${}$ \\
\vspace{0.1in} \textbf{Sergiu I. Vacaru}\thanks{%
emails: sergiu.vacaru@gmail.com and sergiuvacaru@mail.fresnostate.edu ;
\newline
\textit{Address for post correspondence:\ } 140 Morehampton Rd, Donnybrook,
Dublin 4, Ireland, D04 N2C0}\\
{\small \textit{Physics Department, California State University Fresno,
Fresno, CA 93740, USA}} \\
{\small and } \ {\small \textit{\ Project IDEI, University "Al. I. Cuza"
Ia\c si, Romania}} }
\maketitle

\begin{abstract}
New geometric and analytic methods for generating exact and parametric
solutions in generalized Einstein-Finsler like gravity theories and
nonholonomic Ricci soliton models are reviewed and developed. We show how
generalizations of the Schwarzschild - (anti) de Sitter metric can be
constructed for modified gravity theories with arbitrary modified dispersion
relations, MDRs, and Lorentz invariance violations, LIVs. Such theories can
be geometrized on cotangent Lorentz bundles (phase spaces) as models of
relativistic Finsler-Lagrange-Hamilton spaces. There are considered two
general classes of solutions for gravitational stationary vacuum phase space
configurations and nontrivial (effective) matter sources or cosmological
constants. Such solutions describe nonholonomic deformations of conventional
higher dimension black hole, BH, solutions with general dependence on
effective four dimensional, 4-d, momentum type variables. For the first
class, we study physical properties of Tangherlini like BHs in phase spaces
with generic dependence on an energy coordinate/ parameter. We investigate
also BH configurations on base spacetime and in curved cofiber spaces when
the BH mass and the maximal speed of light determine naturally a cofiber
horizon. For the second class, the solutions are constructed with Killing
symmetry on an energy type coordinate. There are analysed the conditions
when generalizations of Beckenstein-Hawking entropy (for solutions with
conventional horizons) and/or Grigory Perelman's W-entropy (for more general
generic off-diagonal solutions) can be defined for phase space stationary
configurations.

\vskip5pt


\vskip4pt


\end{abstract}

\tableofcontents

%\vskip10pt

\section{ Introduction and preliminaries}

Geometric and phenomenological physical models with modified dispersion
relations, MDRs, are studied in quantum gravity, QG, and modified gravity
theories, MGTs, encoding global and/or local Lorentz invariance violations,
LIVs, noncommutative effects, string and generalized Finsler modifications
etc., see \cite%
{vap97,vnp97,amelino98,mavromatos11,mavromatos13a,kostelecky11,kostelecky16,capoz,nojod1,nojod2,castro07,magueijo04}
and references therein. Black holes, BHs, consist a most suitable and
important example which allows us to study possible physical implications
and impose physical constraints on MGTs. One may therefore use a series of
previous results on quantum corrections and brane-world black holes with
special classes of MDRs, for instance, with even powers of the energy
parameter) \cite{nozari,sefiedgar}. Nevertheless, it is not clear if BH
objects may exist for theories with general MDRs and LIVs and when nonlinear
dispersion relations are derived for QG and MGTs. In general, MDRs result in
a generic off-diagonal nonlinear phase space dynamics modelled on (co)
tangent Lorentz bundles, when the coefficients of (generalized) metrics and
connections depend additionally on velocity/ momentum variables. There are
necessary more advanced geometric, analytic and numeric methods for
constructing solutions of systems of nonlinear partial differential
equations, PDEs, related to modified Einstein equations and generalized
metrics and connections. Another important question is that if new classes
of stationary and quasi-stationary\footnote{%
such phase space solutions transforms into stationary ones for projections
on a base spacetime Lorentz manifold} solutions can be constructed, what
type of new physics is described, for instance, by phase space generalized
BH configurations? To characterize physical properties of such locally
anisotropic BH and cosmological models it would be enough to develop the
standard (Bekenstein-Hawking) entropy and BH thermodynamics \cite%
{bek1,bek2,haw1,haw2} or we have to elaborate on more general concepts of
geometric spacetime and phase space thermodynamics and kinetics (for
instance, Perelman type entropies \cite{perelman1,perelman2,perelman3} and
generalized Einstein-Vlasov systems \cite{vrfdif,vvrfthbh})? We argue that a
more general geometric and statistical thermodynamics interpretation of
general relativity, GR, and MGTs is possible by developing the concepts of
the so-called W- and F-entropies, see reviews of results and motivations in
\cite{ventrlf,vnrflnc,muen01,rajpsuprf}.

We develop the anholonomic frame deformation method, AFDM, for constructing
quasi-stationary solutions of generalized Einstein equations on nonholonomic
cotangent Lorentz bundles and associated models of Finsler-Lagrange-Hamilton
gravity and nonholonomic Ricci soliton configurations. For reviews on AFDM
and applications, we cite our partner works \cite{v18a,v18b,v18c} and
references therein. In this paper, we consider that readers are familiar
with basic concepts of mathematical relativity and differential geometry of
bundle spaces. Such non Riemanninan modified gravity theories can be
canonically constructed using lifts and nonholonomic deformations of
geometric objects considered in general relativity, GR, and Lorentz
spacetime manifolds.

Our purpose is to prove that generalized phase space static and stationary
Schwarz\-schild - de Sitter and/or vacuum like BH like solutions can be
constructed for MGTs with general MDRs characterized by an indicator
function $\varpi (x^{i},E,\overrightarrow{\mathbf{p}},m;\ell _{P})$.%
\footnote{%
On a cotangent bundle $T^{\ast }V$ on a four dimensional, 4-d, Lorentz
manifold $V$, the local coordinates are labeled $\ ^{\shortmid }u=(x,p)=\{\
^{\shortmid }u^{\alpha }=(x^{i},p_{a})\}.$ Such dual phase space coordinates
can be related via Legendre transforms to velocity-type
coordinates/variables on $TV,$ when $\ u=(x,v)=\{u^{\alpha
}=(x^{i},v^{a})\}. $ Unfortunately, we have to elaborate on very
sophisticate systems of labels for frames and coordinates if it is necessary
to construct physically important exact solutions of systems of nonlinear
partial differential equations with variables in (co) tangent bundles. This
imposes us to consider various classes of indices and geometric objects with
abstract left/right, tilde and hat labels and boldface symbols. Such systems
and conventions on notations were introduced in our previous partner works
\cite{v18a,v18b,v18c}. In this paper, the notations for geometric and
physical objects are used in an intuitive geometric form with labels on
symbols and indices adapted to conventional shell by shell 2+2 and 3+1
decompositions of spacetime and phase space dimensions.} In any point $%
x_{0}=\{x_{0}^{i};i=1,2,3,4\}\in V$ with $x^{4}=t$ being a time like
coordinate, we can consider
\begin{equation}
c^{2}\overrightarrow{\mathbf{p}}^{2}-E^{2}+c^{4}m^{2}=\varpi (x_{0}^{i},E,%
\overrightarrow{\mathbf{p}},m;\ell _{P}),  \label{mdrg}
\end{equation}%
where an \textbf{indicator} of modifications $\varpi (...)$ encodes possible
contributions of generalized physical theories with LIVs etc. We can always
chose a corresponding local system of coordinates both on base and co-fiber
coordinates when certain MDRs are parameterized in such a form. For quantum
theories, such modifications are proportional to the Planck length $\ell
_{P}.$

The motivation behind the study of BH solutions with MDRs seam from the fact
that in such conventional extra dimension gravity theories (in our case,
with additional conventional velocity/momentum type coordinates) it is
encoded a more rich information on nonlinear classical and quantum
interactions. Such models are with noncompact phase space (co) fiber
dimensions (for realistic physical models, one should be considered a
maximal speed of light with possible warped and trapped configurations) and
supposed to explain important scenarios in modern cosmology and particle
physics. An indicator $\varpi $ defines certain fundamental nonholonomic
(equivalently, anholonomic, i.e. non-integrable) geometric structures which
are typical for generalized relativistic models of Finsler-Lagrange-Hamilton
spaces.

In this work, the geometric and physical models are defined by three
fundamental and canonical geometric and physical objects on $T^{\ast }%
\mathbf{V}$: 1) the nonlinear connection, N-connection, structure $\
^{\shortmid }\widetilde{\mathbf{N}}[\varpi ]$ and respective N-adapted dual)
frame structure $\ ^{\shortmid }\widetilde{\mathbf{e}}_{\alpha}[\varpi ]$
and $\ ^{\shortmid }\widetilde{\mathbf{e}}^{\alpha }[\varpi ]$ as
functionals of $\varpi $; 2) the distinguished metric, d-metric, structure $%
\ ^{\shortmid }\widetilde{\mathbf{g}}_{\alpha \beta }[\varpi ]=\{\widetilde{g%
}_{jk}(x,p), \ ^{\shortmid }\widetilde{g}^{ab}(x,p)\},$ adapted to the
N-connection structure; and 3) the Cartan, $\widetilde{\mathbf{D}}[\varpi ],$
or canonical, $\widehat{\mathbf{D}}(x,p),$ d-connections, both uniquely
defined by distortions of the Levi-Civita, LC, which is a metric compatible
and torsionless linear connection, $\nabla ,$ see details in next section
and \cite{v18a,v18b,v18c}. We emphasize that we have to work with
generalized Finsler like geometric and physical objects of type 1)-3) and
not only with metric type geometries if MDRs (\ref{mdrg}) are considered in
a MGT. Up to frame/coordinate transforms, we can establish such an
equivalence of geometric data%
\begin{equation}
(\ ^{\shortmid }\mathbf{N};\ ^{\shortmid }\mathbf{e}_{\alpha },\ ^{\shortmid
}\mathbf{e}^{\alpha };\ \ ^{\shortmid }\mathbf{g}_{\alpha \beta
})\longleftrightarrow (\widetilde{H},\ ^{\shortmid }\widetilde{\mathbf{N}};\
^{\shortmid }\widetilde{\mathbf{e}}_{\alpha },\ ^{\shortmid }\widetilde{%
\mathbf{e}}^{\alpha };\ \ ^{\shortmid }\widetilde{g}^{ab},\ \ ^{\shortmid }%
\widetilde{g}_{ab};\widetilde{\mathbf{D}})\longleftrightarrow (\ \
_{s}^{\shortmid }\mathbf{N};\ \ ^{\shortmid }\mathbf{e}_{\alpha _{s}},\ \
^{\shortmid }\mathbf{e}^{\alpha _{s}};\ \ \ ^{\shortmid }\mathbf{g}_{\alpha
_{s}\beta _{s}};\ _{s}\widehat{\mathbf{D}}),  \notag
\end{equation}%
where tilde values are defined by canonical geometric objects and the left
low, or up, labels like '$\ ^{\shortmid }$' are used for geometric objects
determined by a Hamilton space with nondegenerate Hessian for a conventional
Hamiltonian $\widetilde{H}(x,p).$\footnote{%
For a MDR (\ref{mdrg}), we can introduce an effective Hamiltonian $%
H(p):=E=\pm (c^{2}\overrightarrow{\mathbf{p}}^{2}+c^{4}m^{2}-\varpi (E,%
\overrightarrow{\mathbf{p}},m;\ell _{P}))^{1/2}.$ In general, such values
depend also on base curves spacetime coordinates, for instance, for local
constructions in GR. We put "tilde" on a symbol in order to emphasize that
such a value is canonically defined by a generating function. Boldface
symbols are used for geometric/ physical objects which are
constructed/defined in a form adapted to a N-connection structure.} The left
label $s=1,2,3,4$ is used for a conventional nonholonomic diadic formalism
with (2+2)+(2+2) splitting of the 8-d total bundle space $\ _{s}T^{\ast }%
\mathbf{V}$ over a nonholonomic Lorentz manifold $\mathbf{V}$. Such a diadic
formalism and the canonical shell connection (s-connection, $\ _{s}\widehat{%
\mathbf{D}}$) allow a very general decoupling and integration via generating
functions, effective generating sources of $\ ^{\shortmid }\widehat{\Upsilon
}_{\alpha _{s}\beta _{s}},$ and integration functions of generalized
Einstein equations for MGTs with MDRs,%
\begin{equation}
\ ^{\shortmid }\widehat{\mathbf{R}}_{\alpha _{s}\beta _{s}}[\
_{s}^{\shortmid }\widehat{\mathbf{D}}]=\ ^{\shortmid }\widehat{\Upsilon }%
_{\alpha _{s}\beta _{s}}.  \label{meinsteqtbcand}
\end{equation}%
In this formula, $\ ^{\shortmid }\widehat{\mathbf{R}}_{\alpha _{s}\beta
_{s}} $ is the Ricci tensor for a shell adapted canonical d-connection $\
_{s}^{\shortmid }\widehat{\mathbf{D}}.$

Generalize Finsler like BH solutions have been studied in a series of our
works \cite{gvvepjc14,bubuianucqg17,vacaruepjc17}, see also and references
therein, on commutative and noncommutative, supersymmetric, fractional,
string / brane, nonholonomic and/or Finsler like generalizations of the GR
theory. For certain classes of nonholonomic deformations, such generic
off-diagonal solutions can define black ellipsoid / toroid / wormhole
configurations self-consistently embedded in higher dimension (super) spaces
with noncommutative and/or Finsler like variables \cite%
{v1998bh,v2001bh,v2002bh,v2003be1,v2003be2,v2007bh,v2010abh,v2010bbh,v2011fbr,v2013abe,v2013bbe, v2014abh,v2014bbh}%
. Similar constructions can be performed on tangent bundles $T\mathbf{V}$
endowed with structures of Lagrange spaces (Finsler spaces consisting some
particular homogeneous examples) and higher order and/or nonholonomic jet
and/or algebroid generalizations.

In this paper, there are two main goals:\ The first one is to prove that
stationary and quasi-stationary generalizations of the Schwarzschild metric
with extra dimensional momentum like variables can be constructed as exact
and parametric solutions of modified Einstein equations on $T^{\ast }\mathbf{%
V.}$ We consider that such a study of BHs with MDRs may bring about new
information on MGTs, QG and possible generalized uncertainty principle etc.
Former our results and the AFDM are reviewed in the directions of research
7, 9-12, 14, 18-20 discussed in Appendix B4 of \cite{v18b}. The second main
goal is to study how a statistical thermodynamic geometric approach can be
elaborated for stationary configurations and BH solutions on generalized
spacetimes and phase spaces. Here we emphasize that analogous constructions
involving modifications and generalizations of the Bekenstein-Hawking
definitions for BH entropy and spacetime thermodynamics \cite%
{bek1,bek2,haw1,haw2} (see, for instance, further developments and
alternative constructions in \cite%
{jacobson,padman,verlinde,yang,emparan,pappas16}) can be elaborated only for
very special subclasses to solutions. Such higher symmetry generalized
metrics are with conventional horizons when a corresponding hypersurface
area can be computed (for instance, black elipsoid/torus, holographic and/or
emergent configurations in phase spaces) for physical objects imbedded into
certain phase space backgrounds with high symmetry and flat space asymptotic
structure. In our previous works \cite%
{vrfijmpa,vacvis,valexiu,ventrlf,vejtp1,vejtp2,vnonsymmet,vnhrf,vnrflnc,vrfdif,bubvcjp,vrevrf,vvrfthbh, muen01,vmedit,vrajpootjet,rajpsuprf,vprocmg}
(see also a review of directions of research 10 and 17 in Appendix B4 of
\cite{v18b}), we proved that for generic off-diagonal exact solutions in GR
and MGTs with nonholonomic, noncommutative, supersymmetric, fractional and
Finsler like variables, we can elaborate a more general approach to the
geometric and statistical thermodynamics of gravitational fields using G.
Perelman's concepts of W- and F-entropy \cite{perelman1,perelman2,perelman3}%
. On mathematics of Ricci flows and certain physical applications, we cite
\cite%
{friedan1,friedan2,friedan3,hamilt1,hamilt2,hamilt3,monogrrf1,monogrrf2,monogrrf3,tsey}%
).

This work is organized as follow: In section \ref{s2}, we study generalized
vacuum phase space and Schwarzschild - de Sitter metrics with generic
dependence of coefficients on an energy type variable. There are constructed
exact and parametric solutions for BHs in energy depending phase
backgrounds. As typical examples, we construct and analyse some important
physical properties of nonholonomic deformations of Tangherlini's BHs \cite%
{tangherlini63} phase analogs which can be considered on (co) tangent
Lorentz bundles. There are studied also double BH metrics describing black
hole configurations both on a base spacetime and in a (co) fiber with
explicit energy dependence. Certain important nonlinear symmetries for BH
stationary phase space deformations are analysed. Section \ref{s3} is
devoted to exact solutions with energy type Killing symmetry describing
phase space stationary BHs. This defines another class of generalizations of
Tangherlini's solutions which depend generically on some momentum type
variables and posses an energy type fixed parameter. In general, all
stationary and BH phase space configurations are characterized by respective
relativistic Finsler-Lagrange-Hamilton symmetries as we show in section \ref%
{s4}. MDRs result in a a more rich non-Riemannian phase space geometry of
BHs which can be constructed as small parametric deformations of well known
higher dimension solutions \cite{tangherlini63,pappas16,emparan}. Such
metrics and other type BH, wormhole, solitonic and cosmological
configurations can be self-consistently imbedded in nontrivial vacuum and
nonvacuum backgrounds of MGTs with MDRs. Section \ref{s5} is devoted to a
research on the entropy of BHs in gravity theories with MDRs. We analyse and
develop on two definitions of gravitational entropy for stationary phase
configurations, BHs and generalized Ricci solitions (following
Bekenstein-Hawking and G. Perelman ideas). Concluding remarks are presented
in section \ref{s6}.

\section{BHs with dependence on an energy type variable}

\label{s2}

\subsection{Geometric preliminaries}

We follow the conventions for indices and coordinates of geometric objects
on $T\mathbf{V}$ and $T^{\ast }\mathbf{V}$ enabled with nonholonomic $%
(3+1)+(3+1)$ and/or diadic $(2+2)+(2+2)$ splitting in details in \cite{v18c}%
\footnote{\label{fncoordinatec}\ For local coordinates on a 8-d tangent
bundle $TV,$ we consider \ $u^{\alpha }=(x^{i},v^{a}),$ (or in brief, $%
u=(x,v)),$ when $i,j,k,...=1,2,3,4$ and $a,b,c,...=5,6,7,8;$ and for
cumulative indices $\alpha ,\beta ,...=1,2,...8.$ Similarly, on a cotangent
bundle $T^{\ast }V,$ we write $\ ^{\shortmid }u^{\alpha }=(x^{i},p_{a}),$
(in brief, $\ ^{\shortmid }u=(x,p)),$ where $x=\{x^{i}\}$ are considered as
coordinates for a base Lorentz manifold $V.$ The coordinate $x^{4}=t$ is
considered as time like one and $p_{8}=E$ is an energy type one. If
necessary, we shall work with 3+1 decompositions when, for instance, $x^{%
\grave{\imath}},$ for $\grave{\imath}=1,2,3,$ are used for space
coordinates; and $p_{\grave{a}},$ for $\grave{a}=5,6,7,$ are used for
momentum like coordinates.
\par
To construct generic off-diagonal exact solutions is useful to introduce
\textbf{diadic indices} with a conventional (2+2)+(2+2) splitting. Such
indices are labeled as follow: $\alpha _{1}=(i_{1}),$ $\alpha
_{2}=(i_{1},i_{2}=a_{2}),\beta _{2}=(j_{1},j_{2}=b_{2});\alpha
_{3}=(i_{3},a_{3}),\beta _{3}=(j_{3},b_{3}),...;\alpha
_{4}=(i_{4},a_{4}),\beta _{4}=(j_{4},b_{4}),$ for $i_{1},j_{1}=1,2;$ $%
i_{3},j_{3}=1,2,3,4;$ $i_{4},j_{4}=1,2,3,4,5,6;$ and $%
a_{2},b_{2}=3,4;a=(a_{3},a_{4}),b=(b_{3},b_{4}),$ for $a_{3},b_{3}=5,6$ and $%
a_{4},b_{4}=7,8,$ etc. A diadic splitting can be adapted to the splitting of
$h$-space into 2-d horizontal and vertical subspaces, $\ _{1}h$ and $\ \
_{2}v,$ of (co) vertical spaces $v$ into $\ ^{3}v$ and $\ ^{4}v$ into
conventional four 2-d shells labeled with left up or low abstract indices
like $\ ^{s}v,$ or $\alpha _{s}=(i_{s},a_{s})$ for $s=1,2,3,4$ referring to
ordered shells. We shall put shell labels on the left up, or left low, to
symbols for certain geometric objects if it will be necessary. Indices can
be contracted on corresponding shells using ordered diadic groups $\alpha
_{2}=(i_{1},a_{2})=1,2,3,4;\alpha _{3}=(\alpha
_{2},a_{3})=1,2,3,4,5,6;\alpha =\alpha _{4}=(\alpha _{3},a_{4})=1,2,...,8.$
In diadic form, the shell coordinates split in the form $%
x^{i}=(x^{i_{1}},y^{a_{2}}),v^{a}=(v^{a_{3}},v^{a_{4}});p_{a}=(p_{a_{3}},p_{a_{4}}).
$ We shall write also $\ _{s}u=\{$ $u^{\alpha _{s}}=(x^{i_{s}},v^{a_{s}})\}$
and $\ _{s}^{\shortmid }u=\{\ ^{\shortmid }u^{\alpha
_{s}}=(x^{i_{s}},p_{a_{s}})\}$ for cumulative indices on corresponding $s$%
-shell. For certain classes of solutions, we shall adapt that system of
notations in order to consider higher dimensional spherical variables.},
when boldface symbols are used for spaces and geometric/ physical objects
enabled with respective nonlinear connection, N-connection, structures $%
\mathbf{N}$ and $\ \ ^{\shortmid }\mathbf{N.}$ Here we note that the
N-connections are defined by certain Whitney sums $\oplus $ of conventional $%
h$- and $v$--distributions, or $h$ and $cv$--distributions, (for 4+4
splitting), when
\begin{equation}
\mathbf{N}:T\mathbf{TV}=hTV\oplus vTV\mbox{ or }\ \ ^{\shortmid }\mathbf{N}:T%
\mathbf{T}^{\ast }\mathbf{V}=hT^{\ast }V\oplus vT^{\ast }V.  \label{ncon}
\end{equation}%
Additionally, (2+2)+(2+2) splitting can be stated if diadic decompositions
are considered both for the base and fiber spaces when the N-connections are
defined and labelled by a low left label $s=1,2,3,4;$
\begin{eqnarray}
\mbox{ for a "shell" splitting } \ _{s}\mathbf{N}:\ _{s}T\mathbf{TV} &=&\
^{1}hTV\oplus \ ^{2}hTV\oplus \ ^{3}vTV\oplus \ ^{4}vTV\mbox{ and }  \notag
\\
\ _{s}^{\shortmid }\mathbf{N}:\ \ _{s}T\mathbf{T}^{\ast }\mathbf{V} &=&\
^{1}hT^{\ast }V\oplus \ ^{2}hT^{\ast }V\oplus \ ^{3}vT^{\ast }V\oplus \
^{4}vT^{\ast }V,  \label{ncon2}
\end{eqnarray}
Such diadic splitting are important for decoupling (modified) Einstein
equations and generating exact solutions in explicit form as we proved in
\cite{v18c,v18a,v18b}\footnote{\label{fnotnconcoef2} We consider splitting
into conventional 2-dim nonholonomic distributions of $TTV$ and $TT^{\ast }V$
if%
\begin{eqnarray*}
\dim (\ ^{1}hTV) &=&\dim (\ ^{2}hTV)=\dim (\ ^{3}vTV)=(\ ^{4}vTV)=2%
\mbox{
and } \\
\dim (\ ^{1}hT^{\ast }V) &=&\dim (\ ^{2}hT^{\ast }V)=\dim (\ ^{3}vT^{\ast
}V)=(\ ^{4}vT^{\ast }V)=2,
\end{eqnarray*}%
where left up labels like $\ ^{1}h,\ ^{3}v$ etc. state that [using
nonholonomic (equivalently, anholonomic and/or non-integrable)
distributions] the respective 8-d total spaces split into 2-d shells 1,2,3
and 4.
\par
We can consider local formulas for (\ref{ncon}) when $\mathbf{N}=N_{i}^{a}%
\frac{\partial }{\partial v^{a}}\otimes dx^{i}$ and/or $\ ^{\shortmid }%
\mathbf{N}=\ ^{\shortmid }N_{ia}\frac{\partial }{\partial p_{a}}\otimes
dx^{i},$ with respective N-connection coefficients $\mathbf{N}=\{N_{i}^{a}\}$
or $\ ^{\shortmid }\mathbf{N}=\{\ ^{\shortmid }N_{ia}\};$ the up label bar $%
\ "^{\shortmid }"$ will be used in order to emphasize that certain
geometric/physical objects are defined on cotangent bundles. For a
nonholonomic diadic splitting with N-connections (\ref{ncon2}) there are
used such parameterizations of coefficients:
\begin{eqnarray*}
\ _{s}\mathbf{N} &=&%
\{N_{i_{1}}^{a_{2}}(x^{i_{1}},y^{a_{2}}),N_{i_{1}}^{a_{3}}(x^{i_{1}},y^{a_{2}},v^{b_{3}}),N_{i_{1}}^{a_{4}}(x^{i_{1}},y^{a_{2}},v^{b_{3}},v^{b_{4}}),
\\
&&N_{i_{2}}^{a_{3}}(x^{i_{1}},y^{a_{2}},v^{b_{3}}),N_{i_{2}}^{a_{4}}(x^{i_{1}},y^{a_{2}},v^{b_{3}},v^{b_{4}}),N_{a_{3}}^{a_{4}}(x^{i_{1}},y^{a_{2}},v^{b_{3}},v^{b_{4}})\}%
\mbox{ and } \\
\ \ _{s}^{\shortmid }\mathbf{N} &=&\{\ ^{\shortmid
}N_{i_{1}}^{i_{2}}(x^{i_{1}},y^{a_{2}}),\ ^{\shortmid
}N_{i_{1}a_{3}}(x^{i_{1}},y^{a_{2}},p_{b_{3}}),\ ^{\shortmid
}N_{i_{1}a_{4}}(x^{i_{1}},y^{a_{2}},p_{b_{3}},p_{b_{4}}), \\
&&\ ^{\shortmid }N_{i_{2}a_{3}}(x^{i_{1}},y^{a_{2}},p_{b_{3}}),\ ^{\shortmid
}N_{i_{2}a_{4}}(x^{i_{1}},y^{a_{2}},p_{b_{3}},p_{b_{4}}),\ ^{\shortmid }N_{\
a_{4}}^{a_{3}}(x^{i_{1}},y^{a_{2}},p_{b_{3}},p_{b_{4}})\}.
\end{eqnarray*}%
}.

For a dual Lorentz bundle enabled with nonholonomic diadic structure, $_{s}%
\mathbf{T}^{\ast }\mathbf{V,}$ we can parameterize a prime\textbf{\ }metric $%
\ ^{\shortmid }\mathbf{\mathring{g}}$ in the form
\begin{eqnarray}
\ \ ^{\shortmid }\mathbf{\mathring{g}} &=&\ _{s}^{\shortmid }\mathbf{%
\mathring{g}}=\ ^{\shortmid }\mathring{g}_{\alpha _{s}\beta
_{s}}(x^{i_{s}},p_{a_{s}})d\ ^{\shortmid }u^{\alpha _{s}}\otimes d\
^{\shortmid }u^{\beta _{s}}=\ ^{\shortmid }\mathbf{\mathring{g}}_{\alpha
_{s}\beta _{s}}(\ _{s}^{\shortmid }u)\ ^{\shortmid }\mathbf{\mathbf{%
\mathring{e}}}^{\alpha _{s}}\mathbf{\otimes \ ^{\shortmid }\mathbf{\mathring{%
e}}}^{\beta _{s}}  \label{primedm} \\
&=&\ ^{\shortmid }\mathring{g}_{i_{s}j_{s}}(x^{k_{s}})e^{i_{s}}\otimes
e^{j_{s}}+\ ^{\shortmid }\mathbf{\mathring{g}}%
^{a_{s}b_{s}}(x^{i_{s}},p_{a_{s}})\ ^{\shortmid }\mathbf{\mathring{e}}%
_{a_{s}}\otimes \ ^{\shortmid }\mathbf{\mathring{e}}_{b_{s}},\mbox{ for }
\notag \\
\ ^{\shortmid }\mathbf{\mathring{e}}_{\alpha _{s}} &=&(\ ^{\shortmid }%
\mathbf{\mathring{e}}_{i_{s}}=\partial _{i_{s}}-\ ^{\shortmid }\mathring{N}%
_{i_{s}}^{b_{s}}(\ ^{\shortmid }u)\partial _{b_{s}},\ \ ^{\shortmid }{e}%
_{a_{s}}=\partial _{a_{s}})\mbox{ and }\ ^{\shortmid }\mathbf{\mathring{e}}%
^{\alpha _{s}}=(dx^{i_{s}},\mathbf{\mathring{e}}^{a_{s}}=dy^{a_{s}}+\
^{\shortmid }\mathring{N}_{i_{s}}^{a_{s}}(\ ^{\shortmid }u)dx^{i_{s}}).
\notag
\end{eqnarray}%
In our works, we label prime metrics and related geometric objects like
connections, frames etc. with a small circle on the left/right/up of
corresponding symbols. Prime metrics $\ _{s}\mathbf{\mathring{g}=\{\
^{\shortmid }\mathbf{\mathring{g}}_{\alpha _{s}\beta _{s}}(\ _{s}^{\shortmid
}u)\}}$ on $_{s}\mathbf{T}^{\ast }\mathbf{V}$ can be parameterized in
similar forms using respective coordinates and omitting the label \ "$%
^{\shortmid }$". A prime s-metric $\ ^{\shortmid }\mathbf{\mathring{g}}$ (%
\ref{primedm}) may be or not a solution of certain gravitational field
equations in a MGT or GR with phase space extension. For MDRs with a nonzero
$\varpi (x^{i},E,\overrightarrow{\mathbf{p}},m;\ell _{P})$ (\ref{mdrg}), we
obtain nonholonomic deformations to nonlinear quadratic elements determined
by \textbf{target} s-metrics of type $\ \ ^{\shortmid }\mathbf{g}=\
_{s}^{\shortmid }\mathbf{g.}$ A set of $\eta $-polarization, or
gravitational polarization, functions (here we omit the Einstein convention
on repeating indices because they are not 'up-low' type)
\begin{equation*}
\ _{s}^{\shortmid }\mathbf{\mathring{g}}\rightarrow \ _{s}^{\shortmid }%
\mathbf{g}=[\ ^{\shortmid }g_{\alpha _{s}}=\ ^{\shortmid }\eta _{\alpha
_{s}}\ ^{\shortmid }\mathring{g}_{\alpha _{s}},\ ^{\shortmid
}N_{i_{s-1}}^{a_{s}}=\ \ ^{\shortmid }\eta _{i_{s-1}}^{a_{s}}\ ^{\shortmid }%
\mathring{N}_{i_{s-1}}^{a_{s}}],
\end{equation*}%
are defined as transforms of a prime s-metric (\ref{primedm}) with
respective N-connection coefficients into certain classes of exact/
parametric solutions of generalized Einstein equations with MDRs (\ref%
{meinsteqtbcand}).

In N-adapted coefficient form, the nonholonomic transforms of prime to
target s-metrics in terms of $\eta $-polarizations can be parameterized
\begin{eqnarray}
\ \ _{s}^{\shortmid }\mathbf{\mathring{g}} &\rightarrow &\ _{s}^{\shortmid }%
\mathbf{g}=\ ^{\shortmid }g_{i_{s}}(x^{k_{s}})dx^{i_{s}}\otimes dx^{i_{s}}+\
^{\shortmid }g_{a_{s}}(x^{i_{s}},p_{b_{s}})\ ^{\shortmid }\mathbf{e}%
^{a_{s}}\otimes \ ^{\shortmid }\mathbf{e}^{a_{s}}  \label{dmpolariz} \\
&=&\ ^{\shortmid }\eta _{i_{k}}(x^{i_{1}},y^{a_{2}},p_{a_{3}},p_{a_{4}})\
^{\shortmid }\mathring{g}%
_{i_{s}}(x^{i_{1}},y^{a_{2}},p_{a_{3}},p_{a_{4}})dx^{i_{s}}\otimes dx^{i_{s}}
\notag \\
&&+\ ^{\shortmid }\eta _{b_{s}}(x^{i_{1}},y^{a_{2}},p_{a_{3}},p_{a_{4}})\
^{\shortmid }\mathring{g}_{b_{s}}(x^{i_{1}},y^{a_{2}},p_{a_{3}},p_{a_{4}})\
^{\shortmid }\mathbf{e}^{b_{s}}[\eta ]\otimes \ ^{\shortmid }\mathbf{e}%
^{b_{s}}[\eta ],  \notag \\
\ ^{\shortmid }\mathbf{e}^{\alpha _{s}}[\eta ] &=&(dx^{i_{s}},\ ^{\shortmid }%
\mathbf{e}^{a_{s}}=dy^{a_{s}}+\ ^{\shortmid }\eta
_{i_{s}}^{a_{s}}(x^{i_{1}},y^{a_{2}},p_{a_{3}},p_{a_{4}})\ ^{\shortmid }%
\mathring{N}%
_{i_{s}}^{a_{s}}(x^{i_{1}},y^{a_{2}},p_{a_{3}},p_{a_{4}})dx^{i_{s}}).  \notag
\end{eqnarray}%
The s-coefficients of metrics and N-elongated dual basis $\ ^{\shortmid }%
\mathbf{e}^{\alpha _{s}}[\eta ]$ used in (\ref{dmpolariz}) are respectively
defined by formulas
\begin{eqnarray}
\ ^{\shortmid }g_{i_{1}}(x^{k_{1}}) &=&\ ^{\shortmid }\eta
_{k_{1}}(x^{i_{1}},y^{a_{2}},p_{a_{3}},p_{a_{4}})\ ^{\shortmid }\mathring{g}%
_{k_{1}}(x^{i_{1}},y^{a_{2}},p_{a_{3}},p_{a_{4}}),  \notag \\
\ ^{\shortmid }g_{b_{2}}(x^{i_{1}},y^{3}) &=&\ ^{\shortmid }\eta
_{b_{2}}(x^{i_{1}},y^{a_{2}},p_{a_{3}},p_{a_{4}})\ ^{\shortmid }\mathring{g}%
_{b_{1}}(x^{i_{1}},y^{a_{2}},p_{a_{3}},p_{a_{4}}),  \notag \\
\ ^{\shortmid }g_{a_{3}}(x^{i_{2}},p_{6}) &=&\ ^{\shortmid }\eta
^{a_{3}}(x^{i_{1}},y^{b_{2}},p_{b_{3}},p_{b_{4}})\ ^{\shortmid }\mathring{g}%
^{a_{3}}(x^{i_{1}},y^{b_{2}},p_{b_{3}},p_{b_{4}}),  \notag \\
\ ^{\shortmid }g_{a_{4}}(x^{i_{3}},E) &=&\ ^{\shortmid }\eta
^{a_{4}}(x^{i_{1}},y^{b_{2}},p_{b_{3}},p_{b_{4}})\ ^{\shortmid }\mathring{g}%
^{a_{4}}(x^{i_{1}},y^{a_{2}},p_{a_{3}},p_{a_{4}})  \notag \\
\ ^{\shortmid }N_{i_{1}}^{a_{2}}(x^{k_{1}},y^{3}) &=&\eta
_{i_{1}}^{a_{2}}(x^{i_{1}},y^{b_{2}},p_{b_{3}},p_{b_{4}})\ ^{\shortmid }%
\mathring{N}_{i_{1}}^{a_{2}}(x^{i_{1}},y^{b_{2}},p_{b_{3}},p_{b_{4}}),
\notag \\
\ ^{\shortmid }N_{i_{2}}^{a_{3}}(x^{k_{1}},y^{b_{2}},p_{6}) &=&\ ^{\shortmid
}\eta _{i_{2}a_{3}}(x^{i_{1}},y^{b_{2}},p_{b_{3}},p_{b_{4}})\ ^{\shortmid }%
\mathring{N}_{i_{2}a_{3}}(x^{i_{1}},y^{b_{2}},p_{b_{3}},p_{b_{4}}),  \notag
\\
\ ^{\shortmid }N_{i_{3}}^{a_{4}}(x^{k_{1}},y^{b_{2}},p_{a_{3}},E) &=&\eta
_{i_{3}a_{4}}(x^{i_{1}},y^{b_{2}},p_{b_{3}},p_{b_{4}})\mathring{N}%
_{i_{3}a_{4}}(x^{i_{1}},y^{b_{2}},p_{b_{3}},p_{b_{4}}).  \label{coeftargpol}
\end{eqnarray}%
Multiples of type $\ ^{\shortmid }\eta \ ^{\shortmid }\mathring{g}$ in (\ref%
{coeftargpol}) may depend, in principle, on extra shell coordinates.
Nevertheless, such products are subjected to the condition that the target
s-metrics (with the coefficients in the left sides) are adapted to the shell
coordinates ordered form $s=1,2,3,4.$

Here we note that for any prescribed prime s-metric $\ _{s}^{\shortmid }%
\mathbf{\mathring{g}}$ we can consider as generating functions a subclass of
$\eta $-polarizations $\ ^{\shortmid }\eta _{4}(x^{i_{1}},y^{3}),\
^{\shortmid }\eta ^{5}(x^{i_{1}},y^{a_{1}},p_{6}),\ ^{\shortmid }\eta
^{7}(x^{i_{1}},y^{a_{1}},p_{a_{2}},E)$ which should be defined from the
condition that the target s-metric $\ _{s}^{\shortmid }\mathbf{g}$ is a
quasi-stationary solution of (\ref{meinsteqtbcand}). Variants of (3+1)
decompositions on the base space and in the typicall fiber can be
additionally prescribed if we wont to define conventional space and time
and, respectively, momentum and energy, variables in certain forms which
also adapted to corresponding N-connection structures. To generate BH
solutions, there will be considered also spherical coordinate systems of
respective 3-d, 5-d and 6-d space and phase spaces of Euclidean signature
and embedding such configurations in 8-d $T^{\ast }\mathbf{V.}$

\subsection{Prime and target metrics for nonholonomic deformations of BHs}

We consider on $T^{\ast }\mathbf{V}$ a prime metric $\ ^{\shortmid }\mathbf{%
\mathring{g}}=\{\ ^{\shortmid }\mathbf{\mathring{g}}_{\alpha _{s}\beta
_{s}}\}$ (\ref{primedm}) which in corresponding local coordinates describe
an embedding or it is an analogous of the Tangherlini BH solution for $4+m$
dimensional spacetimes $(m=1,2,...),$ see \cite{tangherlini63} and \cite%
{pappas16}. In futher sections a "math ring, or double ring" labels like $%
\mathring{g}$ and/or $^{\circ \circ }g$ \ will be used in order to
emphasized that certain classes of solutions involve a prime metric with
well defined and important physical properties like a BH solution, an
imbedding of such a lower dimensional metric, or certain well defined limits
of more general classes of solutions.

\subsubsection{Cotangent bundle analogs and imbedding of prime 6-d
Tangherlini solutions}

In this subsection, an extra dimensional index $m^{\prime }=5,6$ is used for
coordinates $p_{m}$ in cofiber spaces. We can multiply momenta to an
additional constant and use total space coordinates of the same dimension as
the base space coordinates, $[x^{i}]=[p_{a}].$ For spherical symmetry
coordinates on the third shell $s=3,$ the radial coordinate%
\begin{equation*}
\ ^{\shortmid }r=\sqrt{(x^{1})^{2}+(x^{2})^{2}+(y^{3})^{2}+(p_{5^{\prime
}})^{2}+(p_{6^{\prime }})^{2}}
\end{equation*}%
is defined for a 5-d phase subspace with signature $(+++++)$ and some
Cartezian coordinates $(x^{1},x^{2},y^{3},p_{5^{\prime }},p_{6^{\prime }}).$
The prime quadratic line element is parameterized
\begin{eqnarray}
ds^{2} &=&\ \ ^{\shortmid }\mathring{g}_{\alpha _{s}\beta
_{s}}(x^{i_{s}},p_{a_{s}})d\ ^{\shortmid }u^{\alpha _{s}}d\ ^{\shortmid
}u^{\beta _{s}}=\ ^{\shortmid }\mathbf{\mathring{g}}_{\alpha _{s}\beta
_{s}}(\ _{s}^{\shortmid }u)\ ^{\shortmid }\mathbf{\mathbf{\mathring{e}}}%
^{\alpha _{s}}\mathbf{\ ^{\shortmid }\mathbf{\mathring{e}}}^{\beta _{s}}
\notag \\
&=&h^{-1}(\ ^{\shortmid }r)d(\ ^{\shortmid }r)^{2}-h(\ ^{\shortmid
}r)dt^{2}+(\ ^{\shortmid }r)^{2}d\Omega _{4}^{2}+(dp_{7})^{2}-dE^{2}
\label{pmtang}
\end{eqnarray}%
In these formulas,
\begin{equation*}
h(\ ^{\shortmid }r)=1-\frac{\ ^{\shortmid }\mu }{(\ ^{\shortmid }r)^{3}}-%
\frac{\ ^{\shortmid }\kappa _{6}^{2}\ ^{\shortmid }\Lambda (\ ^{\shortmid
}r)^{2}}{10},
\end{equation*}%
the area of the $5$-dimensional unit sphere is given by
\begin{equation}
d\Omega _{4}^{2}=d\theta _{3}^{2}+\sin ^{2}\theta _{3}\left( d\theta
_{2}^{2}+\sin ^{2}\theta _{2}(d\theta _{1}^{2}+\sin ^{2}\theta _{1}d\varphi
^{2})\right) ,  \label{5usph}
\end{equation}%
when the shell $s=3$ coordinates are $(p_{5}=\theta _{2},p_{6}=\theta _{3})$
and shell $s=4$ coordinates are $(p_{7},p_{8}=E).$ Identifying $x^{1}=\
^{\shortmid }r,$ $x^{3}=\theta =\theta _{1},y^{3}=\varphi ,y^{4}=t$ on the
shell $s=2,$ we obtain instead of (\ref{5usph}) a 2-d unite sphere for a
base spacetime $\mathbf{V,}$%
\begin{equation*}
d\Omega _{2}^{2}=d\Omega ^{2}=d\theta ^{2}+\sin ^{2}\theta d\varphi
^{2}=(dx^{2})^{2}+\sin ^{2}(x^{2})(dy^{3})^{2}
\end{equation*}%
considered for a 4-d Lorentz manifold. For some fixed coordinates $%
(p_{7},E), $ the prime metric $\ ^{\shortmid }\mathbf{\mathring{g}}_{\alpha
_{s}\beta _{s}}$ (\ref{pmtang}) describes a 6-d Schwarzschild-de Sitter
phase space with effective cosmological constant $\ ^{\shortmid }\Lambda $
and BH mass $\ ^{\shortmid }M$ $\ $through the relation
\begin{equation*}
\ ^{\shortmid }\mu =\frac{\ ^{\shortmid }\kappa _{6}^{2}\ ^{\shortmid }M}{4}%
\frac{\Gamma \lbrack 5/2]}{\pi ^{5/2}}.
\end{equation*}%
Depending on the values of the parameters $\ ^{\shortmid }M$ and $\
^{\shortmid }\Lambda ,$ this metric may have two, one, or zero horizons
corresponding to the real, positive roots of the equation $h(\ ^{\shortmid
}r)=0.$ For $\ ^{\shortmid }\widehat{\Upsilon }_{\alpha _{s}\beta _{s}}$
determined by $\ ^{\shortmid }\Lambda ,$ a metric (\ref{pmtang}) may define
a solution of gravitational field equations for MGTs with MDRs (\ref%
{meinsteqtbcand}).

In a similar form, we can consider prime metrics and embedding in $T^{\ast }%
\mathbf{V,}$ i.e. $\ ^{\shortmid }\mathbf{\mathring{g}=\{}\ ^{\shortmid }%
\mathbf{\mathring{g}}_{\alpha _{s}\beta _{s}}\},$ for other types higher
dimension BH solutions studied, for instance, in \cite{myers,emparan}. In
this work the prime and target metrics will be (quasi) stationary ones of
signature $(+++-;+++-),$ which is different from standard higher dimension
BH solutions with signature $(+++-+++...+)$. In principle, such a$\
^{\shortmid }\mathbf{\mathring{g}}_{\alpha _{s}\beta _{s}}$maybe or not a
solution of certain gravitational field equations but BH target
configurations $\ _{s}^{\shortmid }\mathbf{g=\{\ _{s}^{\shortmid }\mathbf{g}}%
_{\alpha _{s}\beta _{s}}\mathbf{\}}$ positively can be generated if ceratin
"small" nonholonomic deformations $\ \ _{s}^{\shortmid }\mathbf{\mathring{g}}%
\rightarrow \ _{s}^{\shortmid }\mathbf{g}$ are considered for a BH $\
^{\shortmid }\mathbf{\mathring{g}}$ or embedding in $T^{\ast }\mathbf{V}$
and respective relativistic Finsler-Lagrange-Hamilton model.

\subsubsection{Target stationary 8-d phase space s-metrics}

We parameterize the quadratic line element for a target stationary s-metric $%
\ _{s}^{\shortmid }\mathbf{g}$ written the form
\begin{eqnarray}
&&ds^{2} =\ ^{\shortmid }g_{i_{s}}(x^{k_{s}})(dx^{i_{s}})^{2}+\ ^{\shortmid
}g_{a_{s}}(x^{i_{s}},p_{b_{s}})(\ ^{\shortmid }\mathbf{e}^{a_{s}})^{2}
\label{ans1} \\
&&=g_{1}(\ ^{\shortmid }r,\theta )(d\ ^{\shortmid }r)^{2}+g_{2}(\
^{\shortmid }r,\theta )d\theta ^{2}+g_{3}(\ ^{\shortmid }r,\theta ,\varphi
)(\delta \varphi )^{2}+g_{4}(\ ^{\shortmid }r,\theta ,\varphi )(\delta
t)^{2}+\ ^{\shortmid }g^{5}(\ ^{\shortmid }r,\theta ,\varphi ,\theta
_{3})(\delta \theta _{2})^{2}  \notag \\
&&+\ ^{\shortmid }g^{6}(\ ^{\shortmid }r,\theta ,\varphi ,\theta
_{3})(\delta \theta _{3})^{2}+\ ^{\shortmid }g^{7}(\ ^{\shortmid }r,\theta
,\varphi ,\theta _{2},\theta _{3},E)(\delta p_{7})^{2}+\ ^{\shortmid
}g^{8}(\ ^{\shortmid }r,\theta ,\varphi ,\theta _{2},\theta _{3},E)(\delta
E)^{2},  \notag
\end{eqnarray}%
where the coefficients are with respect to N-elongated dual frames $\
^{\shortmid }\mathbf{e}^{\alpha },$%
\begin{eqnarray*}
\ ^{\shortmid }\mathbf{e}^{1} &=&d\ ^{\shortmid }r,\ ^{\shortmid }\mathbf{e}%
^{2}=d\theta ,\ ^{\shortmid }\mathbf{e}^{3}=\delta \varphi =d\varphi
+N_{1}^{3}(\ ^{\shortmid }r,\theta ,\varphi )d(\ ^{\shortmid }r)+N_{2}^{3}(\
^{\shortmid }r,\theta ,\varphi )d\theta , \\
\ ^{\shortmid }\mathbf{e}^{4} &=&\delta t=dt+N_{1}^{4}(\ ^{\shortmid
}r,\theta ,\varphi )d(\ ^{\shortmid }r)+N_{2}^{4}(\ ^{\shortmid }r,\theta
,\varphi )d\theta , \\
\ ^{\shortmid }\mathbf{e}^{5} &=&\delta \theta _{2}=d\theta _{2}+N_{1}^{5}(\
^{\shortmid }r,\theta ,\varphi ,\theta _{3})d(\ ^{\shortmid }r)+N_{2}^{5}(\
^{\shortmid }r,\theta ,\varphi ,\theta _{3})d\theta + \\
&&N_{3}^{5}(\ ^{\shortmid }r,\theta ,\varphi ,\theta _{3})d\varphi
+N_{4}^{5}(\ ^{\shortmid }r,\theta ,\varphi ,\theta _{3})dt, \\
\ ^{\shortmid }\mathbf{e}^{6} &=&\delta \theta _{3}=d\theta _{3}+N_{1}^{6}(\
^{\shortmid }r,\theta ,\varphi ,\theta _{3})d(\ ^{\shortmid }r)+N_{2}^{6}(\
^{\shortmid }r,\theta ,\varphi ,\theta _{3})d\theta + \\
&&N_{3}^{6}(\ ^{\shortmid }r,\theta ,\varphi ,\theta _{3})d\varphi
+N_{4}^{6}(\ ^{\shortmid }r,\theta ,\varphi ,\theta _{3})dt, \\
\ ^{\shortmid }\mathbf{e}^{7} &=&\delta p_{7}=dp_{7}+N_{1}^{7}(\ ^{\shortmid
}r,\theta ,\varphi ,\theta _{3},E)d(\ ^{\shortmid }r)+N_{2}^{7}(\
^{\shortmid }r,\theta ,\varphi ,\theta _{3},E)d\theta +N_{3}^{7}(\
^{\shortmid }r,\theta ,\varphi ,\theta _{3},E)d\varphi + \\
&&N_{4}^{7}(\ ^{\shortmid }r,\theta ,\varphi ,\theta _{3},E)dt+N_{5}^{7}(\
^{\shortmid }r,\theta ,\varphi ,\theta _{3},E)d\theta _{2}+N_{6}^{7}(\
^{\shortmid }r,\theta ,\varphi ,\theta _{3},E)d\theta _{3}, \\
\ ^{\shortmid }\mathbf{e}^{8} &=&\delta E=dE+N_{1}^{8}(\ ^{\shortmid
}r,\theta ,\varphi ,\theta _{3},E)d(\ ^{\shortmid }r)+N_{2}^{8}(\
^{\shortmid }r,\theta ,\varphi ,\theta _{3},E)d\theta +N_{3}^{8}(\
^{\shortmid }r,\theta ,\varphi ,\theta _{3},E)d\varphi + \\
&&N_{4}^{8}(\ ^{\shortmid }r,\theta ,\varphi ,\theta _{3},E)dt+N_{5}^{8}(\
^{\shortmid }r,\theta ,\varphi ,\theta _{3},E)d\theta _{2}+N_{6}^{8}(\
^{\shortmid }r,\theta ,\varphi ,\theta _{3},E)d\theta _{3},
\end{eqnarray*}

Nonholonomic transforms of prime to target s-metrics, $\ \ _{s}^{\shortmid }%
\mathbf{\mathring{g}}$ (\ref{pmtang}) $\rightarrow \ _{s}^{\shortmid }%
\mathbf{g}$ (\ref{ans1}) can be parameterized in terms of gravitational $%
\eta $-polarization functions following Convention 5.1 with formulas (66)
and (67) in \cite{v18c}. The N-adapted coefficients of s-metrics are related
by such formulas:%
\begin{eqnarray}
\ ^{\shortmid }g_{i_{1}}(x^{k_{1}}) &=&\ ^{\shortmid }g_{i_{1}}(\
^{\shortmid }r,\theta )=\ ^{\shortmid }\eta
_{i_{1}}(x^{k_{1}},y^{a_{2}},p_{a_{3}},p_{a_{4}})\ ^{\shortmid }\mathring{g}%
_{i_{1}}(x^{k_{1}},y^{a_{2}},p_{a_{3}},p_{a_{4}}),\mbox{ where }i_{1}=1,2
\notag \\
\mbox{ and }\ ^{\shortmid }\mathring{g}_{1} &=&h^{-1}(\ ^{\shortmid }r),\
^{\shortmid }\mathring{g}_{2}=(\ ^{\shortmid }r)^{2}\sin ^{2}\theta _{3}\sin
^{2}\theta _{2}\mbox{ for }\ ^{\shortmid }\eta _{1}(\ ^{\shortmid }r,\theta
),\ ^{\shortmid }\eta _{2}(\ ^{\shortmid }r,\theta ,\theta _{2},\theta _{3});
\label{smetrpolar1}
\end{eqnarray}%
\begin{eqnarray*}
\ ^{\shortmid }g_{b_{2}}(x^{i_{1}},y^{3}) &=&\ ^{\shortmid }g_{b_{2}}(\
^{\shortmid }r,\theta ,\varphi )=\ ^{\shortmid }\eta
_{b_{2}}(x^{i_{1}},y^{a_{2}},p_{a_{3}},p_{a_{4}})\ ^{\shortmid }\mathring{g}%
_{b_{1}}(x^{i_{1}},y^{a_{2}},p_{a_{3}},p_{a_{4}}),\mbox{ where }a_{2}=3,4 \\
\mbox{ and }\ ^{\shortmid }\mathring{g}_{3} &=&(\ ^{\shortmid }r)^{2}\sin
^{2}\theta _{3}\sin ^{2}\theta _{2}\sin ^{2}\theta ,\ ^{\shortmid }\mathring{%
g}_{4}=-h(\ ^{\shortmid }r)\mbox{ for }\ ^{\shortmid }\eta _{3}(\
^{\shortmid }r,\theta ,\varphi ,\theta _{2},\theta _{3}),\ ^{\shortmid }\eta
_{4}(\ ^{\shortmid }r,\theta ,\varphi );
\end{eqnarray*}%
\begin{eqnarray*}
\ ^{\shortmid }g_{a_{3}}(x^{i_{2}},p_{6}) &=&\ ^{\shortmid }g_{a_{3}}(\
^{\shortmid }r,\theta ,\varphi ,\theta _{2})=\ ^{\shortmid }\eta
^{a_{3}}(x^{i_{1}},y^{b_{2}},p_{b_{3}},p_{b_{4}})\ ^{\shortmid }\mathring{g}%
^{a_{3}}(x^{i_{1}},y^{b_{2}},p_{b_{3}},p_{b_{4}}),\mbox{ where }a_{3}=5,6 \\
\mbox{ and }\ ^{\shortmid }\mathring{g}^{5} &=&(\ ^{\shortmid }r)^{2}\sin
^{2}\theta _{3},\ ^{\shortmid }\mathring{g}^{6}=(\ ^{\shortmid }r)^{2}%
\mbox{
for }\ ^{\shortmid }\eta ^{5}(\ ^{\shortmid }r,\theta ,\varphi ,\theta
_{3}),\ \ ^{\shortmid }\eta ^{6}(\ ^{\shortmid }r,\theta ,\varphi ,\theta
_{3});
\end{eqnarray*}%
\begin{eqnarray*}
\ ^{\shortmid }g_{a_{4}}(x^{i_{3}},E) &=&\ ^{\shortmid }g_{a_{4}}(\
^{\shortmid }r,\theta ,\varphi ,\theta _{2},\theta _{3},E)=\ ^{\shortmid
}\eta ^{a_{4}}(x^{i_{1}},y^{b_{2}},p_{b_{3}},p_{b_{4}})\ ^{\shortmid }%
\mathring{g}^{a_{4}}(x^{i_{1}},y^{a_{2}},p_{a_{3}},p_{a_{4}}),\mbox{ where }%
a_{4}=7,8 \\
\mbox{ and }\ ^{\shortmid }\mathring{g}^{7} &=&1,\ ^{\shortmid }\mathring{g}%
^{8}=-1\mbox{ for }\ ^{\shortmid }\eta ^{7}(\ ^{\shortmid }r,\theta
_{3},E),\ ^{\shortmid }\eta ^{8}(\ ^{\shortmid }r,E);
\end{eqnarray*}%
and, for N-connection coefficients,
\begin{eqnarray}
\ ^{\shortmid }N_{i_{1}}^{a_{2}}(x^{k_{1}},y^{3}) &=&N_{i_{1}}^{a_{2}}(\
^{\shortmid }r,\theta ,\varphi )=\eta
_{i_{1}}^{a_{2}}(x^{i_{1}},y^{b_{2}},p_{b_{3}},p_{b_{4}})\ ^{\shortmid }%
\mathring{N}_{i_{1}}^{a_{2}}(x^{i_{1}},y^{b_{2}},p_{b_{3}},p_{b_{4}}),
\label{ncoefpolar1} \\
\ ^{\shortmid }N_{a_{3}i_{2}}(x^{k_{1}},y^{b_{2}},p_{6}) &=&\ ^{\shortmid
}N_{a_{3}i_{2}}(\ ^{\shortmid }r,\theta ,\varphi ,\theta _{3})=\ ^{\shortmid
}\eta _{a_{3}i_{2}}(x^{i_{1}},y^{b_{2}},p_{b_{3}},p_{b_{4}})\ ^{\shortmid }%
\mathring{N}_{a_{3}i_{2}}(x^{i_{1}},y^{b_{2}},p_{b_{3}},p_{b_{4}}),  \notag
\\
\ ^{\shortmid }N_{a_{4}i_{3}}(x^{k_{1}},y^{b_{2}},p_{a_{3}},E) &=&\
^{\shortmid }N_{a_{4}i_{3}}(\ ^{\shortmid }r,\theta ,\varphi ,\theta
_{2},\theta _{3},E)=\ ^{\shortmid }\eta
_{a_{4}i_{3}}(x^{i_{1}},y^{b_{2}},p_{b_{3}},p_{b_{4}})\ ^{\shortmid }%
\mathring{N}_{a_{4}i_{3}}(x^{i_{1}},y^{b_{2}},p_{b_{3}},p_{b_{4}}),  \notag
\end{eqnarray}%
where the N-connection coefficients $\ ^{\shortmid }\mathring{N}%
_{i_{1}}^{a_{2}},\ ^{\shortmid }\mathring{N}_{a_{3}i_{2}},\ ^{\shortmid }%
\mathring{N}_{a_{4}i_{3}}$ are nonzero in arbitrary local coordinates and
vanish in spherical coordinates for a prime s-metric (\ref{pmtang}).

For this class of target stationary s-metrics with Killing symmetry on $%
\partial /\partial p_{7},$ it is important that the N-adapted coefficients
are parameterized on coordinates $(\ ^{\shortmid }r,\theta ,\varphi ,\theta
_{2},\theta _{3},E)$ as it is respectively stated in above formulas. In
principle, the coefficients of $\ _{s}^{\shortmid }\mathbf{\mathring{g}}$
and $\eta $-polarizations may depend on all phase space coordinates
(including dependencies on time like variables etc. which can be always
introduced by arbitrary frame and coordinate transforms). Nevertheless,
respective products of such coefficients have to satisfy the conditions (\ref%
{smetrpolar1}) and (\ref{ncoefpolar1}) with respect to certain N-adapted
frames in order to generate target s-metrics which will posses a decoupling
property of modified Einstein equations.

\subsection{Off-diagonal ansatz for radial and energy dependence}

The $\eta $-polarization functions are subjected to the condition that
nonholonomic deformations
\begin{equation*}
\ _{s}^{\shortmid }\mathbf{\mathring{g}}\rightarrow \ _{s}^{\shortmid }%
\mathbf{g}= [\ ^{\shortmid }g_{\alpha _{s}}= \ ^{\shortmid }\eta _{\alpha
_{s}}\ ^{\shortmid }\mathring{g}_{\alpha _{s}},\
^{\shortmid}N_{i_{s-1}}^{a_{s}}= \ ^{\shortmid }\eta _{i_{s-1}}^{a_{s}}\
^{\shortmid } \mathring{N}_{i_{s-1}}^{a_{s}}]
\end{equation*}
generate target metrics $\ _{s}^{\shortmid }\mathbf{g}$ as solutions of
gravitational field equations with MDRs (\ref{meinsteqtbcand}). In this
section, we construct in explicit form nonholonomic deformations of 8-d
imbedding of prime 6-d Tangherlini like solutions (\ref{pmtang}) to
stationary phase BH configurations with explicit dependence on energy type
coordinate $E.$

\subsubsection{Configurations with nontrivial shell matter sources}

Such exact generic off-diagonal solutions can be generated following the
conditions of Corollary 5.2 with formulas (66) and (67) in \cite{v18c}. In
spherical variables with additional $p_{7}$ and $E$ coordinates, such
stationary configurations with Killing symmetry on $\partial /\partial p_{7}$
and explicit dependence on $E$ are defined by such quadratic line elements:

The quasi-stationary phase configurations on cotangent Lorentz bundles in
terms of $\eta $-polarization functions are described by an asatz of type (%
\ref{ans1}),
\begin{eqnarray}
&&ds^{2}=g_{\alpha _{s}\beta _{s}}du^{\alpha _{s}}du^{\beta _{s}}=e^{\psi (\
^{\shortmid }r,\theta )}[(d\ ^{\shortmid }r)^{2}+(\ ^{\shortmid
}r)^{2}(d\theta )^{2}]  \label{offdiagpolf} \\
&&-\frac{[\partial _{\varphi }(\ ^{\shortmid }\eta _{4}\ ^{\shortmid }%
\mathring{g}_{4})]^{2}}{|\int d\varphi (\ _{2}^{\shortmid }\widehat{\Upsilon
})\partial _{\varphi }(\ ^{\shortmid }\eta _{4}\ ^{\shortmid }\mathring{g}%
_{4})|\ (\ ^{\shortmid }\eta _{4}\ ^{\shortmid }\mathring{g}_{4})}\{d\varphi
+\frac{\partial _{i_{1}}[\int d\varphi (\ _{2}^{\shortmid }\widehat{\Upsilon
})\ \partial _{\varphi }(\ ^{\shortmid }\eta _{4}\ ^{\shortmid }\mathring{g}%
_{4})]}{(\ _{2}^{\shortmid }\widehat{\Upsilon })\partial _{\varphi }(\
^{\shortmid }\eta _{4}\ ^{\shortmid }\mathring{g}_{4})}dx^{i_{1}}\}^{2}+
\notag \\
&&(\ ^{\shortmid }\eta _{4}\ ^{\shortmid }\mathring{g}_{4})\{dt+[\
_{1}n_{k_{1}}+\ _{2}n_{k_{1}}\int d\varphi \frac{\lbrack \partial _{\varphi
}(\ ^{\shortmid }\eta _{4}\ ^{\shortmid }\mathring{g}_{4})]^{2}}{|\int
dy^{3}(\ _{2}^{\shortmid }\widehat{\Upsilon })\partial _{\varphi }(\
^{\shortmid }\eta _{4}\ ^{\shortmid }\mathring{g}_{4})|\ (\ ^{\shortmid
}\eta _{4}\ ^{\shortmid }\mathring{g}_{4})^{5/2}}]dx^{\acute{k}_{1}}\}+
\notag
\end{eqnarray}%
\begin{eqnarray*}
&&(\ ^{\shortmid }\eta ^{5}\ ^{\shortmid }\mathring{g}^{5})\{d\theta _{2}+[\
_{1}n_{k_{2}}+\ _{2}n_{k_{2}}\int d\theta _{3}\frac{[\frac{\partial }{%
\partial \theta _{3}}(\ ^{\shortmid }\eta ^{5}\ ^{\shortmid }\mathring{g}%
^{5})]^{2}}{|\int d\theta _{3}(\ _{3}^{\shortmid }\widehat{\Upsilon })\
\frac{\partial }{\partial \theta _{3}}(\ ^{\shortmid }\eta ^{5}\ ^{\shortmid
}\mathring{g}^{5})|\ (\ ^{\shortmid }\eta ^{5}\ ^{\shortmid }\mathring{g}%
^{5})^{5/2}}]dx^{k_{2}}\} \\
&&-\frac{[\frac{\partial }{\partial \theta _{3}}(\ ^{\shortmid }\eta ^{5}\
^{\shortmid }\mathring{g}^{5})]^{2}}{|\int d\theta _{3}\ (\ _{3}^{\shortmid }%
\widehat{\Upsilon })\frac{\partial }{\partial \theta _{3}}(\ ^{\shortmid
}\eta ^{5}\ ^{\shortmid }\mathring{g}^{5})\ |\ (\ ^{\shortmid }\eta ^{5}\
^{\shortmid }\mathring{g}^{5})}\{d\theta _{3}+\frac{\partial _{i_{2}}[\int
d\theta _{3}(\ _{3}^{\shortmid }\widehat{\Upsilon })\frac{\partial }{%
\partial \theta _{3}}(\ ^{\shortmid }\eta ^{5}\ ^{\shortmid }\mathring{g}%
^{5})]}{(\ _{3}^{\shortmid }\widehat{\Upsilon })\frac{\partial }{\partial
\theta _{3}}(\ ^{\shortmid }\eta ^{5}\ ^{\shortmid }\mathring{g}^{5})}%
dx^{i_{2}}\}^{2}+
\end{eqnarray*}%
\begin{eqnarray*}
&&(\ ^{\shortmid }\eta ^{7}\ ^{\shortmid }\mathring{g}^{7})\{dp_{7}+[\
_{1}n_{k_{3}}+\ _{2}n_{k_{3}}\int dE\frac{[\partial _{E}(\ ^{\shortmid }\eta
^{7}\ ^{\shortmid }\mathring{g}^{7})]^{2}}{|\int dE\ (\ _{4}^{\shortmid }%
\widehat{\Upsilon })\partial _{E}(\ ^{\shortmid }\eta ^{7}\ ^{\shortmid }%
\mathring{g}^{7})|\ (\ ^{\shortmid }\eta ^{7}\ ^{\shortmid }\mathring{g}%
^{7})^{5/2}}]dx^{k_{3}}\} \\
&&-\frac{[\partial _{E}(\ ^{\shortmid }\eta ^{7}\ ^{\shortmid }\mathring{g}%
^{7})]^{2}}{|\int dE\ (\ _{4}^{\shortmid }\widehat{\Upsilon })\partial
_{E}(\ ^{\shortmid }\eta ^{7}\ ^{\shortmid }\mathring{g}^{7})\ |\ (\
^{\shortmid }\eta ^{7}\ ^{\shortmid }\mathring{g}^{7})}\{dE+\frac{\partial
_{i_{3}}[\int dE(\ _{4}^{\shortmid }\widehat{\Upsilon })\ \partial _{E}(\
^{\shortmid }\eta ^{7}\ ^{\shortmid }\mathring{g}^{7})]}{(\ _{4}^{\shortmid }%
\widehat{\Upsilon })\partial _{E}(\ ^{\shortmid }\eta ^{7}\ ^{\shortmid }%
\mathring{g}^{7})}dx^{i_{3}}\}^{2},
\end{eqnarray*}%
where $_{1}n_{k_{1}}(\ ^{\shortmid }r,\theta ),$ $\ _{2}n_{k_{1}}(\
^{\shortmid }r,\theta ),\ _{1}n_{k_{2}}(\ ^{\shortmid }r,\theta ,\varphi ),\
_{2}n_{k_{2}}(\ ^{\shortmid }r,\theta ,\varphi ),\ _{1}n_{k_{3}}(\
^{\shortmid }r,\theta ,\varphi ,\theta _{2},\theta _{3}),\ _{2}n_{k_{3}}(\
^{\shortmid }r,\theta ,\varphi ,\theta _{2},\theta _{3}),$ are integration
functions on respective shells. The coefficients of such s--metrics are
written in terms of $\eta $-polarization functions and determined by
generating functions $[\psi (\ ^{\shortmid }r,\theta ),^{\shortmid }\eta
_{4}(\ ^{\shortmid }r,\theta ,\varphi ),\ ^{\shortmid }\eta ^{5}(\
^{\shortmid }r,\theta ,\varphi ,\theta _{3}),$ $\ ^{\shortmid }\eta ^{7}(\
^{\shortmid }r,\theta ,\varphi ,\theta _{2},\theta _{3},E)];$ generating
sources $\ _{1}^{\shortmid }\widehat{\Upsilon }(\ ^{\shortmid }r,\theta ),\
_{2}^{\shortmid }\widehat{\Upsilon }(\ ^{\shortmid }r,\theta ,\varphi ),\
_{3}^{\shortmid }\widehat{\Upsilon }(\ ^{\shortmid }r,\theta ,\varphi
,\theta _{3}),$ \newline
$\ _{4}^{\shortmid }\widehat{\Upsilon }(\ ^{\shortmid }r,\theta ,\varphi
,\theta _{2},\theta _{3},E)];$ and prime s-metric coefficients $\
^{\shortmid }\mathbf{\mathring{g}}_{\alpha _{s}\beta _{s}}$ (\ref{pmtang})\
following such formulas%
\begin{eqnarray*}
\ ^{\shortmid }\eta _{1}\ ^{\shortmid }\mathring{g}_{1} &=&\ ^{\shortmid
}\eta _{2}\ ^{\shortmid }\mathring{g}_{2}=e^{\psi (x^{k_{1}})},\ ^{\shortmid
}\eta _{3}\ ^{\shortmid }\mathring{g}_{3}=-\frac{[\partial _{\varphi }(\
^{\shortmid }\eta _{4}\ ^{\shortmid }\mathring{g}_{4})]^{2}}{|\int d\varphi
\partial _{\varphi }[(\ _{2}^{\shortmid }\widehat{\Upsilon })(\ ^{\shortmid
}\eta _{4}\ ^{\shortmid }\mathring{g}_{4})]|\ (\ ^{\shortmid }\eta _{4}\
^{\shortmid }\mathring{g}_{4})}, \\
\ ^{\shortmid }\eta ^{6}\ ^{\shortmid }\mathring{g}^{6} &=&-\frac{[\partial
^{6}(\ ^{\shortmid }\eta ^{5}\ ^{\shortmid }\mathring{g}^{5})]^{2}}{|\int
dp_{6}(\ _{3}^{\shortmid }\widehat{\Upsilon })\ \partial ^{6}[(\ ^{\shortmid
}\eta ^{5}\ ^{\shortmid }\mathring{g}^{5})]\ |\ (\ ^{\shortmid }\eta ^{5}\
^{\shortmid }\mathring{g}^{5})},\ ^{\shortmid }\eta ^{8}\ ^{\shortmid }%
\mathring{g}^{8}=-\frac{[\partial _{E}(\ ^{\shortmid }\eta ^{7}\ ^{\shortmid
}\mathring{g}^{7})]^{2}}{|\int dE[(\ _{4}^{\shortmid }\widehat{\Upsilon }%
)\partial _{E}(\ ^{\shortmid }\eta ^{7}\ ^{\shortmid }\mathring{g}^{7})\ |\
(\ ^{\shortmid }\eta ^{7}\ ^{\shortmid }\mathring{g}^{7})};
\end{eqnarray*}%
\begin{eqnarray}
\ ^{\shortmid }\eta _{i_{1}}^{3}\ ^{\shortmid }\mathring{N}_{i_{1}}^{3} &=&%
\frac{\partial _{i_{1}}\ \int d\varphi (\ _{2}^{\shortmid }\widehat{\Upsilon
})\ \partial _{\varphi }(\ ^{\shortmid }\eta _{4}\ ^{\shortmid }\mathring{g}%
_{4})}{(\ _{2}^{\shortmid }\widehat{\Upsilon })\ \partial _{\varphi }(\
^{\shortmid }\eta _{4}\ ^{\shortmid }\mathring{g}_{4})},
\label{noffdiagpolf} \\
\ \ ^{\shortmid }\eta _{k_{1}}^{4}\ ^{\shortmid }\mathring{N}_{k_{1}}^{4}
&=&\ _{1}n_{k_{1}}+\ _{2}n_{k_{1}}\int d\varphi \frac{\lbrack \partial
_{\varphi }(\ ^{\shortmid }\eta _{4}\ ^{\shortmid }\mathring{g}_{4})]^{2}}{%
|\int d\varphi (\ _{2}^{\shortmid }\widehat{\Upsilon })\partial _{\varphi
}(\ ^{\shortmid }\eta _{4}\ ^{\shortmid }\mathring{g}_{4})|\ (\ ^{\shortmid
}\eta _{4}\ ^{\shortmid }\mathring{g}_{4})^{5/2}},  \notag
\end{eqnarray}%
\begin{eqnarray*}
\ ^{\shortmid }\eta _{k_{2}5}\ ^{\shortmid }\mathring{N}_{k_{2}5} &=&\
_{1}n_{k_{2}}+\ _{2}n_{k_{2}}\int dp_{6}\frac{[\partial ^{6}(\ ^{\shortmid
}\eta ^{5}\ ^{\shortmid }\mathring{g}^{5})]^{2}}{|\int dp_{6}\ (\
_{3}^{\shortmid }\widehat{\Upsilon })\partial ^{6}(\ ^{\shortmid }\eta ^{5}\
^{\shortmid }\mathring{g}^{5})|\ (\ ^{\shortmid }\eta ^{5}\ ^{\shortmid }%
\mathring{g}^{5})^{5/2}}, \\
\ ^{\shortmid }\eta _{i_{2}6}\ ^{\shortmid }\mathring{N}_{i_{2}6} &=&\frac{%
\partial _{i_{2}}\ \int dp_{6}(\ _{3}^{\shortmid }\widehat{\Upsilon })\
\partial ^{6}(\ ^{\shortmid }\eta ^{5}\ ^{\shortmid }\mathring{g}^{5})}{\
_{3}^{\shortmid }\widehat{\Upsilon }\ \partial ^{6}(\ ^{\shortmid }\eta
^{5}\ ^{\shortmid }\mathring{g}^{5})},
\end{eqnarray*}%
\begin{eqnarray*}
\ ^{\shortmid }\eta _{k_{3}7}\ ^{\shortmid }\mathring{N}_{k_{3}7} &=&\
_{1}n_{k_{3}}+\ _{2}n_{k_{3}}\int dE\frac{[\partial _{E}(\ ^{\shortmid }\eta
^{7}\ ^{\shortmid }\mathring{g}^{7})]^{2}}{|\int dE\ (\ _{4}^{\shortmid }%
\widehat{\Upsilon })\partial _{E}(\ ^{\shortmid }\eta ^{7}\ ^{\shortmid }%
\mathring{g}^{7})|\ (\ ^{\shortmid }\eta ^{7}\ ^{\shortmid }\mathring{g}%
^{7})^{5/2}}, \\
\ ^{\shortmid }\eta _{i_{3}8}\ ^{\shortmid }\mathring{N}_{i_{3}8} &=&\frac{%
\partial _{i_{3}}\ \int dE(\ _{4}^{\shortmid }\widehat{\Upsilon })\ \partial
_{E}(\ ^{\shortmid }\eta ^{7}\ ^{\shortmid }\mathring{g}^{7})}{\
_{4}^{\shortmid }\widehat{\Upsilon }\ \partial _{E}(\ ^{\shortmid }\eta
^{7}\ ^{\shortmid }\mathring{g}^{7})}.
\end{eqnarray*}

A solution of type (\ref{offdiagpolf}) defines nonholonomic deformations of
the 6-d Tangherlini BH (\ref{pmtang}) into a 8-d phase space. It contains a
nontrivial nonholonomic induced torsion which can be constrained to Levi
Civita, LC, configurations by imposing additional constraints on integration
and generating functions, see details in \cite{v18c} and subsection \ref%
{exlc}. For nonsingular small deformations on a parameter $\varepsilon ,$
such a s-metric describes a BH embedded self-consistently in a locally
anisotropic polarized phase space media. We note that the N-connection
coefficients (\ref{noffdiagpolf}) result in generic off-diagonal phase space
metrics written in higher dimension spherical coordinates. Such solutions
can not be diagonalized for nontrivial anholonomy coefficients.
Nevertheless, we can choose respective subclasses of generating and
integration functions when nonholonomic deformations may result in certain
diagonal configurations. Such generalized Tangherlini type solutions
describe certain BH configurations with phase space degrees of freedom and
contributions from effective and matter field sources on shall. \ All such
sources result in MDRs of type (\ref{mdrg}).

\subsubsection{Stationary vacuum off-diagonal configurations}

\label{ssvacuum}Let us consider an example of target vacuum solution of the
generalized Einstein equations (\ref{meinsteqtbcand}) with zero source
describing a stationary phase space as in Definition 5.4 and Consequence 5.4
of \cite{v18c} (other types solutions can be generated similarly as in
section 5.5 of that work).

Type 1 vacuum off-diagonal quasi-stationary phase space configurations are
defined by the conditions $\partial _{\varphi }g_{4}=0$ but $g_{4}\neq 0,$ $%
\ \partial _{\varphi }g_{3}\neq 0$ and $g_{3}\neq 0;\frac{\partial }{%
\partial \theta _{3}}(\ ^{\shortmid }g^{5})=0$ but $\ ^{\shortmid }g^{5}\neq
0,$ $\frac{\partial }{\partial \theta _{3}}(\ ^{\shortmid }g^{6})\neq 0$ and
$\ ^{\shortmid }g^{6}\neq 0;$ and $\partial _{E}(\ ^{\shortmid }g^{7})=0$
but $\ ^{\shortmid }g^{7}\neq 0,$ $\partial _{E}(\ ^{\shortmid }g^{8})\neq 0$
and $\ ^{\shortmid }g^{8}\neq 0.$ The target quadratic line element can be
parameterized in a shell adapted form
\begin{equation*}
d\ ^{\shortmid
}s_{v1,8d}^{2}=ds_{v,s1}^{2}+ds_{v1,s2}^{2}+ds_{v1,s3}^{2}+ds_{v1,s4}^{2},%
\mbox{ where }
\end{equation*}%
\begin{eqnarray}
ds_{v,s1}^{2} &=&e^{\ ^{0}\psi (\ ^{\shortmid }r,\theta )}[(d\ ^{\shortmid
}r)^{2}+(\ ^{\shortmid }r)^{2}(d\theta )^{2}],\   \label{vacuumphase1} \\
ds_{v1,s2}^{2} &=&g_{3}(\ ^{\shortmid }r,\theta ,\varphi )[d\varphi
+w_{k_{1}}(\ ^{\shortmid }r,\theta ,\varphi )dx^{k_{1}}]^{2}  \notag \\
&&+g_{4}(\ ^{\shortmid }r,\theta )[dt+(\ _{1}n_{k_{1}}(\ ^{\shortmid
}r,\theta )+\ _{2}n_{k_{1}}(\ ^{\shortmid }r,\theta )\int d\varphi
/g_{3})dx^{k_{1}}]^{2},  \notag
\end{eqnarray}%
\begin{eqnarray*}
ds_{v1,s3}^{2} &=&\ \ ^{\shortmid }g^{5}(\ ^{\shortmid }r,\theta ,\varphi
)[d\theta ^{2}+(\ _{1}n_{k_{2}}(\ ^{\shortmid }r,\theta ,\varphi )+\
_{2}n_{k_{2}}(\ ^{\shortmid }r,\theta ,\varphi )\int d\theta _{3}/\ \
^{\shortmid }g^{6})dx^{k_{2}}]^{2} \\
&&+\ \ ^{\shortmid }g^{6}(\ ^{\shortmid }r,\theta ,\varphi ,\theta
_{3})[d\theta _{3}+\ ^{\shortmid }w_{k_{2}}(\ ^{\shortmid }r,\theta ,\varphi
,\theta _{3})dx^{k_{2}}]^{2}, \\
ds_{v1,s4}^{2} &=&\ \ ^{\shortmid }g^{7}(\ ^{\shortmid }r,\theta ,\varphi
,\theta _{2},\theta _{3})[dp_{7}+(\ _{1}n_{k_{3}}(\ ^{\shortmid }r,\theta
,\varphi ,\theta _{2},\theta _{3})+\ _{2}n_{k_{3}}(\ ^{\shortmid }r,\theta
,\varphi ,\theta _{2},\theta _{3})\int dE/\ \ ^{\shortmid
}g^{8})dx^{k_{3}}]^{2} \\
&&+\ \ ^{\shortmid }g^{8}(\ ^{\shortmid }r,\theta ,\varphi ,\theta
_{2},\theta _{3},E)[dE+\ ^{\shortmid }w_{k_{3}}(\ ^{\shortmid }r,\theta
,\varphi ,\theta _{2},\theta _{3},E)dx^{k_{3}}]^{2}.
\end{eqnarray*}%
This vacuum phase solution posses Killing symmetries on $\partial _{t}$ and $%
\partial ^{7}$ being determined by arbitrary generating functions
\begin{eqnarray*}
g_{3}(\ ^{\shortmid }r,\theta ,\varphi ) &=&\ ^{\shortmid }\eta _{3}\
^{\shortmid }\mathring{g}_{3},g_{4}(\ ^{\shortmid }r,\theta )=\ ^{\shortmid
}\eta _{4}\ ^{\shortmid }\mathring{g}_{4}\mbox{ and }w_{k_{1}}(\ ^{\shortmid
}r,\theta ,\varphi )=\eta _{k_{1}}^{3}\ ^{\shortmid }\mathring{N}%
_{k_{1}}^{3}; \\
\ \ ^{\shortmid }g^{5}(\ ^{\shortmid }r,\theta ,\varphi ) &=&\ ^{\shortmid
}\eta ^{5}\ ^{\shortmid }\mathring{g}^{5},\ \ ^{\shortmid }g^{6}(\
^{\shortmid }r,\theta ,\varphi ,\theta _{3})=\ ^{\shortmid }\eta ^{6}\
^{\shortmid }\mathring{g}^{6}\mbox{ and }\ ^{\shortmid }w_{k_{2}}(\
^{\shortmid }r,\theta ,\varphi ,\theta _{3})=\ ^{\shortmid }\eta _{k_{2}6}\
^{\shortmid }\mathring{N}_{k_{2}6}; \\
\ \ ^{\shortmid }g^{7}(\ ^{\shortmid }r,\theta ,\varphi ,\theta _{2},\theta
_{3}) &=&\ ^{\shortmid }\eta ^{7}\ ^{\shortmid }\mathring{g}^{7},\ \
^{\shortmid }g^{8}(\ ^{\shortmid }r,\theta ,\varphi ,\theta _{2},\theta
_{3},E)=\ ^{\shortmid }\eta ^{8}\ ^{\shortmid }\mathring{g}^{8} \\
&&\mbox{ and }\ ^{\shortmid }w_{k_{3}}(\ ^{\shortmid }r,\theta ,\varphi
,\theta _{2},\theta _{3},E)=\ ^{\shortmid }\eta _{k_{3}8}\ ^{\shortmid }%
\mathring{N}_{k_{3}8};
\end{eqnarray*}%
and when the integration functions can be parameterized
\begin{eqnarray*}
\ _{1}n_{k_{1}}(\ ^{\shortmid }r,\theta ),\ _{2}n_{k_{1}}(\ ^{\shortmid
}r,\theta )\sim \ ^{\shortmid }\mathring{N}_{k_{1}}^{4};\ _{1}n_{k_{2}}(\
^{\shortmid }r,\theta ,\varphi ),\ _{2}n_{k_{2}}(\ ^{\shortmid }r,\theta
,\varphi )\sim \ ^{\shortmid }\mathring{N}_{k_{2}5}; && \\
\ _{1}n_{k_{3}}(\ ^{\shortmid }r,\theta ,\varphi ,\theta _{2},\theta _{3}),\
_{2}n_{k_{3}}(\ ^{\shortmid }r,\theta ,\varphi ,\theta _{2},\theta _{3})\sim
\ ^{\shortmid }\mathring{N}_{k_{3}7}, &&
\end{eqnarray*}%
where $\ ^{0}\psi (\ ^{\shortmid }r,\theta )$ is a solution of 2-d Laplace
equation. Such generating and integration functions can be prescribed to be
singular or of necessary smooth class. If the $\eta $-polarizations are
smooth, a corresponding solution (\ref{vacuumphase1}) describes a 6-d
Schwarzschild-de Sitter phase space BH with effective cosmological constant $%
\ ^{\shortmid }\Lambda $ and BH mass $\ ^{\shortmid }M$ defined by data $\
_{s}^{\shortmid }\mathbf{\mathring{g}}$ (\ref{pmtang}) nonholonomically
deformed and embedded self-consistently in a vacuum phase space aether.
Re-defining the coordinates and for small $\varepsilon $-parametric
deformations (see details in Theorem 5.1 in \cite{v18c}), we can always
prove that this class a solutions are similar to extra dimensional BHs but
with certain polarizations of constants and horizons to dependencies on
momentum type coordinates. Here we note the value of $\ ^{\shortmid }\Lambda
$ should be chosen respectively for different types for Finsler gravity and
nonholonomic gravity models, for instacne, as in the early works \cite%
{v1998bh,v2001bh} (for off-diagonal corrections to BH solutions) or in more
recent approaches with induced cosmological constants in a class of
Finsler-Randers and DGP gravity theories  \cite{basil1,basil2}.

\subsubsection{Diagonal s-metrics with energy depending polarization
functions}

We can generate exact solutions with diagonal phase spaces following the
Definition 5.3 and Corollary 5.3 in \cite{v18c} if the N-connection
coefficients are zero. A s-metric is diagonal on a shell $s$ if there are
satisfied the conditions $\ ^{\shortmid }N_{i_{s-1}}^{a_{s}}=\ ^{\shortmid
}\eta _{i_{s-1}a_{s}}\ ^{\shortmid }\mathring{N}_{i_{s-1}a_{s}}=0.$ Choosing
a prime diagonal metric, for instance $\ _{s}^{\shortmid }\mathbf{\mathring{g%
}}$ (\ref{pmtang}) in diagonal coordinates and prescribing a corresponding
subclass of data $(\ ^{\shortmid }\eta _{4},\ ^{\shortmid }\eta ^{5},\
^{\shortmid }\eta ^{7}),$ we generate diagonal configurations. Such exact
solutions depend on the type of generating and polarization functions
(singular, or smooth ones) used for explicit constructions.

By straightforward computations using spherical phase space coordinates, we
can prove that there are generated diagonal quasi-stationary phase space
configurations if the $\eta $-polarization functions for a s-metric (\ref%
{offdiagpolf}) are chosen in the form%
\begin{eqnarray}
\ ^{\shortmid }\eta _{4} &=&(\ ^{\shortmid }\mathring{g}_{4})^{-1}\int
d\varphi \ \ ^{\shortmid }\eta _{4}^{[1]}(\varphi )/(\ _{2}^{\shortmid }%
\widehat{\Upsilon })+\ \ ^{\shortmid }\eta _{4}^{[2]}(\ ^{\shortmid
}r,\theta ),  \label{diagsmeta} \\
&&\mbox{ for integration functions }\ ^{\shortmid }\eta _{4}^{[1]}(\varphi
)\ \mbox{ and }\ ^{\shortmid }\eta _{4}^{[2]}(\ ^{\shortmid }r,\theta ),%
\mbox{ where }\ _{1}n_{k_{1}}=\ _{2}n_{k_{1}}=0;  \notag \\
\ \ ^{\shortmid }\eta ^{5} &=&(\ ^{\shortmid }\mathring{g}^{5})^{-1}\int
d\theta _{3}\ \ ^{\shortmid }\eta _{\lbrack 1]}^{5}(\theta _{3})/(\
_{3}^{\shortmid }\widehat{\Upsilon })+\ \ ^{\shortmid }\eta _{\lbrack
2]}^{5}(\ ^{\shortmid }r,\theta ,\varphi ),  \notag \\
&&\mbox{ for integration functions }\ \ ^{\shortmid }\eta _{\lbrack
1]}^{5}(\theta _{3})\ \mbox{ and }\ \ ^{\shortmid }\eta _{\lbrack 2]}^{5}(\
^{\shortmid }r,\theta ,\varphi ),\mbox{ where }\ _{1}n_{k_{2}}=\
_{2}n_{k_{2}}=0;  \notag \\
\ ^{\shortmid }\eta ^{7} &=&(\ ^{\shortmid }\mathring{g}^{7})^{-1}\int dE\ \
^{\shortmid }\eta _{\lbrack 1]}^{7}(E)/(\ _{4}^{\shortmid }\widehat{\Upsilon
})+\ \ ^{\shortmid }\eta _{\lbrack 2]}^{7}(\ ^{\shortmid }r,\theta ,\varphi
,\theta _{2},\theta _{3}),  \notag \\
&&\mbox{ for integration functions }\ \ ^{\shortmid }\eta _{\lbrack
1]}^{7}(E)\ \mbox{ and }\ \ ^{\shortmid }\eta _{\lbrack 2]}^{7}(\
^{\shortmid }r,\theta ,\varphi ,\theta _{2},\theta _{3}),\mbox{ where }\
_{1}n_{k_{3}}=\ _{2}n_{k_{3}}=0.  \notag
\end{eqnarray}

Diagonal stationary configurations consist a special class of phase space
nonlinear and nonholonomically systems when the generalized gravitational
dynamics is defined by diagonal s-metrics self-consistently embedded into
diagonal phase spaces backgrounds. The dependence on the energy type
coordinate $E$ is determined by a generating function $\ ^{\shortmid }\eta
^{7}(\ ^{\shortmid }r,\theta ,\varphi ,\theta _{2},\theta _{3},E).$ In
particular, we can generate exact solutions for a $\ ^{\shortmid }\eta
^{7}(E)$ or certain re-defined systems of coordinates resulting in
dependencies on an energy type parameter. Such "rainbow" type models are
known due to \cite{magueijo04} where $\ E$ is considered as a free
parameter. In our approach, dependencies on $E$ and other momentum type
coordinates are defined in a self-consistent form by solutions of
generalized Einstein equations for MGTs with MDR and/or respective
Finsler-Lagrange-Hamilton variables. We can consider $\eta $-polarization
functions (\ref{offdiagpolf}) with decompositions on a small parameter $%
\varepsilon $ following the conditions of Consequence 5.3 and formulas (77)
in \cite{v18c}. Formulas with decompositions on small parameters allow an
obvious physical interpretation of such generalized BH solutions using
analogies from GR and extra dimensions with small deformations of horizons
and nonlinear polarizations of physical constants. In a more general
context, we can consider that certain $\eta $-polarization functions define
a topologically nontrivial phase space structure with possible filaments,
anisotropies, fractional configurations etc. which may model BHs, for
instance, when the sources $\ _{s}^{\shortmid }\widehat{\Upsilon }$ are
approximated to cosmological constants.

\subsubsection{Levi-Civita off-diagonal energy phase space solutions}

\label{exlc}The stationary solutions constructed above are with
nonholonomically induced torsions. We can impose additional constraints on
the generating and integration functions and extract zero torsion
configurations following the conditions of Consequence 4.2 and formulas (55)
in \cite{v18c}.

The quadratic line nonlinear elements for stationary LC-configurations with
spherical phase space symmetry and generic dependence on $E$ can be
parameterized in such a form:
\begin{eqnarray}
ds_{LCst}^{2} &=&\check{g}_{\alpha \beta }(x^{k_{3}},E)du^{\alpha }du^{\beta
}=e^{\psi (\ ^{\shortmid }r,\theta )}[(d\ ^{\shortmid }r)^{2}+(\ ^{\shortmid
}r)^{2}(d\theta )^{2}]+  \label{lctarget} \\
&&\frac{[\partial _{\varphi }(\ _{2}^{\shortmid }\check{\Psi})]^{2}}{4(\
_{2}^{\shortmid }\widehat{\Upsilon }[\ _{2}^{\shortmid }\check{\Psi}%
])^{2}\{g_{4}^{[0]}-\int d\varphi \partial _{\varphi }[(\ _{2}^{\shortmid }%
\check{\Psi})^{2}]/4(\ _{2}^{\shortmid }\widehat{\Upsilon })\}}\{d\varphi
+[\partial _{i_{1}}(\ _{2}^{\shortmid }\check{A})]dx^{i_{1}}\}+  \notag \\
&&\{g_{4}^{[0]}-\int d\varphi \frac{\partial _{\varphi }[(\ _{2}^{\shortmid }%
\check{\Psi})^{2}]}{4(\ _{2}^{\shortmid }\widehat{\Upsilon }[\
_{2}^{\shortmid }\check{\Psi}])}\}\{dt+\partial _{i_{1}}[\
^{2}n(x^{k_{1}})]dx^{i_{1}}\}+  \notag
\end{eqnarray}%
\begin{eqnarray*}
&&\{g_{5}^{[0]}-\int d\theta _{3}\frac{\frac{\partial }{\partial \theta _{3}}%
[(\ _{3}^{\shortmid }\Psi )^{2}]}{4(\ _{3}^{\shortmid }\widehat{\Upsilon }[\
_{3}^{\shortmid }\check{\Psi}])}\}\{d\theta _{2}+\partial _{i_{2}}[\
^{3}n(x^{k_{2}})]dx^{i_{2}}\}+ \\
&&\frac{[\frac{\partial }{\partial \theta _{3}}(\ _{3}^{\shortmid }\Psi
)]^{2}}{4(\ _{3}^{\shortmid }\widehat{\Upsilon }[\ _{3}^{\shortmid }\check{%
\Psi}])^{2}\{g_{5}^{[0]}-\int d\theta _{3}\frac{\partial }{\partial \theta
_{3}}[(\ _{3}^{\shortmid }\Psi )^{2}]/4(\ _{3}^{\shortmid }\widehat{\Upsilon
}[\ _{3}^{\shortmid }\check{\Psi}])\}}\{d\theta _{3}+[\partial _{i_{2}}(\
_{3}^{\shortmid }\check{A})]dx^{i_{2}}\}^{2}+ \\
&&\{g_{7}^{[0]}-\int dE\frac{\partial _{E}[(\ _{4}^{\shortmid }\Psi )^{2}]}{%
4(\ _{4}^{\shortmid }\widehat{\Upsilon }[\ _{4}^{\shortmid }\check{\Psi}])}%
\}\{dp_{7}+\partial _{i_{3}}[\ ^{4}n(x^{k_{3}})]dx^{i_{3}}\}- \\
&&\frac{[\partial _{E}(\ _{4}^{\shortmid }\Psi )]^{2}}{4(\ _{4}^{\shortmid }%
\widehat{\Upsilon }[\ _{4}^{\shortmid }\check{\Psi}])^{2}\{g_{7}^{[0]}-\int
dE\partial _{E}[(\ _{4}^{\shortmid }\Psi )^{2}]/4(\ _{4}^{\shortmid }%
\widehat{\Upsilon }[\ _{4}^{\shortmid }\check{\Psi}])\}}\{dE+[\partial
_{i_{3}}(\ _{4}^{\shortmid }\check{A})]dx^{i_{3}}\}^{2}.
\end{eqnarray*}

In above formulas, there are considered generating functions, generating
sources and integration functions with respective functional dependence $%
[...]$:
\begin{eqnarray*}
s=2: &&\ _{2}^{\shortmid }\Psi =\ _{2}^{\shortmid }\check{\Psi}(\
^{\shortmid }r,\theta ,\varphi ),\partial _{\varphi }(\partial _{i_{1}}\
_{2}^{\shortmid }\check{\Psi})=\partial _{i_{1}}\partial _{\varphi }(\
_{2}^{\shortmid }\check{\Psi}), \\
&& \check{w}_{i_{1}} = \partial _{i_{1}}(\ _{2}^{\shortmid }\check{\Psi}%
)/\partial _{\varphi }(\ _{2}^{\shortmid }\check{\Psi})=\partial _{i_{1}}(\
_{2}^{\shortmid }\check{A});n_{i_{1}}=\partial _{i_{1}}[\ ^{2}n(\
^{\shortmid }r,\theta )]; \\
&&\ _{2}^{\shortmid }\widehat{\Upsilon }(\ ^{\shortmid }r,\theta ,\varphi
)=\ _{2}^{\shortmid }\widehat{\Upsilon }[\ _{2}^{\shortmid }\check{\Psi}],%
\mbox{ or }\ _{2}^{\shortmid }\widehat{\Upsilon }=const; \\
s=3: &&\ \ _{3}^{\shortmid }\Psi =\ _{3}^{\shortmid }\check{\Psi}(\
^{\shortmid }r,\theta ,\varphi ,\theta _{3}),\frac{\partial }{\partial
\theta _{3}}[\partial _{i_{2}}(\ _{3}^{\shortmid }\check{\Psi})]=\partial
_{i_{2}}\frac{\partial }{\partial \theta _{3}}(\ _{3}^{\shortmid }\check{\Psi%
}); \\
&& \check{w}_{i_{2}} = \partial _{i_{2}}(\ _{3}^{\shortmid }\Psi )/\frac{%
\partial }{\partial \theta _{3}}(\ _{3}^{\shortmid }\Psi )=\partial
_{i_{2}}(\ _{3}^{\shortmid }\check{A});\ n_{i_{2}}=\partial _{i_{2}}[\
^{3}n(\ ^{\shortmid }r,\theta ,\varphi )]; \\
&&\ _{3}^{\shortmid }\widehat{\Upsilon }(\ ^{\shortmid }r,\theta ,\varphi
,\theta _{3})=\ _{3}^{\shortmid }\widehat{\Upsilon }[\ _{3}^{\shortmid }%
\check{\Psi}],\mbox{ or }\ _{3}^{\shortmid }\widehat{\Upsilon }=const; \\
s=4: &&\ \ _{4}^{\shortmid }\Psi =\ _{4}^{\shortmid }\check{\Psi}(\
^{\shortmid }r,\theta ,\varphi ,\theta _{3},E),[\partial _{i_{3}}\partial
_{E}(\ _{4}^{\shortmid }\check{\Psi})]=\partial _{i_{3}}\partial _{E}(\
_{4}^{\shortmid }\check{\Psi}); \\
&&\check{w}_{i_{3}} =\partial _{i_{3}}(\ _{4}^{\shortmid }\Psi )/\partial
_{E}(\ _{4}^{\shortmid }\Psi )=\partial _{i_{3}}(\ _{4}^{\shortmid }\check{A}%
);\ n_{i_{3}}=\partial _{i_{3}}[\ ^{4}n(\ ^{\shortmid }r,\theta ,\varphi
,\theta _{3})]; \\
&&\ _{4}^{\shortmid }\widehat{\Upsilon }(\ ^{\shortmid }r,\theta ,\varphi
,\theta _{3},E)=\ _{4}^{\shortmid }\widehat{\Upsilon }[\ _{4}^{\shortmid }%
\check{\Psi}],\mbox{ or }\ _{4}^{\shortmid }\widehat{\Upsilon }=const.
\end{eqnarray*}

We note that quasi-stationary s-metrics of type (\ref{lctarget}) are generic
off-diagonal if there are nontrivial anholonomy relations for respective
N-adapted bases on a shell. Such exact solutions possess a Killing symmetry
on $\partial ^{7}$ for respective canonical N-adapted frames and coordinate
systems, i.e. when the s-metrics do not depend on coordinate $p_{7}.$ In
principle, we can always extract LC-configurations by considering respective
subclasses of generating functions and (effective) sources and integration
functions. Other type conditions (additional or special ones) allows us to
construct diagonal LC-configurations. In general, such stationary solutions
may be not stabile. But we can chose special classes of nonholonomic
constraints which may stabilize BH nonholonomic LC-configurations.

\subsection{Spacetime and cofiber phase space double BH configurations}

On $T^{\ast }\mathbf{V,}$ we can construct BHs which are different from the
Tangherlini type one (\ref{pmtang}) and nonholonomic deformations studied in
previous subsections.

\subsubsection{Prime phase metrics for double Schwarzschild - de Sitter BHs}

Let us consider two 4-d spherical systems of local coordinates on a base
manifold $\mathbf{V}$ and a typical fiber of $T^{\ast }\mathbf{V,}$
parameterized respectively%
\begin{equation*}
x^{1}=r,x^{2}=\theta ,y^{3}=\varphi ,y^{4}=t;p_{5}=\ ^{p}r,p_{6}=\
^{p}\theta ,p_{7}=\ ^{p}\varphi ,p_{8}=\ E,
\end{equation*}%
\begin{equation*}
\mbox{ where } r=\sqrt{(x^{1^{\prime }})^{2}+(x^{2^{\prime
}})^{2}+(y^{3^{\prime }})^{2}},\ ^{p}r=\sqrt{(p_{5^{\prime
}})^{2}+(p_{6^{\prime }})^{2}+(p_{7^{\prime }})^{2}}.
\end{equation*}%
For such coordinates, prime indices are used for some Cartezian coordinates
and $\theta ,\varphi $ and $\ ^{p}\theta ,\ ^{p}\varphi $ are respective
angular ones (we use a left label 'p' in order to distinguish such spherical
type coordinated from the previous type ones used in previous subsections).

A prime quadratic line element%
\begin{eqnarray}
ds^{2} &=&\ _{\circ \circ }^{\shortmid }g_{\alpha \beta }du^{\alpha
}du^{\beta }=  \label{prm2} \\
&&f^{-1}(r)dr^{2}+r^{2}d\Omega ^{2}-f(r)dt^{2}+\ ^{p}f^{-1}(\ ^{p}r)d(\
^{p}r)^{2}+(\ ^{p}r)^{2}d\ ^{p}\Omega ^{2}-\ ^{p}f(\ ^{p}r)dE^{2},  \notag
\end{eqnarray}%
\begin{equation*}
\mbox{with }f(r)=1-\frac{\mu }{r}-\frac{\kappa ^{2}\Lambda r^{2}}{3}%
\mbox{
and }\ ^{p}f(\ ^{p}r)=1-\frac{\ ^{p}\mu }{\ ^{p}r}-\frac{(\ ^{p}\kappa
)^{2}(\ ^{p}\Lambda )(\ ^{p}r)^{2}}{3}
\end{equation*}%
defines an exact solution of modified Einstein equations (\ref%
{meinsteqtbcand}) with sources $\ ^{\shortmid }\widehat{\Upsilon }_{\alpha
_{s}\beta _{s}}=[\ ^{\shortmid }\widehat{\Upsilon }_{ij}=\Lambda ,\
^{\shortmid }\widehat{\Upsilon }_{ab}=\ ^{p}\Lambda ].$ For a $\ _{\circ
\circ }^{\shortmid }g_{\alpha \beta }$ (\ref{prm2}) (we use the left label "$%
\circ \circ $" in order to emphasize that this is a phase metric with two BH
configurations), the areas of respective 2-dimensional unite spheres are
given by
\begin{eqnarray*}
d\Omega ^{2} &=&d\theta ^{2}+\sin ^{2}\theta d\varphi ^{2}=(dx^{2})^{2}+\sin
^{2}(x^{2})(dy^{3})^{2}, \\
d\ ^{p}\Omega ^{2} &=&d(\ ^{p}\theta )^{2}+\sin ^{2}(\ ^{p}\theta )d(\
^{p}\varphi )^{2}=(dp_{6})^{2}+\sin ^{2}(p_{6})(dp_{7})^{2}.
\end{eqnarray*}%
Such an exact solution defines on the base Lorentz manifold $\mathbf{V}$ a
Schwarzschild - de Sitter spacetime determined by $f(r)$ with standard BH
mass $M$ via formula $\mu =\frac{\kappa ^{2}M\Gamma (3/2)}{2\pi ^{3/2}},$
cosmological constant $\Lambda $ and $\kappa ^{2}$ determined by the Newton
constant in GR, see details in \cite{pappas16}. There is also an analogous
Schwarzschild - de Sitter configuration on a typical cofiber space
determined by $\ ^{p}f(\ ^{p}r),$ where the constants $\ ^{p}\mu ,\
^{p}\Lambda ,\ ^{p}\kappa $ are integration parameters which should be
determined experimentally or computed for certain generalized MGTs. For
instance, we can take $\ ^{p}\Lambda =0$ and consider that $\ ^{p}\mu $
determines the horizon, i.e. the maximal momentum determined by the maximal
speed of light $c$ for a BH with mass $M.$ Such a double BH configuration
describe, for instance, a generalized Schwarzschild phase space with
additional cofiber singularity and horizon configurations. Here we note that
there are two types of singularies in (\ref{prm2}), on the base manifold and
in the cofiber space with horizons similar to those ofr the Schwarzschild -
de Sitter BHs. There are not such solutions, for instance, for the
Finsler-Randers models  \cite{basil1,basil2} even a nontrivial $\
^{p}\Lambda $ can be considered as iduced by certain Finsler background
fluctuations. To generate BH configurations in a (co) fiber space we need a
more rich Finsler structure with canonical d-connections than in the case of
minimal extensions of GR by Randers nonlinear quadratic elements.

\subsubsection{Nonholonomic deformations of phase space double BH s-metrics}

We can use $\ _{\circ \circ }^{\shortmid }g_{\alpha \beta }$ (\ref{prm2}) as
a prime metric instead of 6-d Tangherlini prime metric $\ _{s}^{\shortmid }%
\mathbf{\mathring{g}}$ (\ref{pmtang}). A new class of generic off-diagonal
solutions can be constructed. Such phase space metrics are very different
from those defined by (\ref{offdiagpolf}) and correlate BH configurations on
cofibers with BH solutions on the base spacetime manifold.

The quasi-stationary phase configurations on cotangent Lorentz bundles
written in terms of $\eta $-polarization functions are described by an
ansatz of type (\ref{ans1}),
\begin{eqnarray}
&&ds^{2}=g_{\alpha _{s}\beta _{s}}du^{\alpha _{s}}du^{\beta _{s}}=e^{\psi
(r,\theta )}[(dr)^{2}+r^{2}(d\theta )^{2}]  \label{offdiagpolf2bh} \\
&&-\frac{[\partial _{\varphi }(\ ^{\shortmid }\eta _{4}\ \ _{\circ \circ
}^{\shortmid }g_{4})]^{2}}{|\int d\varphi (\ _{2}^{\shortmid }\widehat{%
\Upsilon })\partial _{\varphi }(\ ^{\shortmid }\eta _{4}\ \ _{\circ \circ
}^{\shortmid }g_{4})|\ (\ ^{\shortmid }\eta _{4}\ \ _{\circ \circ
}^{\shortmid }g_{4})}\{d\varphi +\frac{\partial _{i_{1}}[\int d\varphi (\
_{2}^{\shortmid }\widehat{\Upsilon })\ \partial _{\varphi }(\ ^{\shortmid
}\eta _{4}\ _{\circ \circ }^{\shortmid }g_{4})]}{(\ _{2}^{\shortmid }%
\widehat{\Upsilon })\partial _{\varphi }(\ ^{\shortmid }\eta _{4}\ \ _{\circ
\circ }^{\shortmid }g_{4})}dx^{i_{1}}\}^{2}+  \notag \\
&&(\ ^{\shortmid }\eta _{4}\ \ _{\circ \circ }^{\shortmid }g_{4})\{dt+[\
_{1}n_{k_{1}}+\ _{2}n_{k_{1}}\int d\varphi \frac{\lbrack \partial _{\varphi
}(\ ^{\shortmid }\eta _{4}\ \ _{\circ \circ }^{\shortmid }g_{4})]^{2}}{|\int
dy^{3}(\ _{2}^{\shortmid }\widehat{\Upsilon })\partial _{\varphi }(\
^{\shortmid }\eta _{4}\ \ _{\circ \circ }^{\shortmid }g_{4})|\ (\
^{\shortmid }\eta _{4}\ \ _{\circ \circ }^{\shortmid }g_{4})^{5/2}}]dx^{%
\acute{k}_{1}}\}+  \notag
\end{eqnarray}%
\begin{eqnarray*}
&&(\ ^{\shortmid }\eta ^{5}\ \ _{\circ \circ }^{\shortmid }g^{5})\{d(\
^{p}r)+[\ _{1}n_{k_{2}}+\ _{2}n_{k_{2}}\int d(\ ^{p}\theta )\frac{[\frac{%
\partial }{\partial (\ ^{p}\theta )}(\ ^{\shortmid }\eta ^{5}\ \ _{\circ
\circ }^{\shortmid }g^{5})]^{2}}{|\int d(\ ^{p}\theta )(\ _{3}^{\shortmid }%
\widehat{\Upsilon })\ \frac{\partial }{\partial (\ ^{p}\theta )}(\
^{\shortmid }\eta ^{5}\ \ _{\circ \circ }^{\shortmid }g^{5})|\ (\
^{\shortmid }\eta ^{5}\ \ _{\circ \circ }^{\shortmid }g^{5})^{5/2}}%
]dx^{k_{2}}\} \\
&&-\frac{[\frac{\partial }{\partial (\ ^{p}\theta )}(\ ^{\shortmid }\eta
^{5}\ \ _{\circ \circ }^{\shortmid }g^{5})]^{2}}{|\int d(\ ^{p}\theta )\ (\
_{3}^{\shortmid }\widehat{\Upsilon })\frac{\partial }{\partial (\ ^{p}\theta
)}(\ ^{\shortmid }\eta ^{5}\ \ _{\circ \circ }^{\shortmid }g^{5})\ |\ (\
^{\shortmid }\eta ^{5}\ \ _{\circ \circ }^{\shortmid }g^{5})}\{d(\
^{p}\theta )+\frac{\partial _{i_{2}}[\int d(\ ^{p}\theta )(\ _{3}^{\shortmid
}\widehat{\Upsilon })\frac{\partial }{\partial (\ ^{p}\theta )}(\
^{\shortmid }\eta ^{5}\ \ _{\circ \circ }^{\shortmid }g^{5})]}{(\
_{3}^{\shortmid }\widehat{\Upsilon })\frac{\partial }{\partial (\ ^{p}\theta
)}(\ ^{\shortmid }\eta ^{5}\ \ _{\circ \circ }^{\shortmid }g^{5})}%
dx^{i_{2}}\}^{2}
\end{eqnarray*}%
\begin{eqnarray*}
&&+(\ ^{\shortmid }\eta ^{7}\ \ _{\circ \circ }^{\shortmid }g^{7})\{d\
^{p}\varphi +[\ _{1}n_{k_{3}}+\ _{2}n_{k_{3}}\int dE\frac{[\partial _{E}(\
^{\shortmid }\eta ^{7}\ \ _{\circ \circ }^{\shortmid }g^{7})]^{2}}{|\int dE\
(\ _{4}^{\shortmid }\widehat{\Upsilon })\partial _{E}(\ ^{\shortmid }\eta
^{7}\ \ _{\circ \circ }^{\shortmid }g^{7})|\ (\ ^{\shortmid }\eta ^{7}\ \
_{\circ \circ }^{\shortmid }g^{7})^{5/2}}]dx^{k_{3}}\} \\
&&-\frac{[\partial _{E}(\ ^{\shortmid }\eta ^{7}\ \ _{\circ \circ
}^{\shortmid }g^{7})]^{2}}{|\int dE\ (\ _{4}^{\shortmid }\widehat{\Upsilon }%
)\partial _{E}(\ ^{\shortmid }\eta ^{7}\ \ _{\circ \circ }^{\shortmid
}g^{7})\ |\ (\ ^{\shortmid }\eta ^{7}\ \ _{\circ \circ }^{\shortmid }g^{7})}%
\{dE+\frac{\partial _{i_{3}}[\int dE(\ _{4}^{\shortmid }\widehat{\Upsilon }%
)\ \partial _{E}(\ ^{\shortmid }\eta ^{7}\ \ _{\circ \circ }^{\shortmid
}g^{7})]}{(\ _{4}^{\shortmid }\widehat{\Upsilon })\partial _{E}(\
^{\shortmid }\eta ^{7}\ \ _{\circ \circ }^{\shortmid }g^{7})}%
dx^{i_{3}}\}^{2},
\end{eqnarray*}%
where $_{1}n_{k_{1}}(r,\theta ),$ $\ _{2}n_{k_{1}}(r,\theta ),\
_{1}n_{k_{2}}(r,\theta ,\varphi ),\ _{2}n_{k_{2}}(r,\theta ,\varphi ),\
_{1}n_{k_{3}}(r,\theta ,\varphi ,\ ^{p}r,\ ^{p}\theta ),\
_{2}n_{k_{3}}(r,\theta ,\varphi ,\ ^{p}r,\ ^{p}\theta ),$ are integration
functions on respective shells. The coefficients of such s--metrics for
nonholonomic deformations of double BH pase space solutions are written in
terms of $\eta $-polarization functions and determined by generating
functions $[\psi (r,\theta ),^{\shortmid }\eta _{4}(r,\theta ,\varphi ),\
^{\shortmid }\eta ^{5}(r,\theta ,\varphi ,\theta _{3}),$ $\ ^{\shortmid
}\eta ^{7}(r,\theta ,\varphi ,\ ^{p}r,\ ^{p}\theta ,E)];$ generating sources
$\ _{1}^{\shortmid }\widehat{\Upsilon }(r,\theta ),\ _{2}^{\shortmid }%
\widehat{\Upsilon }(r,\theta ,\varphi ),\ _{3}^{\shortmid }\widehat{\Upsilon
}(r,\theta ,\varphi ,\ ^{p}\theta ),\ _{4}^{\shortmid }\widehat{\Upsilon }%
(r,\theta ,\varphi ,\ ^{p}r,\ ^{p}\theta ,E)].$

We note that formulas for $\eta $-polarization functions in (\ref%
{offdiagpolf2bh}) are similar to (\ref{noffdiagpolf}) but with respective
re-definition of coordinates and coefficients of prime metrics. General
frame and coordinate transforms on$\ T^{\ast }\mathbf{V}$ mix the
conventional spacetime-momentum like variables and it is not clear what
physical meaning may have such solutions for general $\eta $-polarization.
Nevertheless, we can construct nonholonomic deformations of the 6-d
Tangherlini metric, or of double Schwarzschild - de Sitter configurations,
for small parametric $\varepsilon $--deformations. In principle, we can
consider two independent such parameters: one for spacetime deformations and
another one for typical co-fiber configurations. For small parameters, the
BH solutions with nonholonomic deformations possess the same singular
properties as in GR but with certain smail deformations of horizons and
polariations of constants.

\subsection{Stationary configurations with spherical symmetries and BHs in
ener\-gy depending pha\-se spaces}

There are more general classes of exact stationary solutions with 6-d
spherical symmetry nonholonomically deformed on a 8-d $T^{\ast }\mathbf{V}$
then those constructed in previous sections. Such generic off-diagonal
metrics consist explicit examples of nonholonomic stationary phase spaces in
MGT with MDRs defined by Consequence 5.1 and Remark 5.1 in \cite{v18c}. For
simplicity, we construct and analyse solutions of type (\ref{ans1}) when the
conventional radial coordinate depends both on space and fiber momentum
variables. Stationary configurations of type (\ref{offdiagpolf2bh}) (which
may include double BH configurations) can be generated by considering
similar nonholonomic transforms for respective 3+3 spherical symmetries on
the base spacetime and typical fiber.

\subsubsection{Solutions with energy depending generating functions \&
(effective) sources}

Considering certain generated data $[g_{4}(\ ^{\shortmid }r,\theta ,\varphi
),g^{5}(\ ^{\shortmid }r,\theta ,\varphi ,\theta _{3}),g^{7}(\ ^{\shortmid
}r,\theta ,\varphi ,\theta _{2},\theta _{3},E)]$ and nonlinear symmetries
involving generating sources (see next subsection) $\ _{s}^{\shortmid }%
\widehat{\Upsilon }=[\ _{1}^{\shortmid }\widehat{\Upsilon }(\ ^{\shortmid
}r,\theta ),\ _{2}^{\shortmid }\widehat{\Upsilon }(\ ^{\shortmid }r,\theta
,\varphi ),$ $\ _{3}^{\shortmid }\widehat{\Upsilon }(\ ^{\shortmid }r,\theta
,\varphi ,\theta _{3}),$ $\ _{4}^{\shortmid }\widehat{\Upsilon }(\
^{\shortmid }r,\theta ,\varphi ,\theta _{2},\theta _{3},E)],$ we construct
such spherical symmetric nonlinear quadratic elements,
\begin{eqnarray}
ds^{2} &=&g_{\alpha _{s}\beta _{s}}du^{\alpha _{s}}du^{\beta _{s}}=e^{\psi
(\ ^{\shortmid }r,\theta )}[(d\ ^{\shortmid }r)^{2}+(\ ^{\shortmid
}r)^{2}(d\theta )^{2}]-  \notag \\
&& \frac{(\ \partial _{\varphi }g_{4})^{2}}{|\int d\varphi \ \partial
_{\varphi }[(\ _{2}^{\shortmid }\widehat{\Upsilon })g_{4}]|\ g_{4}}%
\{d\varphi +\frac{\partial _{i_{1}}[\int d\varphi (\ _{2}^{\shortmid }%
\widehat{\Upsilon })\ \partial _{\varphi }g_{4}]}{(\ _{2}^{\shortmid }%
\widehat{\Upsilon })\ \ \partial _{\varphi }g_{4}}dx^{i_{1}}\}^{2}+
\label{station6offd} \\
&&g_{4}\{dt+[\ _{1}n_{k_{1}}+\ _{2}n_{k_{1}}\int d\varphi \frac{(\ \partial
_{\varphi }g_{4})^{2}}{|\int d\varphi \ \partial _{\varphi }[(\
_{2}^{\shortmid }\widehat{\Upsilon })g_{4}]|\ [g_{4}]^{5/2}}]dx^{\acute{k}%
_{1}}\}+  \notag
\end{eqnarray}%
\begin{eqnarray*}
&&g^{5}\{d\theta _{2}+[\ _{1}n_{k_{2}}+\ _{2}n_{k_{2}}\int d\theta _{3}\frac{%
[\frac{\partial }{\partial \theta _{3}}(g^{5})]^{2}}{|\int d\theta _{3}\
\frac{\partial }{\partial \theta _{3}}[(\ _{3}^{\shortmid }\widehat{\Upsilon
})g^{5}]|\ [g^{5}]^{5/2}}]dx^{k_{2}}\}- \\
&&\frac{[\frac{\partial }{\partial \theta _{3}}(g^{5})]^{2}}{|\int d\theta
_{3}\ \frac{\partial }{\partial \theta _{3}}[(\ _{3}^{\shortmid }\widehat{%
\Upsilon })g^{5}]\ |\ g^{5}}\{d\theta _{3}+\frac{\partial _{i_{2}}[\int
d\theta _{3}(\ _{3}^{\shortmid }\widehat{\Upsilon })\ \frac{\partial }{%
\partial \theta _{3}}(g^{5})]}{(\ _{3}^{\shortmid }\widehat{\Upsilon })\frac{%
\partial }{\partial \theta _{3}}(g^{5})}dx^{i_{2}}\}^{2}+ \\
&&g^{7}\{dp_{7}+[\ _{1}n_{k_{3}}+\ _{2}n_{k_{3}}\int dE\frac{[\partial
_{E}(g^{7})]^{2}}{|\int dE\ \partial _{E}[(\ _{4}^{\shortmid }\widehat{%
\Upsilon })g^{7}]|\ [g^{7}]^{5/2}}]dx^{k_{3}}\}- \\
&&\frac{[\partial _{E}(g^{7})]^{2}}{|\int dE\ \partial _{E}[(\
_{4}^{\shortmid }\widehat{\Upsilon })g^{7}]\ |\ g^{7}}\{dE+\frac{\partial
_{i_{3}}[\int dE(\ _{4}^{\shortmid }\widehat{\Upsilon })\ \partial
_{E}(g^{7})]}{(\ _{4}^{\shortmid }\widehat{\Upsilon })\partial _{E}(g^{7})}%
dx^{i_{3}}\}^{2}.
\end{eqnarray*}%
Such solutions are also determined by generating functions\newline
$\ _{1}n_{k_{1}}(\ ^{\shortmid }r,\theta ),\ _{2}n_{k_{1}}(\ ^{\shortmid
}r,\theta );\ _{1}n_{k_{2}}(\ ^{\shortmid }r,\theta ,\varphi ),\
_{2}n_{k_{2}}(\ ^{\shortmid }r,\theta ,\varphi );\ _{1}n_{k_{3}}(\
^{\shortmid }r,\theta ,\varphi ,\theta _{3}),\ _{2}n_{k_{3}}(\ ^{\shortmid
}r,\theta ,\varphi ,\theta _{3}).$

The class of solutions (\ref{station6offd}) is nonsingular if the values $\
_{s}^{\shortmid }\widehat{\Upsilon }$ are not zero and the generating and
integration functions are not singular. If there are encoded nonholonomic
deformations of certain prime singular metrics, such target solutions (in
general) are singular and describe BHs with phase space off-diagonal
stationary configurations. With singular generating and integration
functions, we can transform BH solutions into nonsingular ones and
inversely. We can apply the AFDM for vacuum phase space or vacuum Lorentz
manifold configurations as in subsection \ref{ssvacuum}.

\subsubsection{Nonlinear spherical symmetries of generating functions and
sources}

A large class of exact solutions in MGTs with MDRs which are constructed
following the AFDM possess a new type of nonlinear symmetries stated by
Theorem 4.3 with formulas (60) in \cite{v18c}. For nonholonomic deformations
of the 6-d Tangherlini metric $\ _{s}^{\shortmid }\mathbf{\mathring{g}}$ (%
\ref{pmtang}), such symmetries are stated for respective shells when the
generated functions, $\ _{s}^{\shortmid }\Psi ,$ and sources, $\
_{s}^{\shortmid }\widehat{\Upsilon },$ are transformed into equivalent data
for other types generating functions, $\ _{3}^{\shortmid }\Phi ,$ with
effective cosmological constants, $\ _{s}^{\shortmid }\Lambda :$%
\begin{eqnarray*}
&&[\ _{2}^{\shortmid}\Psi(\ ^{\shortmid}r,\theta ,\varphi),\ _{2}^{\shortmid}%
\widehat{\Upsilon}(\ ^{\shortmid}r,\theta,\varphi );\ _{3}^{\shortmid}\Psi(\
^{\shortmid}r,\theta,\varphi,\theta _{3}),\ _{3}^{\shortmid}\widehat{\Upsilon%
}(\ ^{\shortmid }r,\theta,\varphi ,\theta _{3}); \\
&&\ _{4}^{\shortmid}\Psi(\ ^{\shortmid}r,\theta,\varphi ,\theta _{2},\theta
_{3},E),\ _{4}^{\shortmid}\widehat{\Upsilon}(\
^{\shortmid}r,\theta,\varphi,\theta _{2},\theta _{3},E)] \\
&&\rightarrow \left[ \ _{2}^{\shortmid }\Phi (\ ^{\shortmid }r,\theta
,\varphi ),\ _{2}^{\shortmid }\Lambda ;\ _{3}^{\shortmid }\Phi (\
^{\shortmid }r,\theta ,\varphi ,\theta _{3}),\ _{3}^{\shortmid }\Lambda ;\
_{4}^{\shortmid }\Phi (\ ^{\shortmid }r,\theta ,\varphi ,\theta _{2},\theta
_{3},E),\ _{4}^{\shortmid }\Lambda \right] .
\end{eqnarray*}%
Such transforms are determined by formulas {\small
\begin{eqnarray}
s=2: &&\frac{\partial _{\varphi }[(\ _{2}^{\shortmid }\Psi )^{2}]}{\
_{2}^{\shortmid }\widehat{\Upsilon }}=\frac{\partial _{\varphi }[(\
_{2}^{\shortmid }\Phi )^{2}]}{\ _{2}^{\shortmid }\Lambda },
\label{nonltransf} \\
&&\mbox{ i.e. }\ (\ _{2}^{\shortmid }\Phi )^{2}=\ _{2}^{\shortmid }\Lambda
\int d\varphi (\ _{2}^{\shortmid }\widehat{\Upsilon })^{-1}\partial
_{\varphi }[(\ _{2}^{\shortmid }\Psi )^{2}]\mbox{ and/or }(\ _{2}^{\shortmid
}\Psi )^{2}=(\ _{2}^{\shortmid }\Lambda )^{-1}\int d\varphi (\
_{2}^{\shortmid }\widehat{\Upsilon })\partial _{\varphi }[(\ _{2}^{\shortmid
}\Phi )^{2}].  \notag
\end{eqnarray}%
\begin{eqnarray}
s=3: &&\frac{\frac{\partial }{\partial \theta _{3}}[(\ _{3}^{\shortmid }\Psi
)^{2}]}{\ _{3}^{\shortmid }\widehat{\Upsilon }}=\frac{\frac{\partial }{%
\partial \theta _{3}}[(\ _{3}^{\shortmid }\Phi )^{2}]}{\ _{3}^{\shortmid
}\Lambda },  \notag \\
&&\mbox{ i.e. }\ (\ _{3}^{\shortmid }\Phi )^{2}=\ _{3}^{\shortmid }\Lambda
\int d\theta _{3}(\ _{3}^{\shortmid }\widehat{\Upsilon })^{-1}[(\
_{3}^{\shortmid }\Psi )^{2}]\mbox{ and/or }(\ _{3}^{\shortmid }\Psi )^{2}=(\
_{3}^{\shortmid }\Lambda )^{-1}\int d\theta _{3}(\ _{3}^{\shortmid }\widehat{%
\Upsilon })[(\ _{3}^{\shortmid }\Phi )^{2}].  \notag
\end{eqnarray}%
} {\small
\begin{eqnarray}
s=4: &&\frac{\partial _{E}[(\ _{4}^{\shortmid }\Psi )^{2}]}{\
_{4}^{\shortmid }\widehat{\Upsilon }}=\frac{\partial _{E}[(\ _{4}^{\shortmid
}\Phi )^{2}]}{\ _{4}^{\shortmid }\Lambda },  \notag \\
&&\mbox{ i.e. }\ (\ _{4}^{\shortmid }\Phi )^{2}=\ _{4}^{\shortmid }\Lambda
\int dE(\ _{4}^{\shortmid }\widehat{\Upsilon })^{-1}\partial _{E}[(\
_{4}^{\shortmid }\Psi )^{2}]\mbox{ and/or }(\ _{4}^{\shortmid }\Psi )^{2}=(\
_{4}^{\shortmid }\Lambda )^{-1}\int dE(\ _{4}^{\shortmid }\widehat{\Upsilon }%
)\partial _{E}[(\ _{4}^{\shortmid }\Phi )^{2}].  \notag
\end{eqnarray}%
}

There are nonlinear symmetries which are similar to (\ref{nonltransf})
nonholonomic deformations of prime metrics of type $\ _{\circ
\circ}^{\shortmid }g_{\alpha \beta }$ (\ref{prm2}) for double 4-d BH
configurations on spacetime and typical fiber space.

\subsubsection{Off-diagonal solutions with effective constants and zero
torsion}

To extract LC-configurations as in (\ref{lctarget}) the nonlinear symmetries
(\ref{nonltransf}) have to constrained to certain subclasses of generating
functions defined by data
\begin{equation*}
\ \check{g}_{4}(\ ^{\shortmid }r,\theta ,\varphi ),\check{g}^{5}(\
^{\shortmid }r,\theta ,\varphi ,\theta _{3}),\check{g}^{7}(\ ^{\shortmid
}r,\theta ,\varphi ,\theta _{2},\theta _{3},E);
\end{equation*}%
on general results, see Remark 5.1 in \cite{v18c}. The generating functions
(we use inverse hat labels emphasizing that certain additional integrability
conditions are imposed on generating functions) are redefined following
formulas:%
\begin{equation*}
\begin{array}{ccccc}
\mbox{ shell }s=2: &  & \partial _{\varphi }[(\ _{2}^{\shortmid }\check{\Psi}%
)^{2}]=\int d\varphi (\ _{2}^{\shortmid }\widehat{\Upsilon })\partial
_{\varphi }\check{g}_{4}, &  & (\ _{2}^{\shortmid }\check{\Phi})^{2}=-4\
_{2}^{\shortmid }\Lambda \check{g}_{4}; \\
&  &  &  &  \\
\mbox{ shell }s=3: &  & \frac{\partial }{\partial \theta _{3}}[(\
_{3}^{\shortmid }\check{\Psi})^{2}]=\int d\theta _{3}(\ _{3}^{\shortmid }%
\widehat{\Upsilon })\ \frac{\partial }{\partial \theta _{3}}\check{g}^{5}, &
& (\ _{3}^{\shortmid }\check{\Phi})^{2}=-4\ _{3}^{\shortmid }\Lambda \check{g%
}^{6}; \\
&  &  &  &  \\
\mbox{ shell }s=4: &  & \partial _{E}[(\ _{4}^{\shortmid }\check{\Psi}%
)^{2}]=\int dE(\ _{4}^{\shortmid }\widehat{\Upsilon })\ \partial _{E}(\check{%
g}^{7}), &  & (\ _{4}^{\shortmid }\check{\Phi})^{2}=-4\ _{4}^{\shortmid
}\Lambda \check{g}^{8}.%
\end{array}%
\end{equation*}%
Corresponding quadratic line nonlinear elements for exact solutions are
parameterized in the form
\begin{eqnarray}
&& ds_{LC}^{2} =\check{g}_{\alpha _{s}\beta _{s}}du^{\alpha _{s}}du^{\beta
_{s}}=e^{\psi (\ ^{\shortmid }r,\theta )}[(d\ ^{\shortmid }r)^{2}+(\
^{\shortmid }r)^{2}(d\theta )^{2}]  \notag \\
&&-\frac{(\partial _{\varphi }\check{g}_{4})^{2}}{|\int d\varphi (\
_{2}^{\shortmid }\widehat{\Upsilon })(\partial _{\varphi }g_{4})|\ g_{4}}%
\{d\varphi +\partial _{i_{1}}[\ _{2}^{\shortmid }\check{A}(\ ^{\shortmid
}r,\theta ,\varphi )]dx^{i_{1}}\}^{2}+\check{g}_{4}\{dt+\partial _{i_{1}}[\
^{2}n(\ ^{\shortmid }r,\theta )]dx^{i_{1}}\}+  \label{offidiagcosmconstlc} \\
&&\check{g}^{5}\{d\theta _{2}+\partial _{i_{2}}[\
^{2}n(x^{k_{2}})]dx^{i_{2}}\}-\frac{[\frac{\partial }{\partial \theta _{3}}(%
\check{g}^{5})]^{2}}{|\int d\theta _{3}\ (\ _{3}^{\shortmid }\widehat{%
\Upsilon })\frac{\partial }{\partial \theta _{3}}[\check{g}^{5}]\ |\ \check{g%
}^{5}}\{d\theta _{3}+\partial _{i_{2}}[\ _{3}^{\shortmid }\check{A}(\
^{\shortmid }r,\theta ,\varphi ,\theta _{3})]dx^{i_{2}}\}^{2}+  \notag \\
&&\check{g}^{7}\{dp_{7}+\partial _{i_{3}}[\ ^{3}n(x^{k_{3}})]dx^{i_{3}}\}-%
\frac{[\partial _{E}(\check{g}^{7})]^{2}}{|\int dE\ (\ _{4}^{\shortmid }%
\widehat{\Upsilon })\partial _{E}(\check{g}^{7})\ |\ \check{g}^{7}}%
\{dE+\partial _{i_{2}}[\ _{4}^{\shortmid }\check{A}(\ ^{\shortmid }r,\theta
,\varphi ,\theta _{2},\theta _{3},E)]dx^{i_{3}}\}^{2}.  \notag
\end{eqnarray}

Here we note that the effective shell sources $\ _{s}^{\shortmid }\widehat{%
\Upsilon }$ can be absorbed into certain classes of integration functions if
such values do not depend on vertical shell coordinates on respective
shells. The sources $\ _{s}^{\shortmid }\widehat{\Upsilon }$ are encoded
correspondingly in $\ _{s}^{\shortmid }\check{A}$ for (\ref%
{offidiagcosmconstlc}). For nonsingular generating functions and sources,
such solutions define stationary phase space configurations with 6-d
spherical symmetry. If such phase space solutions encode prime BH
configurations, we can consider that this quadratic line nonlinear elements
define new classes of BH solutions with additional cofiber symmetries
determined by MDRs.

\section{Phase space stationary BHs with energy type Killing symmetry}

\label{s3}Applying the AFDM, we can construct various classes of stationary
and BH solutions of modified Einstein equations (\ref{meinsteqtbcand}). For
example, we can generate stationary configurations with a fixed energy phase
coordinate, $E=E_{0}$ (for instance, we can take $E_{0}=0$) and possessing a
Killing symmetry on $\partial /\partial E.$ In this section, we study
nonholonomic deformations of prime BH metrics to target stationary phase
configurations when the s-metric and N-connection coefficients considered
for N-adapted frames depend on space coordinates $(x^{1},x^{2},y^{3})$ and
momentum variables $(p_{5},p_{6},p_{7}) $ but not on $y^{4}=t$ and $p_{8}=E.$

\subsection{7-d Tangherlini like BHs embedded into 8-d phase spaces}

As a prime configuration, we can consider a 7-d Tangherlini BH solution
(instead of a 6-d one (\ref{pmtang})). For a 6-d phase space with signature $%
(++++++),$ we consider a new radial coordinate
\begin{equation*}
\overline{r}=\sqrt{%
(x^{1})^{2}+(x^{2})^{2}+(y^{3})^{2}+(p_{5})^{2}+(p_{6})^{2}+(p_{7})^{2}}.
\end{equation*}%
The prime quadratic line element is parameterized
\begin{eqnarray}
ds^{2} &=&\ \underline{\mathring{g}}_{\alpha _{s}\beta
_{s}}(x^{i_{s}},p_{a_{s}})d\ ^{\shortmid }u^{\alpha _{s}}d\ ^{\shortmid
}u^{\beta _{s}}=\underline{\mathbf{\mathring{g}}}_{\alpha _{s}\beta _{s}}(\
_{s}^{\shortmid }u)\underline{\mathbf{\mathbf{\mathring{e}}}}^{\alpha _{s}}%
\mathbf{\ }\underline{\mathbf{\mathbf{\mathring{e}}}}^{\beta _{s}}  \notag \\
&=&\overline{h}^{-1}(\overline{r})d(\overline{r})^{2}-\overline{h}(\overline{%
r})dt^{2}+(\overline{r})^{2}d\Omega _{5}^{2}-dE^{2},  \label{6usph}
\end{eqnarray}%
where certain symbols are "overlined" in order to emphasize that respective
values are defined by different formulas from those considered in previous
section. In $\underline{\mathbf{\mathring{g}}}_{\alpha _{s}\beta _{s}}$ (\ref%
{6usph}), the 7-d Schwarzschild-de Sitter phase space BH solution is
determined by
\begin{equation*}
\overline{h}(\overline{r})=1-\frac{\ ^{\shortmid }\overline{\mu }}{(%
\overline{r})^{4}}-\frac{\ ^{\shortmid }\kappa _{7}^{2}\ ^{\shortmid }%
\overline{\Lambda }(\overline{r})^{2}}{15},
\end{equation*}%
where the effective gravitational constant $\ ^{\shortmid }\kappa _{7},$
cosmological constant $\ ^{\shortmid }\overline{\Lambda }$ and BH mass $\
^{\shortmid }\overline{\mu }\ $are related through a relation with Gamma
function $\Gamma \lbrack 3],$\ $\ ^{\shortmid }\overline{\mu }=\frac{\
^{\shortmid }\kappa _{7}^{2}\ ^{\shortmid }\overline{M}}{5}\frac{\Gamma
\lbrack 3]}{\pi ^{6}}$. For such solutions, the area of the $5$-dimensional
unit sphere with four angular coordinates $(\theta _{1}=\theta ,\varphi
,\theta _{2},\theta _{3},\theta _{4})$ is given by
\begin{equation*}
d\Omega _{5}^{2}=d\theta _{4}^{2}+\sin ^{2}\theta _{4}\left( d\theta
_{3}^{2}+\sin ^{2}\theta _{3}(d\theta _{2}^{2}+\sin ^{2}\theta _{2}(d\theta
_{1}^{2}+\sin ^{2}\theta _{1}d\varphi ^{2}))\right) ,
\end{equation*}%
when the respective shell $s=3$ coordinates are $(p_{5}=\theta
_{2},p_{6}=\theta _{3})$ and shell $s=4$ coordinates are $(p_{7}=\theta
_{4},p_{8}=E).$

We emphasize that the 7-d BH metric $\underline{\mathbf{\mathring{g}}}%
_{\alpha _{s}\beta _{s}}$ (\ref{6usph}) does not contain as a particular
case certain double BH configurations of type $\ _{\circ \circ }^{\shortmid
}g_{\alpha \beta }$ (\ref{prm2}). Such exact solutions are characterized by
different topological, local and asymptotic properties. This type of phase
space BH metrics can not be constructed in  the Finsler-Randers models \cite%
{basil1,basil2} but exist in other types of generalized
Finsler-Lagrange-Hamilton gravity theories, see examples in \cite%
{v18a,gvvepjc14,bubuianucqg17,vacaruepjc17,v2010bbh,v2013bbe}%
.

\subsection{Nonholonomic deformations of BHs with fixed energy parameter}

Using $\eta $-polarization functions and a prime BH s-metric $\underline{%
\mathbf{\mathring{g}}}_{\alpha _{s}\beta _{s}}$ (\ref{6usph}), we can
generate stationary phase configurations on cotangent Lorentz bundles
described by an ansatz with Killing symmetries on $\partial _{4}=\partial
_{t}$ and $\partial ^{8}=\partial _{E}$ which is different from (\ref{ans1})
being with generic Killing symmetry on $\partial _{4}$ and $\partial ^{7}.$
The AFDM can be applied similarly to solutions (\ref{offdiagpolf}) and (\ref%
{noffdiagpolf}) when the local coordinates, prime metric data and Killing
symmetries are redefined correspondingly. We construct nonholonomic
deformations $\underline{\mathbf{\mathring{g}}}\rightarrow \underline{%
\mathbf{g}}$ defined by such target nonlinear quadratic elements:
\begin{eqnarray}
&&ds^{2}=\underline{g}_{\alpha _{s}\beta _{s}}du^{\alpha _{s}}du^{\beta
_{s}}=e^{\psi (\overline{r},\theta )}[(d\overline{r})^{2}+(\overline{r}%
)^{2}(d\theta )^{2}]  \label{offdiagpolfr} \\
&&-\frac{[\partial _{\varphi }(\ ^{\shortmid }\eta _{4}\ \underline{%
\mathring{g}}_{4})]^{2}}{|\int d\varphi (\ _{2}^{\shortmid }\widehat{%
\Upsilon })\partial _{\varphi }(\ ^{\shortmid }\eta _{4}\ \ \underline{%
\mathring{g}}_{4})|\ (\ ^{\shortmid }\eta _{4}\ \ \underline{\mathring{g}}%
_{4})}\{d\varphi +\frac{\partial _{i_{1}}[\int d\varphi (\ _{2}^{\shortmid }%
\widehat{\Upsilon })\ \partial _{\varphi }(\ ^{\shortmid }\eta _{4}\ \
\underline{\mathring{g}}_{4})]}{(\ _{2}^{\shortmid }\widehat{\Upsilon }%
)\partial _{\varphi }(\ ^{\shortmid }\eta _{4}\ \ \underline{\mathring{g}}%
_{4})}dx^{i_{1}}\}^{2}+  \notag \\
&&(\ ^{\shortmid }\eta _{4}\ \ \underline{\mathring{g}}_{4})\{dt+[\
_{1}n_{k_{1}}+\ _{2}n_{k_{1}}\int d\varphi \frac{\lbrack \partial _{\varphi
}(\ ^{\shortmid }\eta _{4}\ \ \underline{\mathring{g}}_{4})]^{2}}{|\int
dy^{3}(\ _{2}^{\shortmid }\widehat{\Upsilon })\partial _{\varphi }(\
^{\shortmid }\eta _{4}\ \ \underline{\mathring{g}}_{4})|\ (\ ^{\shortmid
}\eta _{4}\ \ \underline{\mathring{g}}_{4})^{5/2}}]dx^{\acute{k}_{1}}\}+
\notag
\end{eqnarray}%
\begin{eqnarray*}
&&(\ ^{\shortmid }\eta ^{5}\ \ \underline{\mathring{g}}^{5})\{d\theta
_{2}+[\ _{1}n_{k_{2}}+\ _{2}n_{k_{2}}\int d\theta _{3}\frac{[\frac{\partial
}{\partial \theta _{3}}(\ ^{\shortmid }\eta ^{5}\ \ \underline{\mathring{g}}%
^{5})]^{2}}{|\int d\theta _{3}(\ _{3}^{\shortmid }\widehat{\Upsilon })\
\frac{\partial }{\partial \theta _{3}}(\ ^{\shortmid }\eta ^{5}\ \
\underline{\mathring{g}}^{5})|\ (\ ^{\shortmid }\eta ^{5}\ \ \underline{%
\mathring{g}}^{5})^{5/2}}]dx^{k_{2}}\} \\
&&-\frac{[\frac{\partial }{\partial \theta _{3}}(\ ^{\shortmid }\eta ^{5}\ \
\underline{\mathring{g}}^{5})]^{2}}{|\int d\theta _{3}\ (\ _{3}^{\shortmid }%
\widehat{\Upsilon })\frac{\partial }{\partial \theta _{3}}(\ ^{\shortmid
}\eta ^{5}\ \ \underline{\mathring{g}}^{5})\ |\ (\ ^{\shortmid }\eta ^{5}\ \
\underline{\mathring{g}}^{5})}\{d\theta _{3}+\frac{\partial _{i_{2}}[\int
d\theta _{3}(\ _{3}^{\shortmid }\widehat{\Upsilon })\frac{\partial }{%
\partial \theta _{3}}(\ ^{\shortmid }\eta ^{5}\ \ \underline{\mathring{g}}%
^{5})]}{(\ _{3}^{\shortmid }\widehat{\Upsilon })\frac{\partial }{\partial
\theta _{3}}(\ ^{\shortmid }\eta ^{5}\ \ \underline{\mathring{g}}^{5})}%
dx^{i_{2}}\}^{2}
\end{eqnarray*}%
\begin{eqnarray*}
&&-\frac{[\frac{\partial }{\partial \theta _{4}}(\ ^{\shortmid }\eta ^{8}\ \
\underline{\mathring{g}}^{8})]^{2}}{|\int d\theta _{4}\ (\ _{4}^{\shortmid }%
\widehat{\Upsilon })\frac{\partial }{\partial \theta _{4}}(\ ^{\shortmid
}\eta ^{8}\ \ \underline{\mathring{g}}^{8})\ |\ (\ ^{\shortmid }\eta ^{8}\ \
\underline{\mathring{g}}^{8})}\{d\theta _{4}+\frac{\partial _{i_{3}}[\int
d\theta _{4}(\ _{4}^{\shortmid }\widehat{\Upsilon })\ \frac{\partial }{%
\partial \theta _{4}}(\ ^{\shortmid }\eta ^{8}\ \ \underline{\mathring{g}}%
^{8})]}{(\ _{4}^{\shortmid }\widehat{\Upsilon })\frac{\partial }{\partial
\theta _{4}}(\ ^{\shortmid }\eta ^{8}\ \ \underline{\mathring{g}}^{8})}%
dx^{i_{3}}\}^{2}+ \\
&&(\ ^{\shortmid }\eta ^{8}\ \ \underline{\mathring{g}}^{8})\{d\theta
_{4}+[\ _{1}n_{k_{3}}+\ _{2}n_{k_{3}}\int d\theta _{4}\frac{[\frac{\partial
}{\partial \theta _{4}}(\ ^{\shortmid }\eta ^{8}\ \ \underline{\mathring{g}}%
^{8})]^{2}}{|\int d\theta _{4}\ (\ _{4}^{\shortmid }\widehat{\Upsilon })%
\frac{\partial }{\partial \theta _{4}}(\ ^{\shortmid }\eta ^{8}\ \
\underline{\mathring{g}}^{8})|\ (\ ^{\shortmid }\eta ^{8}\ \ \underline{%
\mathring{g}}^{8})^{5/2}}]dx^{k_{3}}\},
\end{eqnarray*}%
where $_{1}n_{k_{1}}(\overline{r},\theta ),$ $\ _{2}n_{k_{1}}(\overline{r}%
,\theta ),\ _{1}n_{k_{2}}(\overline{r},\theta ,\varphi ),\ _{2}n_{k_{2}}(%
\overline{r},\theta ,\varphi ),\ _{1}n_{k_{3}}(\overline{r},\theta ,\varphi
,\theta _{2},\theta _{3}),\ _{2}n_{k_{3}}(\overline{r},\theta ,\varphi
,\theta _{2},\theta _{3}),$ are integration functions on respective shells
with new types of radial and angular coordinates. The coefficients of (\ref%
{offdiagpolfr}) are written in terms of $\eta $-polarization functions and
determined by generating functions $[\psi (\overline{r},\theta ),^{\shortmid
}\eta _{4}(\overline{r},\theta ,\varphi ),\ ^{\shortmid }\eta ^{5}(\overline{%
r},\theta ,\varphi ,\theta _{3}),$ $\ ^{\shortmid }\eta ^{8}(\overline{r}%
,\theta ,\varphi ,\theta _{2},\theta _{3},\theta _{4})];$ generating sources
$\ _{1}^{\shortmid }\widehat{\Upsilon }(\overline{r},\theta ),\
_{2}^{\shortmid }\widehat{\Upsilon }(\overline{r},\theta ,\varphi ),$\newline
$\ _{3}^{\shortmid }\widehat{\Upsilon }(\overline{r},\theta ,\varphi ,\theta
_{3}),\ _{4}^{\shortmid }\widehat{\Upsilon }(\overline{r},\theta ,\varphi
,\theta _{2},\theta _{3},\theta _{4})];$ and prime s-metric coefficients $%
\underline{\mathbf{\mathring{g}}}_{\alpha _{s}\beta _{s}}$ (\ref{6usph})\
subjected to such conditions:%
\begin{eqnarray*}
\ ^{\shortmid }\eta _{1}\ \ \underline{\mathring{g}}_{1} &=&\ ^{\shortmid
}\eta _{2}\ \ \underline{\mathring{g}}_{2}=e^{\psi (x^{k_{1}})},\
^{\shortmid }\eta _{3}\ \ \underline{\mathring{g}}_{3}=-\frac{[\partial
_{\varphi }(\ ^{\shortmid }\eta _{4}\ \ \underline{\mathring{g}}_{4})]^{2}}{%
|\int d\varphi \partial _{\varphi }[(\ _{2}^{\shortmid }\widehat{\Upsilon }%
)(\ ^{\shortmid }\eta _{4}\ \ \underline{\mathring{g}}_{4})]|\ (\
^{\shortmid }\eta _{4}\ \ \underline{\mathring{g}}_{4})}, \\
\ ^{\shortmid }\eta ^{6}\ \ \underline{\mathring{g}}^{6} &=&-\frac{[\partial
^{6}(\ ^{\shortmid }\eta ^{5}\ \ \underline{\mathring{g}}^{5})]^{2}}{|\int
dp_{6}(\ _{3}^{\shortmid }\widehat{\Upsilon })\ \partial ^{6}[(\ ^{\shortmid
}\eta ^{5}\ \ \underline{\mathring{g}}^{5})]\ |\ (\ ^{\shortmid }\eta ^{5}\
\ \underline{\mathring{g}}^{5})},\ ^{\shortmid }\eta ^{7}\ \ \underline{%
\mathring{g}}^{7}=-\frac{[\frac{\partial }{\partial \theta _{4}}(\
^{\shortmid }\eta ^{8}\ \ \underline{\mathring{g}}^{8})]^{2}}{|\int d\theta
_{4}[(\ _{4}^{\shortmid }\widehat{\Upsilon })\frac{\partial }{\partial
\theta _{4}}(\ ^{\shortmid }\eta ^{8}\ \ \underline{\mathring{g}}^{8})\ |\
(\ ^{\shortmid }\eta ^{8}\ \ \underline{\mathring{g}}^{8})};
\end{eqnarray*}%
\begin{eqnarray}
\ ^{\shortmid }\eta _{i_{1}}^{3}\ ^{\shortmid }\mathring{N}_{i_{1}}^{3} &=&%
\frac{\partial _{i_{1}}\ \int d\varphi (\ _{2}^{\shortmid }\widehat{\Upsilon
})\ \partial _{\varphi }(\ ^{\shortmid }\eta _{4}\ \ \underline{\mathring{g}}%
_{4})}{(\ _{2}^{\shortmid }\widehat{\Upsilon })\ \partial _{\varphi }(\
^{\shortmid }\eta _{4}\ \ \underline{\mathring{g}}_{4})},
\label{noffdiagpolfr} \\
\ \ ^{\shortmid }\eta _{k_{1}}^{4}\ ^{\shortmid }\mathring{N}_{k_{1}}^{4}
&=&\ _{1}n_{k_{1}}+\ _{2}n_{k_{1}}\int d\varphi \frac{\lbrack \partial
_{\varphi }(\ ^{\shortmid }\eta _{4}\ \ \underline{\mathring{g}}_{4})]^{2}}{%
|\int d\varphi (\ _{2}^{\shortmid }\widehat{\Upsilon })\partial _{\varphi
}(\ ^{\shortmid }\eta _{4}\ \ \underline{\mathring{g}}_{4})|\ (\ ^{\shortmid
}\eta _{4}\ \ \underline{\mathring{g}}_{4})^{5/2}},  \notag
\end{eqnarray}%
\begin{eqnarray*}
\ ^{\shortmid }\eta _{k_{2}5}\ ^{\shortmid }\mathring{N}_{k_{2}5} &=&\
_{1}n_{k_{2}}+\ _{2}n_{k_{2}}\int d\theta _{3}\frac{[\frac{\partial }{%
\partial \theta _{3}}(\ ^{\shortmid }\eta ^{5}\ \ \underline{\mathring{g}}%
^{5})]^{2}}{|\int dp_{6}\ (\ _{3}^{\shortmid }\widehat{\Upsilon })\frac{%
\partial }{\partial \theta _{3}}(\ ^{\shortmid }\eta ^{5}\ \ \underline{%
\mathring{g}}^{5})|\ (\ ^{\shortmid }\eta ^{5}\ \ \underline{\mathring{g}}%
^{5})^{5/2}}, \\
\ ^{\shortmid }\eta _{i_{2}6}\ ^{\shortmid }\mathring{N}_{i_{2}6} &=&\frac{%
\partial _{i_{2}}\ \int d\theta _{3}(\ _{3}^{\shortmid }\widehat{\Upsilon }%
)\ \frac{\partial }{\partial \theta _{3}}(\ ^{\shortmid }\eta ^{5}\ \
\underline{\mathring{g}}^{5})}{\ _{3}^{\shortmid }\widehat{\Upsilon }\ \frac{%
\partial }{\partial \theta _{3}}(\ ^{\shortmid }\eta ^{5}\ \ \underline{%
\mathring{g}}^{5})},
\end{eqnarray*}%
\begin{eqnarray*}
\ ^{\shortmid }\eta _{i_{3}7}\ ^{\shortmid }\mathring{N}_{i_{3}7} &=&\frac{%
\partial _{i_{3}}\ \int d\theta _{4}(\ _{4}^{\shortmid }\widehat{\Upsilon }%
)\ \frac{\partial }{\partial \theta _{4}}(\ ^{\shortmid }\eta ^{8}\ \
\underline{\mathring{g}}^{8})}{\ _{4}^{\shortmid }\widehat{\Upsilon }\ \frac{%
\partial }{\partial \theta _{4}}(\ ^{\shortmid }\eta ^{8}\ \ \underline{%
\mathring{g}}^{8})}, \\
\ ^{\shortmid }\eta _{k_{3}8}\ ^{\shortmid }\mathring{N}_{k_{3}8} &=&\
_{1}n_{k_{3}}+\ _{2}n_{k_{3}}\int d\theta _{4}\frac{[\frac{\partial }{%
\partial \theta _{4}}(\ ^{\shortmid }\eta ^{8}\ \ \underline{\mathring{g}}%
^{8})]^{2}}{|\int d\theta _{4}\ (\ _{4}^{\shortmid }\widehat{\Upsilon })%
\frac{\partial }{\partial \theta _{4}}(\ ^{\shortmid }\eta ^{8}\ \
\underline{\mathring{g}}^{8})|\ (\ ^{\shortmid }\eta ^{8}\ \ \underline{%
\mathring{g}}^{8})^{5/2}}.
\end{eqnarray*}

A solution of type (\ref{offdiagpolfr}) with off-diagonal conditions (\ref%
{noffdiagpolfr}) defines nonholonomic deformations of the 7-d Tangherlini BH
(\ref{6usph}) into a 8-d phase space on $T^{\ast }\mathbf{V.}$ It should be
noted that such stationary solutions are different from all classes of
solutions constructed and considered in \cite{v18c} and above sections
because of fixed energy conditions $p_{8}=E=const$ and Killing symmetry on $%
\partial ^{8}=\partial _{E}.$

Finally we note that new classes of generic off-diagonal solutions in MGT
with MDRs constructed in this section contain (in general) a nontrivial
nonholonomical induced torsion. As in previous section, we can state
additional nonholonomic constraints to LC-configurations or to extract
diagonal metrics. The BH properties can not be preserved for general
nonholonomic deformations on the total phase space. It is important to
analyse additionally if certain topological and singularity conditions allow
projections into a base spacetime black hole configuration. Nevertheless, we
argue that at least for nonsingular small deformations on a parameter $%
\varepsilon ,$ such a s-metric describes a BH embedded self-consistently in
a locally anisotropic polarized phase space media with fixed energy
conditions.

\section{Finsler-Lagrange-Hamilton symmetries of phase space stationary and
BH configurations}

\label{s4} Any BH solution in GR and extra dimension gravity \footnote{%
for instance, the double phase space Schwarzschild solution $\ _{\circ \circ
}^{\shortmid }g_{\alpha \beta }$ (\ref{prm2}) and the Tangherlini 6-d and/or
7-d s-metrics, see $\ _{s}^{\shortmid }\mathbf{\mathring{g}}$ (\ref{pmtang})
and/or $\underline{\mathbf{\mathring{g}}}_{\alpha _{s}\beta _{s}}$ (\ref%
{6usph})} can be nonholonomically extended as exact and/or parametric
solutions in MGTs with MDRs. Such geometric and physical models are
elaborated on (co) tangent Lorentz bundles. Corresponding gravity theories
and their evolution and field equations, and their solutions, can be
described equivalently in Finsler-Lagrange-Hamilton variables, see an
axiomatic approach formulated in section 2 of \cite{v18b}. For
quasi-stationary solutions, such constructions are outlined in section 5.6
and Appendix B of \cite{v18c}.

The goal of this section is to study quasi-stationary Finsler like
symmetries of nonholonomic deformations of BH solutions. As a typical
example of prime metric, we shall consider $\ _{s}^{\shortmid }\mathbf{%
\mathring{g}}$ (\ref{pmtang}) and generalizations with an energy like
variable $E$ using local coordinates $(x^{1}=\ ^{\shortmid }r,x^{3}=\theta
=\theta _{1},y^{3}=\varphi ,y^{4}=t,p_{5}=\theta _{2},p_{6}=\theta
_{3},p_{7},p_{8}=E)$ on a 6-d spherical symmetric phase space $T^{\ast }%
\mathbf{V.}$ In Finsler like variables on total (co) bundles, such
quasi-stationary solutions my depend on a time like variable $t$ and other
types of space and (co) fiber coordinates. Geometric and analytic
constructions can be performed in a more simplified form if the generating
functions are taken with a Killing symmetry on $\partial ^{7}=\partial
/\partial p_{7}$ and (for projections on base spaces) on $\partial _{t}.$

\subsection{Lagrange-Hamilton variables and distortion of connections on
cotangent Lorentz bundles}

Fixing a point $x_{0}^{i^{\prime }}=\{\ ^{\shortmid }r_{0},\theta
_{0},\varphi _{0}\}\in V$ (for $i^{\prime }=1,2,3)$ on a BH configuration
and MDR (\ref{mdrg}), we can construct an effective Hamiltonian
\begin{equation}
H_{0}(p):=E=\pm (c^{2}\overrightarrow{\mathbf{p}}^{2}+c^{4}m^{2}-\varpi
(x_{0}^{i^{\prime }},\overrightarrow{\mathbf{p}},E,m;\ell _{P}))^{1/2}
\label{hamfp}
\end{equation}%
describing the motion of a relativistic point particle propagating in a
typical co-fiber of a $T_{x_{0}^{i^{\prime }}}^{\ast }V$ and $%
\overrightarrow{\mathbf{p}}=(p_{a^{\prime }})=(\theta _{2},\theta
_{3},p_{7}),$ for $a^{\prime }=5,6,7.$ Globalizing the constructions for a
BH with an effective phase space endowed with local coordinates $(\
^{\shortmid }r,\theta ,\varphi ,t,\theta _{2},\theta _{3},p_{7},E),$ we
obtain indicators $\varpi $ depending both on spherical spacetime and phase
space coordinates. In result, the motion of probing particles and linearized
interactions of scalar fields in $T^{\ast }V,$ can be conventionally
described by a Hamiltonian (\ref{hamfp}) generalized as $H(x,p)=H(\
^{\shortmid }r,\theta ,\varphi ,\theta _{2},\theta _{3},p_{7},E)$ for a
stationary relativistic mechanical model.

\subsubsection{BH solutions with associated Lagrange and Hamilton geometries}

The Legendre transforms and the concept of $L$-duality are introduced in
standard form $L\rightarrow H$ following definition
\begin{equation}
H(\ ^{\shortmid }r,\theta ,\varphi ,p_{a^{\prime }}):=p_{a^{\prime
}}v^{a^{\prime }}-L(\ ^{\shortmid }r,\theta ,\varphi ,v^{a^{\prime }}),
\label{legendre}
\end{equation}%
where the 'spherical' velocities $v^{a^{\prime }}$ are defined as solutions
of the equations $p_{a^{\prime }}=\partial L/\partial v^{a^{\prime }}.$ The
inverse Legendre transforms, $H\rightarrow L,$ are constructed
\begin{equation}
L:=p_{a^{\prime }}v^{a^{\prime }}-H,  \label{invlegendre}
\end{equation}%
where $p_{a^{\prime }}$ are solutions of the equations $y^{a^{\prime
}}=\partial H/\partial p_{a^{\prime }}.$

Nonholonomic deformations of BH solutions determined by MDRs (\ref{mdrg})
are characterized by non-Riemannian total phase space geometries with
nonlinear quadratic line elements
\begin{eqnarray}
ds_{L}^{2} &=&L(\ ^{\shortmid }r,\theta ,\varphi ,v^{a^{\prime }}),%
\mbox{
for models on  }TV;  \label{nqe} \\
d\ ^{\shortmid }s_{H}^{2} &=&H(\ ^{\shortmid }r,\theta ,\varphi ,\theta
_{2},\theta _{3},p_{7},E),\mbox{ for models on  }T^{\ast }V.  \label{nqed}
\end{eqnarray}%
Correspondingly, the values $L$ (\ref{nqe}) and $H$ (\ref{nqed}) are called
the Lagrange and Hamilton fundamental (equivalently, generating) functions
for associated mechanical systems. For stationary configurations, such a
relativistic 4-d model is geometrized as a Lagrange space $%
L^{3,1}=(TV,L(x,y))$ with vertical phase space metric structure when $L$ can
be also considered as a generating Lagrange function, $TV\ni (\ ^{\shortmid
}r,\theta ,\varphi ,v^{a^{\prime }})\rightarrow L(\ ^{\shortmid }r,\theta
,\varphi ,v^{a^{\prime }})\in \mathbb{R}.$

We model self-consistent mechanical models for a conventional $L$ considered
as a real valued and differentiable function on $\widetilde{TV}:=TV/\{0\},$
for $\{0\}$ being the null section of $TV,$ and continuous on the null
section of $\pi :TV\rightarrow V.$ Such a model is regular for stationary
configurations if the vertical metric (v-metric, defined as a Hessian
function)
\begin{equation}
\widetilde{g}_{a^{\prime }b^{\prime }}(\ ^{\shortmid }r,\theta ,\varphi
,y^{c^{\prime }}):=\frac{1}{2}\frac{\partial ^{2}L}{\partial y^{a^{\prime
}}\partial y^{b^{\prime }}}  \label{hessls}
\end{equation}%
is non-degenerate, i.e. $\det |\widetilde{g}_{a^{\prime }b^{\prime }}|\neq
0, $ and of constant local Euclidean signature.

A BH configuration with MDRs can be described alternatively by a 4-d
relativistic model of Hamilton space $H^{3,1}=(T^{\ast }V,H(\ ^{\shortmid
}r,\theta ,\varphi ,\theta _{2},\theta _{3},p_{7},E))$ which is determined
by a fundamental Hamilton function on a respective base spacetime Lorentz
manifold $V.$ Such a real valued generating function is defined as a map $%
T^{\ast }V\ni (\ ^{\shortmid }r,\theta ,\varphi ,\theta _{2},\theta
_{3},p_{7},E)\rightarrow H(\ ^{\shortmid }r,\theta ,\varphi ,\theta
_{2},\theta _{3},p_{7},E)\in \mathbb{R}$ for which there are satisfied the
conditions that it is differentiable on $\widetilde{T^{\ast }V}:=T^{\ast
}V/\{0^{\ast }\},$ for $\{0^{\ast }\}$ being the null section of $T^{\ast
}V, $ and continuous on the null section of $\pi ^{\ast }:\ T^{\ast
}V\rightarrow V.$ We define an analogous regular mechanical model if the
co-vertical metric (cv-metric, Hessian),
\begin{equation}
\ \ ^{\shortmid }\widetilde{g}^{a^{\prime }b^{\prime }}(\ ^{\shortmid
}r,\theta ,\varphi ,p_{c^{\prime }}):=\frac{1}{2}\frac{\partial ^{2}H}{%
\partial p_{a^{\prime }}\partial p_{b^{\prime }}}  \label{hesshs}
\end{equation}%
is non-degenerate, i.e. $\det |\ ^{\shortmid }\widetilde{g}^{a^{\prime
}b^{\prime }}|\neq 0,$ and of constant signature.\footnote{%
In our works, we use tilde "\symbol{126}" in order to emphasize that certain
geometric objects are defined canonically by respective Lagrange and/or
Hamilton generating functions. For instance, we write $\widetilde{g}_{ab}$
and $\ ^{\shortmid }\widetilde{g}^{ab}.$ Such (co) vertical metrics may
encode various types of MDRs and LIVs terms etc. Considering general frame/
coordinate transforms on $TV$ and/or $T^{\ast }V,$ we can express any
"tilde" value in a "non-tilde" form. In such cases, we shall write $g_{ab},$
for a v-metric, and $\ ^{\shortmid }g^{ab},$ for a cv-metric.}

Here we note that a model of Finsler phase space is an example of
relativistic Lagrange mechanics when a regular Lagrangian $L=F^{2}$ is
defined by a fundamental (generating) Finsler function subjected to
additional conditions: 1) a fundamental Finsler function $F$ is a real
positive valued one which is differential on $\widetilde{TV}$ and continuous
on the null section of the projection $\pi :TV\rightarrow V$, where $V$ is a
Lorentz manifolds;\ 2) a generating function $F$ satisfies the homogeneity
condition $F(x,\lambda v)=|\lambda |$ $F(x,v),$ for a nonzero real value $%
\lambda ;$ and a Hessian (\ref{hessls}) is defined by $F^{2}$ in such a form
that in any point $(x_{(0)},v_{(0)})$ the v-metric is of signature $(+++-).$
In a similar form, we can define relativistic 4-d Cartan spaces $%
C^{3,1}=(V,C(x,p)),$ when $H=C^{2}(x,p)$ is 1-homogeneous on co-fiber
coordinates $p_{a}$; such Cartan spaces are Finsler ones but on cotangent
bundles when additions conditions for Legendre transforms and possible
almost symplectic structures can be considered, see details and references
in \cite{v18b}. In this work, for simplicity, we shall analyse properties of
BHs with associated Hamilton geometries determined by MDRs.

\subsubsection{N-connections and stationary frames in Lagrange-Hamilton
phase spaces}

Using the formula for MDR (\ref{mdrg}), we can define $L$--dual (related via
Legendre transforms) canonical stationary configurations of Hamilton space $%
\widetilde{H}^{3,1}=(T^{\ast }V,\widetilde{H}(x,p))$ and Lagrange space $%
\widetilde{L}^{3,1}=(TV,\widetilde{L}(x,y)).$ We can consider general
Lagrange or Hamilton variables which may depend explicitly on time and/or
energy like coordinates. In such phase space BH and nonholonomically
deformed configurations, the dynamics of a probing point particle is
described by fundamental generating functions $\widetilde{H}$ and $%
\widetilde{L}$ \ subjected to respective via Hamilton-Jacobi and Lagrange
equations
\begin{equation*}
\frac{dx^{i}}{d\tau }=\frac{\partial \widetilde{H}}{\partial p_{i}}%
\mbox{
and }\frac{dp_{i}}{d\tau }=-\frac{\partial \widetilde{H}}{\partial x^{i}},
\end{equation*}%
\begin{equation*}
\frac{d}{d\tau }\frac{\partial \widetilde{L}}{\partial y^{i}}-\frac{\partial
\widetilde{L}}{\partial x^{i}}=0.
\end{equation*}%
These equations are equivalent to some nonlinear geodesic (semi-spray)
equations
\begin{equation}
\frac{d^{2}x^{i}}{d\tau ^{2}}+2\widetilde{G}^{i}(x,y)=0,\mbox{ for }%
\widetilde{G}^{i}=\frac{1}{2}\widetilde{g}^{ij}(\frac{\partial ^{2}%
\widetilde{L}}{\partial y^{i}}y^{k}-\frac{\partial \widetilde{L}}{\partial
x^{i}}),  \label{ngeqf}
\end{equation}%
where $\widetilde{g}^{ij}$ is inverse to $\widetilde{g}_{ij}$ (\ref{hessls}%
). The value (\ref{ngeqf}) can be used for defining canonical N--connection
structures determined by MDRs in $L$--dual form when
\begin{equation}
\ \widetilde{\mathbf{N}}=\left\{ \widetilde{N}_{i}^{a}:=\frac{\partial
\widetilde{G}}{\partial y^{i}}\right\} \mbox{ and }\ ^{\shortmid }\widetilde{%
\mathbf{N}}=\left\{ \ ^{\shortmid }\widetilde{N}_{ij}:=\frac{1}{2}\left[ \{\
\ ^{\shortmid }\widetilde{g}_{ij},\widetilde{H}\}-\frac{\partial ^{2}%
\widetilde{H}}{\partial p_{k}\partial x^{i}}\ ^{\shortmid }\widetilde{g}%
_{jk}-\frac{\partial ^{2}\widetilde{H}}{\partial p_{k}\partial x^{j}}\
^{\shortmid }\widetilde{g}_{ik}\right] \right\}  \label{canonncon}
\end{equation}%
and $\ \ ^{\shortmid }\widetilde{g}_{ij}$ is inverse to $\ \ ^{\shortmid }%
\widetilde{g}^{ab}$ (\ref{hesshs}).

The canonical N--connections $\widetilde{\mathbf{N}}$ and $\ ^{\shortmid }%
\widetilde{\mathbf{N}}$ from (\ref{canonncon}) define respective systems of
N--adapted (co) frames
\begin{eqnarray}
\widetilde{\mathbf{e}}_{\alpha } &=&(\widetilde{\mathbf{e}}_{i}=\frac{%
\partial }{\partial x^{i}}-\widetilde{N}_{i}^{a}(x,y)\frac{\partial }{%
\partial y^{a}},e_{b}=\frac{\partial }{\partial y^{b}}),\mbox{ on }TV;
\label{cnddapb} \\
\widetilde{\mathbf{e}}^{\alpha } &=&(\widetilde{e}^{i}=dx^{i},\widetilde{%
\mathbf{e}}^{a}=dy^{a}+\widetilde{N}_{i}^{a}(x,y)dx^{i}),\mbox{ on }%
(TV)^{\ast };\mbox{and \ }  \notag \\
\ ^{\shortmid }\widetilde{\mathbf{e}}_{\alpha } &=&(\ ^{\shortmid }%
\widetilde{\mathbf{e}}_{i}=\frac{\partial }{\partial x^{i}}-\ ^{\shortmid }%
\widetilde{N}_{ia}(x,p)\frac{\partial }{\partial p_{a}},\ ^{\shortmid }e^{b}=%
\frac{\partial }{\partial p_{b}}),\mbox{ on }T^{\ast }V;  \label{ccnadap} \\
\ \ ^{\shortmid }\widetilde{\mathbf{e}}^{\alpha } &=&(\ ^{\shortmid
}e^{i}=dx^{i},\ ^{\shortmid }\mathbf{e}_{a}=dp_{a}+\ ^{\shortmid }\widetilde{%
N}_{ia}(x,p)dx^{i})\mbox{ on }(T^{\ast }V)^{\ast }.  \notag
\end{eqnarray}%
The canonical frames and spherical coordinates can be prescribed in such
forms that the coefficients of Finsler like stationary d-metrics and
N-connections do not depend on a respective time like coordinate.

We conclude that BH configurations in GR and extra dimension gravity are
nonholonomically deformed by MDR (\ref{mdrg}) and can be described in
equivalent forms using canonical data $(\widetilde{L},\ \widetilde{\mathbf{N}%
};\widetilde{\mathbf{e}}_{\alpha },\widetilde{\mathbf{e}}^{\alpha })$ and/or
$(\widetilde{H},\ ^{\shortmid }\widetilde{\mathbf{N}};\ ^{\shortmid }%
\widetilde{\mathbf{e}}_{\alpha },\ ^{\shortmid }\widetilde{\mathbf{e}}%
^{\alpha })$ for Lagrange and/or Hamilton spaces. Considering general frame
and coordinate transforms, we omit tilde and work with a general N-splitting
and geometric data $(\mathbf{N};\mathbf{e}_{\alpha },\mathbf{e}^{\alpha })$
and/or $(\ ^{\shortmid }\mathbf{N};\ ^{\shortmid }\mathbf{e}_{\alpha },\
^{\shortmid }\mathbf{e}^{\alpha }).$

\subsubsection{Canonical d-metrics and d-connections for Lagrange-Hamilton
spaces and BHs}

There are canonical d-metric structures $\widetilde{\mathbf{g}}$ and $\
^{\shortmid }\widetilde{\mathbf{g}}$ completely determined by a MDR (\ref%
{mdrg}) and respective "tilde" data,
\begin{eqnarray}
\widetilde{\mathbf{g}} &=&\widetilde{\mathbf{g}}_{\alpha \beta }(x,y)%
\widetilde{\mathbf{e}}^{\alpha }\mathbf{\otimes }\widetilde{\mathbf{e}}%
^{\beta }=\widetilde{g}_{ij}(x,y)e^{i}\otimes e^{j}+\widetilde{g}_{ab}(x,y)%
\widetilde{\mathbf{e}}^{a}\otimes \widetilde{\mathbf{e}}^{a}\mbox{
and/or }  \label{cdms} \\
\ ^{\shortmid }\widetilde{\mathbf{g}} &=&\ ^{\shortmid }\widetilde{\mathbf{g}%
}_{\alpha \beta }(x,p)\ ^{\shortmid }\widetilde{\mathbf{e}}^{\alpha }\mathbf{%
\otimes \ ^{\shortmid }}\widetilde{\mathbf{e}}^{\beta }=\ \ ^{\shortmid }%
\widetilde{g}_{ij}(x,p)e^{i}\otimes e^{j}+\ ^{\shortmid }\widetilde{g}%
^{ab}(x,p)\ ^{\shortmid }\widetilde{\mathbf{e}}_{a}\otimes \ ^{\shortmid }%
\widetilde{\mathbf{e}}_{b},  \label{cdmds}
\end{eqnarray}%
where $\widetilde{\mathbf{e}}_{\alpha }=(\widetilde{\mathbf{e}}_{i},e_{b})$
and$\ ^{\shortmid }\widetilde{\mathbf{e}}_{\alpha }=(\ ^{\shortmid }%
\widetilde{\mathbf{e}}_{i},\ ^{\shortmid }e^{b}).$ Such nonholonomic frame
bases are characterized by corresponding anholonomy relations, for instance,
of type
\begin{equation*}
\lbrack \ ^{\shortmid }\widetilde{\mathbf{e}}_{\alpha },\ ^{\shortmid }%
\widetilde{\mathbf{e}}_{\beta }]=\ ^{\shortmid }\widetilde{\mathbf{e}}%
_{\alpha }\ ^{\shortmid }\widetilde{\mathbf{e}}_{\beta }-\ ^{\shortmid }%
\widetilde{\mathbf{e}}_{\beta }\ ^{\shortmid }\widetilde{\mathbf{e}}_{\alpha
}=\ ^{\shortmid }\widetilde{W}_{\alpha \beta }^{\gamma }\ ^{\shortmid }%
\widetilde{\mathbf{e}}_{\gamma },
\end{equation*}%
with anholonomy coefficients $\ ^{\shortmid }\widetilde{W}_{ib}^{a}=\partial
\ ^{\shortmid }\widetilde{N}_{ib}/\partial p_{a}$ and $\ ^{\shortmid }%
\widetilde{W}_{jia}=\ \mathbf{\ ^{\shortmid }}\widetilde{\Omega }_{ija},$
see explicit definitions and formulas in \cite{v18a}.

The geometry of associated phase geometries is also characterized by
respective Cartan-Lagrange and Cartan-Hamilton d-connections induced
directly by an indicator of MDR and determined by corresponding coefficients
of Lagrange and Hamilton d-metrics (\ref{cdms}) and (\ref{cdmds}). The
coefficients of such values are generated by "tilde" objects with
identifications of d-metric coefficients with corresponding base and (co)
fiber indices:
\begin{eqnarray}
\mbox{ on }T\mathbf{TV},\ \widetilde{\mathbf{D}} &=&\{\widetilde{\mathbf{%
\Gamma }}_{\ \alpha \beta }^{\gamma }=(\widetilde{L}_{jk}^{i},\widetilde{L}%
_{bk}^{a},\widetilde{C}_{jc}^{i},\widetilde{C}_{bc}^{a})\},\mbox{ for }%
\mathbf{[}\widetilde{\mathbf{g}}_{\alpha \beta }=(\widetilde{g}_{jr},%
\widetilde{g}_{ab}),\widetilde{\mathbf{N}}_{i}^{a}=\widetilde{N}_{i}^{a}],
\notag \\
\widetilde{L}_{jk}^{i} &=&\frac{1}{2}\widetilde{g}^{ir}\left( \widetilde{%
\mathbf{e}}_{k}\widetilde{g}_{jr}+\widetilde{\mathbf{e}}_{j}\widetilde{g}%
_{kr}-\widetilde{\mathbf{e}}_{r}\widetilde{g}_{jk}\right) ,\ \widetilde{L}%
_{bk}^{a}\mbox{ as }\widetilde{L}_{jk}^{i},  \notag \\
\ \widetilde{C}_{bc}^{a} &=&\frac{1}{2}\widetilde{g}^{ad}\left( e_{c}%
\widetilde{g}_{bd}+e_{b}\widetilde{g}_{cd}-e_{d}\widetilde{g}_{bc}\right) %
\mbox{ being similar to }\widetilde{C}_{jc}^{i};  \label{carlc}
\end{eqnarray}%
\begin{eqnarray}
\mbox{ on }T\mathbf{T}^{\ast }\mathbf{V},\ \ ^{\shortmid }\widetilde{\mathbf{%
D}} &=&\{\ ^{\shortmid }\widetilde{\mathbf{\Gamma }}_{\ \alpha \beta
}^{\gamma }=(\ ^{\shortmid }\widetilde{L}_{jk}^{i},\ ^{\shortmid }\widetilde{%
L}_{a\ k}^{\ b},\ ^{\shortmid }\widetilde{C}_{\ j}^{i\ c},\ ^{\shortmid }%
\widetilde{C}_{\ j}^{i\ c})\},\mbox{ for }\mathbf{[}\ ^{\shortmid }%
\widetilde{\mathbf{g}}_{\alpha \beta }=(\ ^{\shortmid }\widetilde{g}_{jr},\
^{\shortmid }\widetilde{g}^{ab}),\ ^{\shortmid }\widetilde{\mathbf{N}}%
_{ai}=\ ^{\shortmid }\widetilde{N}_{ai}],  \notag \\
\ ^{\shortmid }\widetilde{L}_{jk}^{i} &=&\frac{1}{2}\ ^{\shortmid }%
\widetilde{g}^{ir}(\ ^{\shortmid }\widetilde{\mathbf{e}}_{k}\ ^{\shortmid }%
\widetilde{g}_{jr}+\ ^{\shortmid }\widetilde{\mathbf{e}}_{j}\ ^{\shortmid }%
\widetilde{g}_{kr}-\ ^{\shortmid }\widetilde{\mathbf{e}}_{r}\ ^{\shortmid }%
\widetilde{g}_{jk}),\ \mbox{ with similar }\ ^{\shortmid }\widetilde{L}_{a\
k}^{\ b},  \notag \\
\ \ \ ^{\shortmid }\widetilde{C}_{\ a}^{b\ c} &=&\frac{1}{2}\ ^{\shortmid }%
\widetilde{g}_{ad}(\ ^{\shortmid }e^{c}\ ^{\shortmid }\widetilde{g}^{bd}+\
^{\shortmid }e^{b}\ ^{\shortmid }\widetilde{g}^{cd}-\ ^{\shortmid }e^{d}\
^{\shortmid }\widetilde{g}^{bc})\mbox{ being similar to }^{\shortmid }%
\widetilde{C}_{\ j}^{i\ c}.  \label{carhc}
\end{eqnarray}%
In Lagrange-Hamilton geometry, we can introduce different types of metric
compatible or incompatible Finsler like d-connections which characterize
phase symmetries of BHs deformed by MDRs, see details in \cite{v18a} and
references therein.

Finally, it should be noted that all above formulas considered for Finsler--Lagrange--Hamilton structures (metrics and and nonlinear/ linear connections and respective torsions and curvatures) can be written in arbitrary frames of references because all geometric constructions are performed on (co) tangent Lorentz bundles.  Such formulas can be written both in abstract or coefficient  forms which are very similar to respective vielbein, tetradic and diadic ones in GR but for some generalize metric-affine and N-connection structures. If we wont to adapt the constructions to a N-connection splitting (\ref{ncon}) and/or (\ref{ncon2}), we should consider a subclass of frame transforms preserving the corresponding h- and v-decompositions, see details in Refs. \cite{v18b,v18c,gvvepjc14}.

\subsection{Phase space BHs in Einstein-Hamilton gravity \& associated
Finsler-Lagrange geometry}

Any BH and s-metric structure can be re-written equivalently in terms of
Finsler like d-metric and d-connection structures using corresponding
N-adapted splitting and frame transforms. We can elaborate on physical
properties of effective Lagrange-Hamilton spaces which are characterized by
different N- and d-connection structures. This results in different models
of associated BH phase space relativistic mechanics. Using nonlinear
symmetries relating generating functions, generating sources and
cosmological constants, we can model various classes of off-diagonal
solutions in a MGT with MDRs as certain classes of Einstein-Hamilton spaces.

Let us consider a nonholonomic diadic structure $\ ^{\shortmid }\mathbf{e}%
_{\alpha _{s}}$ on an open region $U\subset \mathbf{T}^{\ast }\mathbf{V}$.
We can introduce respective vielbein structures, $\ \mathbf{e}_{\ \alpha
_{s}}^{\alpha }\ $\ and $\ ^{\shortmid }\mathbf{e}_{\ \alpha _{s}}^{\alpha
}, $ determined by values of type $H(x,p)$ (\ref{nqed}) and $\widetilde{%
\mathbf{e}}_{\alpha }$ (\ref{ccnadap}) and stated by frame transforms
\begin{equation*}
\mathbf{e}_{\alpha _{s}}=\mathbf{e}_{\ \alpha _{s}}^{\alpha }\ ^{\shortmid }%
\widetilde{\mathbf{e}}_{\alpha },%
\mbox{ for Lagrange (Finsler) variables,
and }\ ^{\shortmid }\mathbf{e}_{\alpha _{s}}=\ ^{\shortmid }\mathbf{e}_{\
\alpha _{s}}^{\alpha }\ ^{\shortmid }\widetilde{\mathbf{e}}_{\alpha },\ %
\mbox{ for Hamilton variables}.
\end{equation*}%
Using such values and dual bases and respective inverse matrices, $\ \mathbf{%
e}_{\alpha ^{\prime }\ }^{\ \alpha _{s}}\ $\ and $\ ^{\shortmid }\mathbf{e}%
_{\alpha ^{\prime }\ }^{\ \alpha _{s}},$ we can always redefine the
geometric data in the form
\begin{equation*}
(\ ^{s}\mathbf{N};\mathbf{e}_{\alpha _{s}},\mathbf{e}^{\alpha
_{s}})\longleftrightarrow (\widetilde{L},\ \widetilde{\mathbf{N}};\widetilde{%
\mathbf{e}}_{\alpha },\widetilde{\mathbf{e}}^{\alpha })\mbox{ and/or }(\
_{s}^{\shortmid }\widetilde{\mathbf{N}};\ ^{\shortmid }\widetilde{\mathbf{e}}%
_{\alpha _{s}},\ ^{\shortmid }\widetilde{\mathbf{e}}^{\alpha
_{s}})\longleftrightarrow (\widetilde{H},\ ^{\shortmid }\widetilde{\mathbf{N}%
};\ ^{\shortmid }\widetilde{\mathbf{e}}_{\alpha },\ ^{\shortmid }\widetilde{%
\mathbf{e}}^{\alpha }).
\end{equation*}

Any BH s-metric and respective off-diagonal metric structure can be
described equivalently as a Hamilton canonical d-metric (\ref{cdmds}) and
inversely. This follows from the possibility to consider necessary type
frame transforms $\ ^{\shortmid }\mathbf{g}_{\alpha _{s}\beta _{s}}=\
^{\shortmid }\mathbf{e}_{\ \alpha _{s}}^{\alpha }\ ^{\shortmid }\mathbf{e}%
_{\ \beta _{s}}^{\beta }\ ^{\shortmid }\widetilde{\mathbf{g}}_{\alpha \beta
} $ and work equivalently with data $\ ^{\shortmid }\mathbf{g}_{\alpha
_{s}\beta _{s}}$ or $\ ^{\shortmid }\widetilde{\mathbf{g}}_{\alpha \beta }.$
Considering distortions $\ _{s}^{\shortmid }\widehat{\mathbf{D}}=\nabla +\
_{s}^{\shortmid }\widehat{\mathbf{Z}}$ $=\ ^{\shortmid }\widetilde{\mathbf{D}%
}+\ _{s}^{\shortmid }\widetilde{\mathbf{Z}},$ we compute $\ _{s}^{\shortmid }%
\widehat{\mathbf{D}}=\ ^{\shortmid }\widetilde{\mathbf{D}}+$ $\
_{s}^{\shortmid }\widehat{\mathbf{Z}},$ for any $\ ^{\shortmid }\mathbf{g=}\
_{s}^{\shortmid }\mathbf{g=}\ ^{\shortmid }\mathbf{\tilde{g}.}$ This allows
us to express the distortions of the Ricci d-tensor of type $\ ^{\shortmid }%
\mathbf{R}_{\alpha _{s}\beta _{s}}$ as distortions
\begin{equation*}
\ ^{\shortmid }\widehat{\mathbf{R}}_{\alpha _{s}\beta _{s}} = \ ^{\shortmid }%
\widetilde{\mathbf{R}}_{\alpha _{s}\beta _{s}}[\ \ ^{\shortmid }\mathbf{%
\tilde{g}},\ \ ^{\shortmid }\widetilde{\mathbf{D}}]+\ ^{\shortmid }%
\widetilde{\mathbf{Z}}_{\alpha _{s}\beta _{s}}[\ \ ^{\shortmid }\mathbf{%
\tilde{g}},\ \ ^{\shortmid }\widetilde{\mathbf{D}}],\ \mbox{ where }\
_{\shortmid }^{e}\widehat{\Upsilon }_{\alpha _{s}\beta _{s}} = -\
^{\shortmid }\widetilde{\mathbf{Z}}_{\alpha _{s}\beta _{s}}.
\end{equation*}
We conclude that the canonical distortion relations $\ _{s}^{\shortmid }%
\widehat{\mathbf{D}}=\ ^{\shortmid }\widetilde{\mathbf{D}}+$ $\
_{s}^{\shortmid }\widehat{\mathbf{Z}}$ result in effective sources of type
\begin{equation*}
\ _{\shortmid }^{e}\widehat{\Upsilon }_{\alpha _{s}\beta _{s}}:=\varkappa (\
\ _{\shortmid }^{e}\widehat{\mathbf{T}}_{\alpha _{s}\beta _{s}}-\frac{1}{2}\
^{\shortmid }\mathbf{g}_{\alpha _{s}\beta _{s}}\ \ \ _{\shortmid }^{e}%
\widehat{\mathbf{T}}), \mbox{ where } \varkappa \ \ \ _{\shortmid }^{e}%
\widehat{\mathbf{T}}_{\alpha _{s}\beta _{s}}=-\ ^{\shortmid }\widehat{%
\mathbf{Z}}_{\alpha _{s}\beta _{s}}[\ _{s}^{\shortmid }\mathbf{g[\
^{\shortmid }\widetilde{\mathbf{g}}}_{\alpha \beta }],\ \ ^{\shortmid }%
\widetilde{\mathbf{D}}].
\end{equation*}

For effective sources, the modified Einstein equations (\ref{meinsteqtbcand}%
) can be modelled equivalently as solutions of generalized Einstein-Hamilton
equations,
\begin{equation*}
\ ^{\shortmid }\widetilde{\mathbf{R}}_{\alpha \beta }[\ ^{\shortmid }%
\widetilde{\mathbf{D}}]=\ ^{\shortmid }\widetilde{\Upsilon }_{\alpha \beta }
\end{equation*}%
where the diadic indices can be omitted. For such an equivalent
representation, the sources are correspondingly redefined by formulas
\begin{equation*}
\ ^{\shortmid }\widehat{\Upsilon }_{\alpha _{s}\beta _{s}}=\ ^{\shortmid }%
\widetilde{\Upsilon }_{\alpha _{s}\beta _{s}}+\ _{\shortmid }^{e}\widehat{%
\Upsilon }_{\alpha _{s}\beta _{s}}\mbox{ and/or }\ \ \ _{\shortmid s}\Lambda
=\ \ _{\shortmid s}^{\phi }\Lambda +\ \ _{\shortmid s}^{e}\widetilde{\Lambda
},
\end{equation*}%
where $\ _{\shortmid s}^{\phi }\Lambda $ is the cosmological constant
associated to matter fields and vacuum fluctuations and the term with tilde
$\ _{\shortmid s}^{e}\widetilde{\Lambda }$ points to possible
contributions to the cosmological constant \ resulting from Hamilton like
degrees of freedom.

\section{Entropies of BHs with MDRs \& Stationary Ricci Solitons}

\label{s5} The stationary solutions in MGTs with MDRs constructed in this
work can be generated for phase space metrics of signature $(+++-;+++-)$ on $%
T^{\ast}\mathbf{V}$ and/or $T\mathbf{V}$. For certain classes of small
parametric deformations (for instance, with spheroid topologies and horizons
for black ellipsoids) of prime metrics describing 6-d, (4+4)-d and 7-d BH
configurations embedded into 8-d phase spaces, we can elaborate on
generalizations of the Bekenstein-Hawking BH entropy and spacetime
thermodynamics \cite{bek1,bek2,haw1,haw2}. We also generated nonholonomic
deformations to stationary generic off-diagonal metrics with generalized
connections and constraints to LC-configurations. Here it should be noted
that applying the anholonomic frame deformation method, AFDM, there were
constructed and studied various classes of more general (non) stationary and
locally anisotropic cosmological solutions in GR and MGTs. Such (non)
commutative, or supersymmetric, fractional and other types Finsler like BH
solutions were studied in a series of works \cite%
{v1998bh,v2001bh,v2002bh,v2003be1,
v2003be2,v2007bh,v2010abh,v2010bbh,v2011fbr,
v2013abe,v2013bbe,v2014abh,v2014bbh}, see also recent works on nonholonomic
superstrings and branes \cite{v18a,gvvepjc14,bubuianucqg17,vacaruepjc17}.
Our conclusion was that the concept of Bekenstein-Hawking entropy and
related thermodynamic characteristics are not applicable for generalized
spacetimes and phase spaces. The motivation is that general exact and
parametric solutions with MDRs are generic off-diagonal and determined by
generating functions and sources and integration functions with generalized
symmetries and depending on various types of space and phase space
variables. Standard and generalized Bekenstein-Hawking definitions are
possible only for subclasses of solutions with conventional horizons, higher
symmetries for duality of gravity and conformal field theories and/or
additional assumptions on related holographic principle etc. In general
form, MDRs and LIVs result in substantial deformations of BH and
cosmological solutions to configurations with nontrivial locally anisotropic
vacuum and non-vacuum configurations with rich geometric structure and/or
locally anisotropic polarizations of physical constants.

In a series of works \cite%
{vrfijmpa,vacvis,valexiu,ventrlf,vejtp1,vejtp2,vnonsymmet,
vnhrf,vnrflnc,vrfdif,bubvcjp,vrevrf,vvrfthbh}, we developed a new
statistical and geometric thermodynamics approach which allows us to
characterize physical properties of generic off-diagonal configurations in
GR and MGTs, see recent results in \cite%
{muen01,vmedit,vrajpootjet,rajpsuprf,vprocmg}. Such gravitational and matter
field geometric flow theories and generalized Ricci solitons can be
elaborated following G. Perelman's definitions of W- and F-entropies \cite%
{perelman1}. The goal of this section is to elaborate on generalized
Bekenstein-Hawking and Perelman thermodynamic models for stationary and BH
configurations with MDRs on phase spaces.

\subsection{Generalized phase space Bekenstein-Hawking entropy}

The formula of the Bekenstein-Hawking entropy $S_{BH}$ for a 4-d BH can be
written in the form
\begin{equation*}
S_{BH}=\frac{c^{3}k_{B}A}{4G\hbar },
\end{equation*}%
where $A$ is the BH surface area (the event horizon), $\hbar $ is the
reduced Planck constant, $c$ is the speed of light, $k_{B}$ is the Boltzmann
constant and G is the gravitational constant. Via such fundamental physical
constants, this formula counts of the BH effective degrees of freedom and
ties together notions from gravitation, thermodynamics and quantum theory.
Various generalizations the Bekenstein-Hawking approach were considered in
MGTs of various dimension and solutions with conventional area of horizons.

\subsubsection{The entropy and temperature of higher dimension Schwarzschild
BHs}

For higher dimensions, the thermodynamics of multidimensional BHs was
studied \cite{konoplya1,konoplya2,gao,yangwen,pourhassan,hollands} and
references therein. The Hawking temperature of a Schwarzschild BH in a
spacetime of dimension $s^{\prime }\geq 4$ is given by formulas (see, for
instance, \cite{pourhassan})
\begin{equation}
T=\frac{s^{\prime }-3}{4\pi }\left( \frac{\Omega _{s^{\prime }-2}}{4}\right)
^{1/(s^{\prime }-2)}S_{0}^{-1/(s^{\prime }-2)},  \label{hawtemp}
\end{equation}%
where $S_{0}$ is the Bekenstein-Hawking entropy is
\begin{equation}
S_{0}=\frac{\Omega _{s^{\prime }-2}(\hat{r}_{0})^{s^{\prime }-2}}{4},
\label{bekhawentr}
\end{equation}%
is determined by the volume of unit -sphere, $\Omega _{s^{\prime }-2}=\frac{%
\pi ^{s^{\prime }/2}}{(s^{\prime }/2)!}$ and horizon radius \newline
$\hat{r}_{0}=(\frac{8\pi \widehat{M}}{(s^{\prime }-2)\Omega _{s^{\prime }-2}}%
)^{1/(s^{\prime }-3)}$. In these formulas, the radial coordinate%
\begin{equation*}
\hat{r}=\sqrt{(x^{1})^{2}+(x^{2})^{2}+(y^{3})^{2}+(y^{5})^{2}+...+(y^{s^{%
\prime }})^{2}}
\end{equation*}%
is defined for a $s^{\prime }-2$ dimensional space like hypersurface endowed
with Cartezian coordinates\newline
$(x^{1},x^{2},y^{3},y^{5},...y^{s^{\prime }}),$ when $y^{4}=t.$ The
quadratic line element is parameterized
\begin{equation*}
ds^{2}=\ \ \widehat{h}^{-1}(\hat{r})d\hat{r}^{2}-\ \widehat{h}(\hat{r}%
)dt^{2}+\hat{r}^{2}d\Omega _{s^{\prime }-2}^{2},
\end{equation*}%
where
\begin{equation*}
\ \ \widehat{h}(\hat{r})=1-\frac{16\pi \widehat{M}}{(s^{\prime }-2)\Omega
_{s^{\prime }-2}(\hat{r})^{s^{\prime }-3}}
\end{equation*}%
is determined by the BH\ mass $\widehat{M}.$

Above formulas can be modified and applied for computing the entropy and
temperature of phase space BHs defined by respective prime metrics $\
^{\shortmid }\mathbf{\mathring{g}}_{\alpha _{s}\beta _{s}}$ (\ref{pmtang}),$%
\ _{\circ \circ }^{\shortmid }g_{\alpha \beta }$ (\ref{prm2}), and/or $%
\underline{\mathbf{\mathring{g}}}_{\alpha _{s}\beta _{s}}$ (\ref{6usph}).

\subsubsection{The Bekenstein-Hawking thermodynamics for 6-d phase space
Schwarzschild BHs}

Considering a prime metric (\ref{pmtang}) with $\ ^{\shortmid }\Lambda =0,$
for $s^{\prime }=6,$ and with trivial extensions on coordinates $(p_{7},E)$
on $\mathbf{T}^{\ast }\mathbf{V,}$ the analogs of BH temperature and entropy
are computed
\begin{equation*}
\ ^{\shortmid }T=\frac{3}{4\pi }\left( \frac{\ ^{\shortmid }\Omega _{4}}{4}%
\right) ^{1/4}\ ^{\shortmid }S_{0}^{-1/4}\mbox{ for }\ ^{\shortmid }S_{0}=%
\frac{\ ^{\shortmid }\Omega _{4}(\ ^{\shortmid }r_{0})^{4}}{4}.
\end{equation*}%
In these formulas, we put a left label $\ $"$^{\shortmid }$" in order to
emphasize that all values are defined for a 6-d phase space BH with the
volume of unit -sphere $\ ^{\shortmid }\Omega _{4}=\frac{\pi ^{3}}{3!}$ and
horizon radius $\ ^{\shortmid }r_{0}=(\frac{2\pi \ ^{\shortmid }M}{%
^{\shortmid }\Omega _{4}})^{1/3}.$ In principle, the values $\ ^{\shortmid
}T $ and $\ ^{\shortmid }S_{0}$ $\ $are given by respective formulas (\ref%
{hawtemp}) and (\ref{bekhawentr}) when the coordinates and physical
constants are redefined for 6-d phase spaces.

The formulas for $\ ^{\shortmid }T$ and $\ ^{\shortmid }S_{0}$ can be
extended for ellipsoidal and/or toroidal configurations (black ellipsoids
and black torus with a conventional hypersurface area) on $E$ as we
considered for various classes of (commutative and noncommutative, or
superstring) nonholonomic and Finsler like generalizations \cite%
{v2003be1,v2003be2,v2007bh,v2010bbh,v2013abe,v2013bbe,gvvepjc14,bubuianucqg17,vacaruepjc17}%
. \ Nevertheless, we can not develop the Bekenstein-Hawking thermodynamics
for more general classes of stationary phase space solutions constructed in
section \ref{s2}. There are attempts to study properties of rainbow BHs \cite%
{sefiedgar,magueijo04,gangop} with some phenomenological dependencies of
physical constants and horizons on $E$ and on a cutting parameter but those
works are not based on finding exact solutions of certain modified Einstein
equation. In our approach, we are able to construct exact solutions for
generalized gravitational field equations in MTS with MDRs and to elaborate
on generalized thermodynamic models of gravitational interactions.

\subsubsection{Bekenstein-Hawking thermodynamics for 7-d phase Schwarzschild
BHs with energy Killing symmetry}

We can consider a prime metric $\underline{\mathbf{\mathring{g}}}_{\alpha
_{s}\beta _{s}}$ (\ref{6usph}) with $\ ^{\shortmid }\overline{\Lambda }=0$
for a Schwarzschild BH in conventional phase space of dimension $s^{\prime
}=7$ with $E=E_{0}=conts$. The formulas (\ref{hawtemp}) and (\ref{bekhawentr}%
) for such phase space configurations with Killing symmetry on $\partial
/\partial E$ allow us to compute the values
\begin{equation*}
\overline{S}_{0}=\frac{\overline{\Omega }_{5}(\overline{r}_{0})^{5}}{4}%
\mbox{ and }\overline{T}=\frac{1}{\pi }\left( \frac{\overline{\Omega }_{5}}{4%
}\right) ^{1/5}\overline{S}_{0}^{-1/5},
\end{equation*}%
where $\overline{\Omega }_{5}=\frac{\pi ^{7/2}}{(7/2)!}$ and $\overline{r}%
_{0}=(\frac{8\pi \overline{M}}{5\overline{\Omega }_{5}})^{1/4}.$

For relativistic stationary configurations on $\mathbf{TV,}$ the horizon
determining such values has to be restricted, warped or trapped by a maximal
speed of light constant. Unfortunately, we can not elaborate on
generalizations of formulas for $\overline{S}_{0}$ and $\overline{T}$ for
new classes of target metrics constructed as nonholonomic deformations of $%
\underline{\mathbf{\mathring{g}}}_{\alpha _{s}\beta _{s}},$ for instance, to
stationary phase configurations of type (\ref{offdiagpolfr}).

\subsubsection{Bekenstein-Hawking phase space thermodynamics for double 4-d
Schwarzschild BHs}

Let us consider a double BH $\ _{\circ \circ }^{\shortmid }g_{\alpha \beta }
$ (\ref{prm2}) defined as an exact solution of modified Einstein equations (%
\ref{meinsteqtbcand}) with zero sources when $\Lambda =0$ and $\ ^{p}\Lambda
=0.$ For such static configurations, we can define two independent
thermodynamical models (one for a base spacetime manifold and another for
typical fiber). The formulas (\ref{hawtemp}) and (\ref{bekhawentr}) for $%
s^{\prime }=4$ result in respective
\begin{eqnarray*}
\mbox{ base spacetime temperature }T &=&\frac{1}{4\pi }\left( \frac{\Omega
_{2}}{4}\right) ^{1/2}S_{0}^{-1/2}; \\
\mbox{ phase fiber  temperature }\ ^{p}T &=&\frac{1}{4\pi }\left( \frac{\
^{p}\Omega _{2}}{4}\right) ^{1/2}\ ^{p}S_{0}^{-1/2},
\end{eqnarray*}%
where respective base and fiber analogs of the Bekenstein-Hawking entropy
are computed
\begin{equation*}
S_{0}=\frac{\Omega _{2}(r_{0})^{2}}{4}\mbox{ and }\ ^{p}S_{0}=\frac{\
^{p}\Omega _{2}(\ ^{p}r_{0})^{2}}{4}.
\end{equation*}%
The volumes of unit -spheres are $\Omega _{2}=\frac{\pi ^{2}}{2!}$ and $\
^{p}\Omega _{2}=\frac{\pi ^{2}}{2!}$ and the respective horizon radiuses are
found $r_{0}=\frac{8\pi M}{2\Omega _{2}}$ and $\ ^{p}r_{0}=\frac{8\pi \ ^{p}M%
}{2\ ^{p}\Omega _{2}}$.

Such formulas can be generalized for black ellipsoid/torus solutions \cite%
{v2003be1,v2003be2,v2007bh,v2010bbh,v2013abe,v2013bbe,gvvepjc14,bubuianucqg17, vacaruepjc17,muen01}%
, but not for general off-diagonal nonholonomic deformations to phase space
stationary configurations of type (\ref{offdiagpolf2bh}).

\subsection{Geometric flows and Perelman's thermodynamics for phase spaces}

We can not apply the concepts of Bekenstein-Hawking entropy and BH
temperature for generic off-diagonal solution in GR and MGTs, for instance,
for stationary and/or cosmological configurations without explicit horizons
and/or assumptions on duality quantum fields - gravity models, holographic
principles etc. In this subsection, we provide an introduction to the theory
of nonholonomic relativistic flows on stationary phase spaces modelled as
cotangent Lorentz bundles $T^{\ast }\mathbf{V}$ of total dimension 8. Such
geometric evolution models are certain dual analogs and
Hamilton-Finsler-Ricci modifications of theories elaborated in \cite%
{vvrfthbh,muen01,vmedit,rajpsuprf,vprocmg}, see also our previous works \cite%
{vrfijmpa,vacvis,valexiu,ventrlf,vejtp1,vejtp2,vnonsymmet,vnhrf,vnrflnc,
vrfdif,bubvcjp,vrevrf}. We generalized for phase spaces the G. Perelman's
definitions of W- and F-entropies \cite{perelman1,perelman2,perelman3}
(rigorous mathematic results on of Ricci flows of Riemanian and K\"{a}hhler
metrics can be found in \cite%
{hamilt1,hamilt2,hamilt3,monogrrf1,monogrrf2,monogrrf3}).

Let us consider a family of nonholonomic 8--d cotangent Lorentz cotangent
bundles $T^{\ast }\mathbf{V}(\tau )$ enabled with corresponding sets of
metrics (and equivalent d-metrics or s-metrics for respective nonholonomic
shell distributions) $\ ^{\shortmid }\mathbf{g}(\tau )=\ ^{\shortmid }%
\mathbf{g}(\tau ,\ ^{\shortmid }u),$ or $\ _{s}^{\shortmid }\mathbf{g}(\tau
)=\ _{s}^{\shortmid }\mathbf{g}(\tau ,\ _{s}^{\shortmid }u),$ of signature $%
(+++-;+++-);$ and N--connection $\ ^{\shortmid }\mathbf{N}(\tau )=\
^{\shortmid }\mathbf{N}(\tau ,\ ^{\shortmid }u),$ or $\ _{s}^{\shortmid }%
\mathbf{N}(\tau )=\ _{s}^{\shortmid }\mathbf{N}(\tau ,\ _{s}^{\shortmid }u),$
parameterized by a positive parameter $\tau ,0\leq \tau \leq \tau _{0}.$ Any
relativistic nonholonomic phase space $T^{\ast }\mathbf{V}$ $\subset T^{\ast
}\mathbf{V}(\tau )$ can be enabled with necessary types double nonholonomic
(2+2)+(2+2) and (3+1)+(3+1) splitting (see \cite{vvrfthbh,muen01,v18c} for
the geometry and evolution of spacetimes enabled with such double
distributions). Additionally to coordinate and index conventions from
footnote \ref{fncoordinatec}, we label the local (3+1)+(3+1) coordinates in
the form
\begin{equation*}
\ ^{\shortmid }u=\{\ ^{\shortmid }u^{\alpha }=\ ^{\shortmid }u^{\alpha
_{s}}=(x^{i_{1}},y^{a_{2}};p_{a_{3}},p_{a_{4}})=(x^{\grave{\imath}%
},u^{4}=y^{4}=t;p_{\grave{a}},p_{8}=E)\}
\end{equation*}%
for $i_{1},j_{1},k_{1},...=1,2;$ $a_{1},b_{1},c_{1},...=3,4;$ $%
a_{2},b_{2},c_{2},...=5,6;$ $a_{3},b_{3},c_{3},...=7,8;$ and $\grave{\imath},%
\grave{j},\grave{k},...=1,2,3,$ respectively, $\grave{a},\grave{b},\grave{c}%
,...=5,6,7$ can be used for corresponding spacelike hyper surfaces on a base
manifold and typical cofiber.

A nonholonomic (3+1)+(3+1) splitting can be chosen in such a form that any
open region on a base Lorentz manifold, $U\subset $ $\mathbf{V,}$ is covered
by a family of 3-d spacelike hypersurfaces $\widehat{\Xi }_{t}$
parameterized by a time like parameter $t.$ On a typical cofiber of $T^{\ast
}\mathbf{V,}$ we can consider similar 3-d hypersurfaces $\ ^{\shortmid }%
\widehat{\Xi }_{E}$ of cofiber signature (+++) and parameterized by an
energy type parameter $E.$ We shall write $\ ^{\shortmid }\widehat{\Xi }=(%
\widehat{\Xi }_{t},\ ^{\shortmid }\widehat{\Xi }_{E})$ $\ $for such
nonholonomic distributions of hypersurfaces with conventional splitting 3+3
of signature (+++;+++) on total phase space. For additional shell
decompositions, we can use also a $s$-label, $\ _{s}^{\shortmid }\widehat{%
\Xi }=(\ _{s}\widehat{\Xi }_{t},\ _{s}^{\shortmid }\widehat{\Xi }%
_{E})\subset \ _{s}T^{\ast }\mathbf{V}.$ In general, there are two generic
different types of geometric phase flow theories when 1) $\tau (\chi )$ is a
re-parametrization of the \textbf{temperature} like parameter used for
labeling 4-d Lorentz spacetimes and their phase space configurations and 2) $%
\tau (t)$ is a \textbf{time} like parameter when (3+3)-d spacelike phase
configurations evolve relativistically on a "redefined" time like
coordinate. We shall elaborate in this section only on theories of type 1.

We consider generalizations of Perelman's functionals \cite{perelman1} using
canonical data $(\ ^{\shortmid }\mathbf{g}(\tau ),\ ^{\shortmid }\widehat{%
\mathbf{D}}(\tau ))$ on cotangent Lorentz bundles (for Lagrange-Finsler
geometric flow evolution, see \cite%
{vnonsymmet,vnhrf,vnrflnc,vrfdif,bubvcjp,vrevrf,valexiu}):
\begin{eqnarray}
\ ^{\shortmid }\widehat{\mathcal{F}} &=&\int_{t_{1}}^{t_{2}}\int_{\widehat{%
\Xi }_{t}}\int_{E_{1}}^{E_{2}}\int_{\ ^{\shortmid }\widehat{\Xi }_{E}}e^{-\
^{\shortmid }\widehat{f}}\sqrt{|\ ^{\shortmid }\mathbf{g}_{\alpha \beta }|}%
d^{8}\ ^{\shortmid }u(\ ^{\shortmid }\widehat{R}+|\ ^{\shortmid }\widehat{%
\mathbf{D}}\ ^{\shortmid }\widehat{f}|^{2}),  \label{ffperelmctl} \\
&&\mbox{ and }  \notag \\
\ ^{\shortmid }\widehat{\mathcal{W}} &=&\int_{t_{1}}^{t_{2}}\int_{\widehat{%
\Xi }_{t}}\int_{E_{1}}^{E_{2}}\int_{\ ^{\shortmid }\widehat{\Xi }_{E}}\
^{\shortmid }\widehat{\mu }\sqrt{|\ ^{\shortmid }\mathbf{g}_{\alpha \beta }|}%
d^{8}\ ^{\shortmid }u[\tau (\ ^{\shortmid }\widehat{R}+|\ \ _{h}^{\shortmid }%
\widehat{D}\ ^{\shortmid }\widehat{f}|+|\ \ _{v}^{\shortmid }\widehat{D}\
^{\shortmid }\widehat{f}|)^{2}+\ ^{\shortmid }\widehat{f}-16],
\label{wfperelmctl}
\end{eqnarray}%
where the normalizing function $\ ^{\shortmid }\widehat{f}(\tau ,\
^{\shortmid }u)$ satisfies $\int_{t_{1}}^{t_{2}}\int_{\widehat{\Xi }%
_{t}}\int_{E_{1}}^{E_{2}}\int_{\ ^{\shortmid }\widehat{\Xi }_{E}}\
^{\shortmid }\widehat{\mu }\sqrt{|\ ^{\shortmid }\mathbf{g}_{\alpha \beta }|}%
d^{8}\ ^{\shortmid }u=1$ for $\ ^{\shortmid }\widehat{\mu }=\left( 4\pi \tau
\right) ^{-8}e^{-\ ^{\shortmid }\widehat{f}},$ when the coeficients $%
16=2\times 8$ is taken for 8-d spaces. Similar functionals can be postulated
for nonholonomic geometric flows on $T\mathbf{V}$ using data $(\mathbf{g}%
(\tau ),\ \widehat{\mathbf{D}}(\tau ))$ and redefined integration measures
and normalized functions (conventionally without "$\ ^{\shortmid }$") on
respective hypersurfaces. For diadic shell decompositions, we can write $\
^{\shortmid }\widehat{\mathcal{F}}=\ _{s}^{\shortmid }\widehat{\mathcal{F}}$
and $\ ^{\shortmid }\widehat{\mathcal{W}}=\ _{s}^{\shortmid }\widehat{%
\mathcal{W}}$ using families of s-metrics and canonical s-connections.
Considering LC-configurations with $\ ^{\shortmid }\widehat{\mathbf{D}}%
_{\mid \ ^{\shortmid }\widehat{\mathbf{T}}=0}=\ ^{\shortmid }\nabla ,$ the
values (\ref{ffperelmctl}) and (\ref{wfperelmctl}) transform respectively
into 8-d phase space versions of the so called Perelman's F-entropy and
W-entropy. It should be noted that $\ ^{\shortmid }\widehat{\mathcal{W}}$ \
do not have a character of entropy for pseudo--Riemannian metrics but can be
treated as a value characterizing relativistic geometric hydrodynamic phase
space flows, see similar details in \cite{vvrfthbh}.

Using N--adapted diadic shell and/or double 3+1 frame and coordinate
transforms of metrics and sources with additional dependence on a flow
parameter, we introduce certain canonical parameterizations which will allow
us to decouple and solve systems of nonlinear PDEs and/or to compute entropy
like values. To define thermodynamic like variables for geometric flow
evolution of stationary configurations, we take {\small
\begin{eqnarray*}
&&\ ^{\shortmid }\mathbf{g} =\ ^{\shortmid }\mathbf{g}_{\alpha ^{\prime
}\beta ^{\prime }}(\tau ,\ _{s}^{\shortmid }u)d\ ^{\shortmid }\mathbf{e}%
^{\alpha ^{\prime }}\otimes d\ ^{\shortmid }\mathbf{e}^{\beta ^{\prime
}}=q_{i}(\tau ,x^{k})dx^{i}\otimes dx^{i}+q_{3}(\tau ,x^{k},y^{3})\mathbf{e}%
^{3}\otimes \mathbf{e}^{3}-\breve{N}^{2}(\tau ,x^{k},y^{3})\mathbf{e}%
^{4}\otimes \mathbf{e}^{4}+ \\
&&\ ^{\shortmid }q^{a_{2}}(\tau ,x^{k},y^{3},p_{b_{2}})\ ^{\shortmid }%
\mathbf{e}_{a_{2}}\otimes \ ^{\shortmid }\mathbf{e}_{a_{2}}+\ ^{\shortmid
}q^{7}(\tau ,x^{k},y^{3},p_{b_{2}},p_{b_{3}})\ ^{\shortmid }\mathbf{e}%
_{7}\otimes \ ^{\shortmid }\mathbf{e}_{7}-\ ^{\shortmid }\breve{N}^{2}(\tau
,x^{k},y^{3},p_{b_{2}},p_{b_{3}})\ ^{\shortmid }\mathbf{e}_{8}\otimes \
^{\shortmid }\mathbf{e}_{8},
\end{eqnarray*}%
} where $\ ^{\shortmid }\mathbf{e}^{\alpha _{s}}$ are N-adapted bases total
space of respective cotangent Lorentz bundles. This ansatz is a general one
for a 8--d phase space metric which can be written as an extension of a
couple of 3--d metrics, $q_{ij}=diag(q_{\grave{\imath}})=(q_{i},q_{3})$ on a
hypersurface $\widehat{\Xi }_{t},$ and $\ ^{\shortmid }q^{\grave{a}\grave{b}%
}=diag(\ ^{\shortmid }q^{\grave{a}})=(\ ^{\shortmid }q^{a_{2}},\ ^{\shortmid
}q^{7})$ on a hypersurface $\ ^{\shortmid }\widehat{\Xi }_{E},$ \ if
\begin{equation}
q_{3}=g_{3},\breve{N}^{2}=-g_{4}\mbox{ and }\ ^{\shortmid }q^{7}=\
^{\shortmid }g^{7},\ ^{\shortmid }\breve{N}^{2}=-\ ^{\shortmid }g^{8},
\label{shift1}
\end{equation}%
where $\breve{N}$ is the lapse function on the base and $\ ^{\shortmid }%
\breve{N}^{2}$ is the lapse function in the fiber.

To provide a statistical analogy for thermodynamical models, we can consider
a partition function $Z=\int \exp (-\beta E)d\omega (E)$ for the canonical
ensemble at temperature $\beta ^{-1}$ being defined by the measure taken to
be the density of states $\omega (E).$ The thermodynamical values are
computed in standard form for the average energy, $\ \left\langle
E\right\rangle :=-\partial \log Z/\partial \beta ,$ the entropy $S:=\beta
\left\langle E\right\rangle +\log Z$ and the fluctuation $\sigma
:=\left\langle \left( E-\left\langle E\right\rangle \right)
^{2}\right\rangle =\partial ^{2}\log Z/\partial \beta ^{2}.$ Using$\
^{\shortmid }\widehat{\mathcal{F}}$ (\ref{ffperelmctl}) and
\begin{equation*}
\breve{Z}=\exp \left\{ \int_{\widehat{\Xi }_{t}}\mu \sqrt{|q_{\grave{\imath}%
\grave{j}}|}d\grave{x}^{3}[-\breve{f}+8]~\right\} ,
\end{equation*}%
we can define such thermodynamic values (see \cite%
{vnhrf,vrfdif,vrevrf,vvrfthbh,muen01} proofs of similar theorems for the
dimension 8):
\begin{eqnarray}
\ _{\shortmid }\widehat{\mathcal{E}}\  &=&-\tau ^{2}\int_{\widehat{\Xi }%
_{t}}\mu \sqrt{|q_{\grave{\imath}\grave{j}}|}d\grave{x}^{3}\left( \mathbf{\ }%
_{\shortmid }\widehat{R}+|\ _{\shortmid }\widehat{\mathbf{D}}\tilde{f}|^{2}-%
\frac{3}{\tilde{\tau}}\right) ,  \label{3dthv} \\
\ _{\shortmid }\widehat{S} &=&-\int_{\widehat{\Xi }_{t}}\mu \sqrt{|q_{\grave{%
\imath}\grave{j}}|}d\grave{x}^{3}\left[ \tau \left( \ \mathbf{\ }_{\shortmid
}\widehat{R}+|\ _{\shortmid }\widehat{\mathbf{D}}\tilde{f}|^{2}\right) +%
\tilde{f}-6\right] ,  \notag \\
\ _{\shortmid }\widehat{\sigma } &=&2\ \tau ^{4}\int_{\widehat{\Xi }_{t}}\mu
\sqrt{|q_{\grave{\imath}\grave{j}}|}d\grave{x}^{3}[|\ _{\shortmid }\widehat{%
\mathbf{R}}_{\grave{\imath}\grave{j}}+\ _{\shortmid }\widehat{\mathbf{D}}_{%
\grave{\imath}}\ _{\shortmid }\widehat{\mathbf{D}}_{\grave{j}}\tilde{f}-%
\frac{1}{2\tau }q_{\grave{\imath}\grave{j}}|^{2}].  \notag
\end{eqnarray}%
These formulas can be considered for 4--d configurations taking the lapse
function $\breve{N}=1$ for N-adapted Gaussian coordinates.

Finally we note that the formulas for generalized Perelman's functionals (%
\ref{ffperelmctl}) and (\ref{wfperelmctl}) and derived thermodynamic values
(\ref{3dthv}) are witten in terms of geometric values like $\ _{\shortmid }%
\widehat{\mathbf{D}}$ and $\mathbf{\ }_{\shortmid }\widehat{R}$ all defined
by a total phase space, or projections on the base spacetime,  metric
structure. \ Such values can be defined and computed in pure geometric forms
in various types of commutative and noncommutative/ supersymmetric Finsler
geometric flow evolution models and MGTs \cite%
{vrfijmpa,vacvis,valexiu,vejtp1,vejtp2,vnonsymmet,vrevrf,vmedit,vrajpootjet,vprocmg}%
. There is also an indirect dependence on underlying theories of gravity
because the metric and (non) linear connection structures are defined
differently in MGTs. The values (\ref{ffperelmctl}), \ (\ref{wfperelmctl})
and (\ref{3dthv}) depend also on the type of solutions (general stationary
ones, BH types, locally cosmological ones) are used in an explicit example
of MGT with MDRs. In certain sense, such formulas generalize similar ones for
the Bekenstein-Hawking thermodynamics which can be defined always for
 BH solutions in various MGTs and GR. The main difference is that
the Perelman functionals can be used for characterizing thermodynamically
geometric flow evolution theories, MGTs and nonholonomic Ricci soliton
configurations, and for various classes of stationary and nonstationary, and
locally anisotropic cosmological solutions, even such geometric
configurations are not characterized by horizons or holographic and/or
duality properties.

\section{Summary and Concluding Remarks}

\label{s6} Study of the black hole, BH, solutions and their thermodynamics
is one of the important topics of theoretical physics with applications in
high energy physics, astrophysics and cosmology, and quantum information
theories. BH thermodynamics provides a relation between quantum physics and
gravitation.

In this work (a partner in a series of papers \cite%
{v18a,v18b,v18c,gvvepjc14,bubuianucqg17}), we addressed both fundamental and
phenomenological questions on BH physics in Modified Gravity Theories, MGTs,
with general modified dispersion relations, MDRs, encoding Lorentz
invariance violations, LIVs. Such theories can be geometrized on (co)
tangent Lorentz bundles with nonlinear quadratic elements and generalized
Finsler like (non) linear connections. In \cite{v18b}, we elaborated an
axiomatic approach to such MGTs which can be modelled effectively as a
generalized nonholonomic Lagrange-Hamilton dynamics. The general decoupling
and integrability of systems of nonlinear partial differential equations,
PDEs, describing such stationary configurations is proven in \cite{v18c}. In
this connection, we constructed in explicit form and studied possible
physical implications two classes of BH solutions with MDRs.

We provided details on generating solutions and studied most important
physical properties of BHs with dependence on an energy type variable (for
the 1-st class of solutions). We considered generalizations of six
dimensional, 6-d, Tangherlini type solutions \cite{tangherlini63} (analogs
of Schwarzschild -- de Sitter BH solutions) and metrics with double BH
configurations on relativistic 8-d phase spaces. For the 2-d class of
solutions, we have demonstrated that 7-d Tangherlini like BHs can be
embedded and nonholonomically generalized to stationary configurations in
8-d phase spaces. There were defined new classes of nonlinear symmetries for
BHs with MDRs. Furthermore, we proved that such solutions are characterized
by respective Finsler-Lagrange-Hamilton symmetries depending on the type of
generating functions and (effective) sources we use.

Another general goal of this article was to study the thermodynamics of BHs
with MDRs and LIVs. We proved that in particular we can elaborate analogs of
Bekenstein-Hawking thermodynamics \cite{bek1,bek2,haw1,haw2} on phase spaces
modelled as (co) tangent Lorentz bundles with conventional horizons (for
instance, of rotoid or toroid symmetry) in the total space. Nevertheless,
such an approach can not applied for more general classes of exact BH and
other type solutions with generic off-diagonal metrics and dependence on
some (or all) spacetime and total phase space coordinates. We conjectured
that we can characterize in a more general statistical thermodynamics form
various classes of exact solutions in general relativity and MGTs if we
develop for relativistic phase spaces \cite{ventrlf,vnrflnc,muen01,rajpsuprf}
the concepts of W- and F-entropy introduced by Grigory Perelman in his
research on geometric flow theory \cite{perelman1,perelman2,perelman3}.

\vskip5pt \textbf{Acknowledgments:}\ This work consists a natural extension
of the research program for the project IDEI in Romania,
PN-II-ID-PCE-2011-3-0256. Author thanks D. Singleton for hosting his adjunct
position at Fresno State University.


\begin{thebibliography}{99}
\bibitem{vap97} S. Vacaru, Locally anisotropic gravity and strings, Ann.
Phys. (NY) 256 (1997) 39-61; arXiv: gr-qc/9604013

\bibitem{vnp97} S. Vacaru, Superstrings in higher order extensions of
Finsler superspaces, Nucl. Phys. B, 434 (1997) 590 -656; arXiv:
hep-th/9611034

\bibitem{amelino98} G. Amelino-Camelia, J. R. Ellis, N. E. Mavromatos, D. V.
Nanopoulos and S. Sarkar, Potential sensitivity of gamma-ray burster
observations to wave dispersion in vacuo, Nature 393 (1998) 763-765; arXiv:
astro-ph/9712103

\bibitem{mavromatos11} N. E. Mavromatos, Quantum-gravity induced Lorentz
violation and dynamical mass generation, Phys. Rev. D 83 (2011) 025018,
arXiv: 1011.3528

\bibitem{mavromatos13a} N. E. Mavromatos and S. Sarkar, CPT-violating
leptogenesis induced by gravitational backgrounds, J, Phys. Conf. Ser. 442
(2013) 012020

\bibitem{kostelecky11} V. Alan Kostelecky and N. Russell, Data tabels for
Lorentz and CPT violations, Rev. Mod. Phys. 83 (2011) 11-31; arXiv: 0801.0287

\bibitem{kostelecky16} V. Alan Kostelecky (ed.), Proceedings, 7th Meeting on
CPT and Lorentz Symmetry (CPT 16): Boomington, Indiana, USA, June 20-24,
2016, Conference CNUM: C16-06-20.5 and Contributions, World Scientific
(Singapore, 2017) 320 pp.

\bibitem{capoz} S. Capozziello and V. Faraoni, Beyond Einstein Gravity
(Springer, Berlin, 2010)

\bibitem{nojod1} S. Nojiri and S. Odintsov, Unified cosmic history in
modified gravity: from F(R) theory to Lorentz non-invariant models, Phys.
Rept. 505 (2011) 59-144; arXiv: 1011.0544

\bibitem{nojod2} S. Nojiri, S. D. Odintsov and V. K. Oikonomou, Modified
gravity theories in nutshell: inflation, bounce and late-time evolution,
Phys. Rept. 692 (2017) 1-104; arXiv: 1705.11098

\bibitem{castro07} C. Castro, On the noncommutative and nonassociative
geometry of octonionic spacetime, modified dispersion relations and grand
unification, J. Math. Phys. 48 (2007) 073517

\bibitem{magueijo04} J. Magueijo and L. Smolin, Gravity's rainbow, Class.
Quant. Grav. 21 (2004) 1725; arXiv: gr-qc/0305055

\bibitem{nozari} K. Nozari and A. S. Sefiedgar, Comparison of approaches to
quantum correction of black hole thermodynamics, Phys. Lett. B 635 (2006)
156-160, arXiv: gr-qc/061116

\bibitem{sefiedgar} A. S. Sefiedgar and H. R. Sepangi, Brane-world black
hole entropy from modified dispersion relations, Phys. Lett. B 692 (2010)
281-285; arXiv: 1007.4904

\bibitem{bek1} J. D. Bekenstein, Black holes and entropy, Phys. Rev. D 7
(1973) 2333-2346

\bibitem{bek2} J. D. Bekenstein, Generalized second law of thermdynamics in
black hole physics, Phys. Rev. D 9 (1974) 3292-3300

\bibitem{haw1} S. W. Hawking, Particle creation by black holes, Commun.
Math. Phys. 43 (1975) 199-220, Erratum: 46 (1976) 2006

\bibitem{haw2} S. W. Hawking, Black holes and thermodynamics, Phys. Rev. D
13 (1976) 191-197

\bibitem{perelman1} G. Perelman, The entropy formula for the Ricci flow and
its geometric applications, arXiv: math. DG/0211159

\bibitem{perelman2} G. Perelman, Ricci flow with surgery on
three--manifolds, arXiv: math.DG/0303109

\bibitem{perelman3} G. Perelman, Finite extintion time for the solutions to
the Ricci flow on certain three-manifolds, arXiv: math.DG/0307245

\bibitem{vrfdif} S. Vacaru, Diffusion and self-organized criticality in
Ricci flow evolution of Einstein and Finsler spaces, Symmetry: Culture and
Science, 23 (2013) 105-124, arXiv: 1010.2021

\bibitem{vvrfthbh} V. Ruchin, O. Vacaru and S. Vacaru, On relativistic
generalization of Perelman's W-entropy and statistical thermodynamic
description of gravitational fileds, Eur. Phys. J. \textbf{C 77} (2017) 184,
arXiv: 1312.2580

\bibitem{ventrlf} S. Vacaru, The entropy of Lagrange-Finsler spaces and
Ricci flows, Rep. Math. Phys. 63 (2009) 95-110; arXiv: math.DG/ 0701621

\bibitem{vnrflnc} S. Vacaru, Spectral functionals, nonholonomic Dirac
operators, and noncommutative Ricci flows, J. Math. Phys. \textbf{\ 50 }
(2009) 073503, arXiv: 0806.3814

\bibitem{muen01} T. Gheorghiu, V. Ruchin, O. Vacaru and S. Vacaru, Geometric
flows and Perelman's thermodynamics for black ellipsoids in $R^{2}$ and
Einstein gravity theories, Annals of Phys. NY \textbf{\ 369} (2016) 1-35,
arXiv: 1602.08512

\bibitem{rajpsuprf} S. Rajpoot and S. Vacaru, On supersymmetric geometric
flows and R2 inflation from scale invariant supergravity, Annals of Physics,
NY, 384 (2017) 20-60; arXiv: 1606.06884

\bibitem{v18a} L. Bubuianu and S. Vacaru, Deforming black hole and
cosmological solutions by quasiperiodic and/or pattern forming structures in
modified and Einstein gravity, Eur. Phys. J. C 78 (2018) 393; arXiv:
1706.02584

\bibitem{v18b} S. Vacaru, On axiomatic formulation of gravity and matter
field theories with MDRs and Finsler-Lagrange-Hamilton geometry on (co)
tangent Lorentz bundles, arXiv: 1801.064444v1

\bibitem{v18c} L. Bubuianu and S. Vacaru, Quasi--stationary solutions in
gravity theories with modified dispersion relations and
Finsler-Lagrange-Hamilton geometry, arXiv: 1806.04500

\bibitem{gvvepjc14} T. Gheorghiu, O. Vacaru and S. Vacaru, Off-diagonal
deformations of Kerr black holes in Einstein and modified massive gravity
and Higher Dimensions, Eur. Phys. J. C 74 (2014) 3152; arXiv: 1312.4844

\bibitem{bubuianucqg17} L. Bubuianu, K. Irwin and S. Vacaru, Heterotic
supergravity with internal almost-Kaehler spaces; instantons for SO(32), or
E8 x E8, gauge groups; and deformed black holes with soliton, quasiperiodic
and/or pattern-forming structures, Class. Quant. Grav. 34 (2017) 075012;
arXiv: 1611.00223

\bibitem{vacaruepjc17} S. Vacaru and K. Irwin, Off-diagonal deformations of
Kerr metrics and black ellipsoids in heterotic supergravity, Eur. Phys. J. C
77 (2017) 17; arXiv: 1608.01980

\bibitem{v1998bh} S. Vacaru, Exact solutions in locally anisotropic gravity
and strings, in: "Particile, Fields and Gravitations", ed. J. Rembielinski,
AIP Conference Proceedings, No. 453, American Institute of Physics,
(Woodbury, New York) 1998, p. 528-537; arXiv: gr-qc/9806080

\bibitem{v2001bh} S. Vacaru, Anholonomic soliton-dilaton and black hole
solutions in general relativity, JHEP, 04 (2001) 009; arXiv: gr-qc/0005025

\bibitem{v2002bh} S. Vacaru and D. Singleton, Warped solitonic deformations
and propagation of black holes in 5D vacuum gravity, Class. Quant. Grav. 19
(2002) 3583-3602; arXiv: hep-th/0112112

\bibitem{v2003be1} S. Vacaru, Horizons and geodesics of black ellipsoids,
Int. J. Mod. Phys. D. 12 (2003) 479-494; arXiv: gr-qc/0206014

\bibitem{v2003be2} S. Vacaru, Perturbations and stability of black
ellipsoids, Int. J. Mod. Phys. D 12 (2003) 461-478; arXiv: gr-qc/0206016

\bibitem{v2007bh} S. Vacaru, Parametric nonholonomic frame transforms and
exact solutions in gravity, Int. J. Geom. Meth. Mod. Phys. 4 (2007)
1285-1334; arXiv: 0704.3986

\bibitem{v2010abh} S. Vacaru, On general solutions in Einstein and high
dimensional gravity, Int. J. Theor. Phys. 49 (2010) 884-913; arXiv:
0909.3949v4

\bibitem{v2010bbh} S. Vacaru, Finsler black holes induced by noncommutative
anholonomic distributions in Einstein gravity, Class. Quant. Grav. 27 (2010)
105003 (19pp); arXiv: 0907.4278

\bibitem{v2011fbr} S. Vacaru, Finsler branes and quantum gravity
phenomenology with Lorentz symmetry violations, Class. Quant. Grav. 28
(2011) 215991; arXiv: 1008.4912

\bibitem{v2013abe} S. Vacaru, Black holes, ellipsoids, and nonlinear waves
in pseudo-Finsler spaces and Einstein gravity, Int. J. Theor. Physics 52
(2013) 1654-1681; arXiv: 0905.4401

\bibitem{v2013bbe} S. Vacaru, Hidden symmetries for ellipsoid-solitonic
deformations of Kerr-Sen black holes and quantum anomalies, Eur. Phys. J. C
73 (2013) 2287; arXiv: 1106.1033

\bibitem{v2014abh} P. Stavrinos, O. Vacaru and S. Vacaru, Off-diagonal
solutions in modified Einstein and Finsler theories on tangent Lorentz
bundles, Int. J. Mod. Phys. D 23 (2014) 1450094; arXiv: 1401.2879

\bibitem{v2014bbh} S. Vacaru, E. Veli Veliev and Enis Yazici, A geometric
method of constructing exact solutions in modified f(R,T) gravity with
Yang-Mills and Higgs Interactions, Int. J. Geom. Meth. Mod. Phys. 11 (2014)
1450088; arXiv: 1411.2849

\bibitem{jacobson} T. Jacobson, Thermodynamics of space-time: The Einstein
equation of state, Phys.\ Rev. Lett. 75 (1995) 1260-1263, arXiv:
gr-qc/9504004

\bibitem{padman} T. Padmanabhan, \ General relativity from a thermdynamic
perspective, Gen. Gel. Grav. 46 (2014) 1673, arXiv: 1302.3253

\bibitem{verlinde} E. P. Verlinde, On the origin of gravity and the laws of
Newton, JHEP 1104 (2011) 029, arXiv: 1001.0785

\bibitem{yang} Hyun Seok Yang and M. Sivakumar, Emergent gravity from
quantized spacetime, Phys. Rev. D82 (2010) 045004, arXiv: 0908.2809

\bibitem{emparan} R. Emparan and H. S. Reall, Black holes in higher
dimensions, Living Rev. Rel. 11 (2008) 6

\bibitem{pappas16} T. Pappas, P. Kanti and N. Pappas, Hawing radiation
spectra for scalar field by a higher-dimensional Schwarzschild-de-Sitter
black hole, Phys. Rev. D 94 (2016) 024035; arXiv: 1604.08617

\bibitem{vrfijmpa} S. Vacaru, Ricci flows and solitonic pp-waves, Int. J.
Mod. Phys. \textbf{\ A21 } (2006) 4899-4912, arXiv: hep-th/ 0602063

\bibitem{vacvis} S. Vacaru and M. Visinescu, Nonholonomic Ricci flows and
running cosmological constant: I. 4D Taub-NUT metrics, Int. J. Mod. Phys.
\textbf{\ A22 } (2007) 1135-1159, arXiv: gr-qc/ 0609085

\bibitem{valexiu} Maria Alexiou, P. Stavrinos and S. Vacaru, Nonholonomic
Ricci Flows of Riemann Metrics and Lagrange-Finsler Geometry, J. Phys. Math.
7 (2016) 2 [14 pages]; see previous versions in arXiv: math.DG/ 0612162

\bibitem{vejtp1} S. Vacaru, Nonholonomic Ricci flows and parametric
deformations of the solitonic pp-waves and Schwarzschild solutions,
Electronic Journal of Theoretical Physics (EJTP) 6, N21 (2009) 63-93;
http://www.ejtp.com/articles/ejtpv6i21p63.pdf; arXiv: 0705.0729 [math-ph]

\bibitem{vejtp2} S. Vacaru, Nonholonomic Ricci flows: exact solutions and
gravity, Electronic Journal of Theoretical Physics (EJTP) 6, N20, (2009)
27---58; http://www.ejtp.com/articles/ejtpv6i20p27.pdf; arXiv: 0705.0728
[math-ph]

\bibitem{vnonsymmet} S. Vacaru, Nonholonomic Ricci flows, exact solutions in
gravity, and symmetric and nonsymmetric metrics, \emph{\ }Int. J. Theor.
Phys.\emph{\ } \textbf{\ 48 } (2009) 579-606, arXiv: 0806.3812

\bibitem{vnhrf} S. Vacaru, Nonholonomic Ricci flows: II. Evolution equations
and dynamics, \emph{\ }J. Math. Phys.\emph{\ } \textbf{\ 49 } (2008) 043504,
arXiv: math.DG/0702598

\bibitem{vrfactrf} S. Vacaru, Fractional nonholonomic Ricci flows, Chaos,
Solitons \& Fractals 45 (2012) 1266-1276; arXiv: 1004.0625 [math.DG]

\bibitem{bubvcjp} L. Bubuianu and S. Vacaru, Dynamical Equations and
Lagrange-Ricci Flow Evolution on Prolongation Lie Algebroids, Can. J. Phys.
(2018), arXiv: 1108.4333v3

\bibitem{vrevrf} S. Vacaru, Metric compatible or noncompatible Finsler-Ricci
flows, Int. J. Geom. Meth. Mod. Phys. 9 (2012) 1250041 (26 pages); arXiv:
1106.4888 [physics.gen-ph]

\bibitem{vmedit} S. Vacaru, Almost Kaehler Ricci flows and Einstein and
Lagrange-Finsler structures on Lie algebroids, \emph{\ }Medit. J. Math.\emph{%
\ } \textbf{\ 12 } (2015) 1397-1427, arXiv: 1306.2813

\bibitem{vrajpootjet} S. Rajpoot and S. Vacaru, Nonholonomic jet
deformations, exact solutions for modified Ricci soliton and Einstein
equations, Int. J. Geom. Meth. Mod. Phys. \textbf{\ 14 } (2016) 1750032,
arXiv: 1411.1861

\bibitem{vprocmg} M. Alexiou, T. Gheorghiu, P. Stavrinos, O. Vacaru, and S.
Vacaru, Nonholonomic Ricci flows and Finsler-Lagrange f(R,F,L)-modified
gravity and dark matter effects, in: Proceedings, 14th Marcel Grossmann
Meeting on Recent Developments in Theoretical and Experimental General
Relativity, Astrophysics, and Relativistic Field Theories (MG14), Rome,
Italy, July 12-18, 2015, Conference C15-07-12 (2017) 2371-2375

\bibitem{friedan1} D. Friedan, Nonlinear models in $2+\varepsilon $
dimensions, PhD Thesis (Berkely) LBL-11517, UMI-81-13038, Aug 1980. 212pp

\bibitem{friedan2} D. Friedan, Nonlinear models in $2+\varepsilon $
dimensions, \emph{\ }Phys. Rev. Lett.\emph{\ } \textbf{45} (1980) 1057-1060

\bibitem{friedan3} D. Friedan, Nonlinear models in $2+\varepsilon $
dimensions, \ Ann. of Physics\emph{\ } \textbf{163} (1985) 318-419

\bibitem{hamilt1} R. S. Hamilton, Three-manifolds with postive Ricci
curvature, \emph{\ }J. Diff. Geom.\emph{\ } \textbf{\ 17 } (1982) 255-306

\bibitem{hamilt2} R. S. Hamilton, The Ricci flow on surfaces, in:
Mathematics and General Relativity, \emph{\ }Contemp. Math. \textbf{\ 71},
p. 237-262, Amer. Math. Soc., Providence, 1988

\bibitem{hamilt3} R. S. Hamilton, in: Surveys in Differential Geometry, vol.
2 (International Press, 1995), pp. 7-136

\bibitem{monogrrf1} H. -D. Cao and H. -P. Zhu, A complete proof of the
Poincar\'{e} and geometrization conjectures - application of the
Hamilton--Perelman theory of the Ricci flow, Asian J. Math. \textbf{10}
(2006) 165-495

\bibitem{monogrrf2} J. W. Morgan and G. Tian, Ricci flow and the Poincar\'{e}
conjecture, AMS, Clay Mathematics Monogaphs, vol. 3 (2007)

\bibitem{monogrrf3} B. Kleiner and J. Lott, Notes on Perelman's papers,
Geometry \& Topology \textbf{12} (2008) 2587-2855

\bibitem{tsey} A. A. Tseytlin, On sigma model RG flow, "central charge"
action and Perelman's entropy, Phys. Rev. \textbf{\ D75 } (2007) 064024,
arXiv: hep-th/0612296

\bibitem{tangherlini63} F. R. Tangherlini, Shwarzschild field in n
dimensions and the dimensionality of space problem, Nuovo Cim. 27 (1963)
636-651

\bibitem{myers} R.\ C. Meyers and M. J. Perry, Black holes in higher
dimensional space-times, Annals Phys. 172 (1986) 304

\bibitem{konoplya1} R. A. Konoplya and A. Zhidenko, Stability of higher
dimensional Reissner-Nordstom-anti-de Sitter black holes, Phys. Rev. D78
(2008) 104017, arXiv: 0809.2048

\bibitem{konoplya2} R. A. Konoplya and A. Zhidenko, Stability of
multidimensional black holes: complete numerical analysis, Nucl. Phys. B 777
(2007) 182-202, arXiv: hep-th/0703231

\bibitem{gao} S. Gao and J. P. S. Lemos, Collapsing and static thin massive
charged dust shells in a Reissner-Nordstr\"{o}m black holes background in
higher dimensions, Int. J. Mod. Phys. A 23 (2008) 2943-2960, arXiv: 0804.0295

\bibitem{yangwen} S.Z. Yang, D. Wen and K. Lin, Fermions tunneling form
higher-dimensional Reissner-Nordstr\"{o}m black hole: Semiclassical and
beyond semiclassical approximation, Adv. High Energy Phys. 2016 (2016)
4647069, arXiv: 0903.1983

\bibitem{pourhassan} B. Pourhassan, K. Kokabi and S. Rangyan, Thermodynamics
of higher dimensional black holes with higher order thermal fluctuations,
Gen. Rel. Gravit. 49 (2017) 144-171, arXiv: 1710.06299

\bibitem{hollands} S. Hollands and A. Ishibashi, Black hole uniquenes
theorems in higher dimensional spacetimes, Class. Quant. Grav. 29, n. 16
(2012), topical review, arXiv: 1206.1164

\bibitem{gangop} S. Gangopandyay and A. Dutta, Constraints on rainbow
gravity functions from black hole thermodynamics, Europ. Phys. Lett. 115
(2016) 50005, arXiv: 1606.08295

\bibitem{basil1} S. Basilakos, A. P. Kouretsis, E. N. Saridakis and P.
Stavrinos, Resembiling dark energy and modified gravity with Finsler-Randers
cosmology, Phys. Rev. D 83 (2013) 123510, arXiv: 1311.5915

\bibitem{basil2} S. Basilakos and P. Stavrinos, Cosmological equivalence
between the Finsler-Randers space-time and the DGP gravity model, Phys. Rev.
D 87 (2013) 043506, arXiv: 1301.4327
\end{thebibliography}
\end{document}